\documentclass[aps,prl,twocolumn]{revtex4-1}
\usepackage[T1]{fontenc}
\usepackage[latin1]{inputenc}
\usepackage{lmodern}
\expandafter\let\csname equation*\endcsname\relax
\expandafter\let\csname endequation*\endcsname\relax
\usepackage{amsmath}
\usepackage{esint}
\usepackage{amssymb}
\usepackage{graphicx}
\usepackage{xcolor}
\usepackage{natbib}
\usepackage{bm}
\usepackage{bbold}
\usepackage[mathscr]{euscript}
\usepackage[cal=boondoxo]{mathalfa}
\usepackage{comment}

\def\ln{\textrm{ln}}

\def\pM{\mathrel{\raise 2pt \hbox{\tiny(}\!\raise 1pt \hbox{+}\settowidth {\dimen03} {+}\hskip-\dimen03 \raise -2.4pt \hbox {$-$} \!\raise 2pt \hbox{\tiny)}}}

\begin{document}
\title{Degradation of phonons in disordered moir\'e superlattices}
\author{H\'ector Ochoa$^{1,2,3}$ and Rafael M. Fernandes$^4$}
\affiliation{$^1$Donostia International Physics Center, 20018 Donostia-San Sebastian, Spain\\
$^2$IKERBASQUE, Basque Foundation for Science, Maria Diaz de Haro 3, 48013 Bilbao, Spain\\
$^3$Department of Physics, Columbia University, New York, NY 10027, USA\\
$^4$School of Physics and Astronomy, University of Minnesota, Minneapolis, MN 55455, USA}

\begin{abstract}
The elastic collective modes of a moir\'e superlattice arise not from vibrations of a rigid crystal but from the relative displacement between the constituent layers. Despite their similarity to acoustic phonons, these modes, called phasons, are not protected by any conservation law. Here, we show that disorder in the relative orientation between the layers and thermal fluctuations associated with their sliding motion degrade the propagation of sound in the moir\'e superlattice. Specifically, the phason modes become overdamped at low energies and acquire a finite gap, which displays a universal dependence on the twist-angle variance. Thus, twist-angle inhomogeneity is manifested not only in the non-interacting electronic structure of moir\'e systems, but also in their phonon-like modes. More broadly, our results have important implications for the electronic properties of twisted moir\'e systems that are sensitive to the electron-phonon coupling.
\end{abstract}
\maketitle

\textit{Introduction}. The discovery of twisted moir\'e systems has opened a new route to investigate correlated-electron and topological effects in highly-tunable narrow bands \cite{Balents_review}, both experimentally \cite{Cao2018a,Cao2018b,Yankowitz1059,Sharpe19,STM_Andrei19,STM_Pasupathy19,Efetov19,STM_Yazdani19,STM_Yazdani19,STM_Perge19,Young19,Pablo_nematics,Pomeranchuk1,Pomeranchuk2} and theoretically \cite{Xu2018a,Po2018,Isobe2018,Kennes2018,Rademaker2018,Dodaro2018,Thomson2018,Lin2018,Guinea2018,Sherkunov2018,Liu2018,Venderbos18,Song2019,Kang2019,Bascones19,Tarnopolsky2019,Roy2019,Nandkishore2019,Balents19,Huang2019,Zhang2019a,Gonzalez2019,Classen19,Uchoa19,Yuan2019magic,Vafek20,MacDonald20,Repellin20,Cenke_nematics,Christos20,Bultinck20,Vafek20_RG,FCZhang20,Cea20,Savary20,Bernevig_TBG3,Bernevig_TBG4,Bernevig_TBG5,ZYMeng21,Wang2021,Potasz2021,Khalaf21,Vafek21,Chichinadze21}. In the case of twisted bilayer graphene (TBG), several works have proposed that the electron-phonon interaction plays an important role in shaping the phase diagram, either by acting in tandem with strong electron-electron correlations or by possibly driving instabilities on its own, such as superconductivity \cite{Wu_etal,Bernevig_phonons,Wu_etal2,Fabrizio19,Lewandoski_etal,Cea_Guinea,moire_nematicity}. Recent experiments in double-gated devices report superconductivity even when the Coulomb interaction is strongly screened \cite{exp1,exp2,exp3}, which might be an indication of electron-phonon coupling playing a prominent role in the emergence of superconductivity. Progress in this problem thus requires the elucidation of the lattice excitations at the energy and length scales of the moir\'e superlattice. While they are inherited from the phonon modes of the individual graphene layers, a full description is complicated by the adhesion forces between them and by the intrisic inhomogeneities of the relative displacement (heterostrain) and relative orientation (twist angle) of the layers. 

To capture the low-energy electronic properties of TBG, one often considers electronic states that live on the sites and bonds of the triangular moir\'e superlattice (or of its dual) \cite{Koshino2018,Kang2018,Yuan2018,Zou2018,Po2019}. Analogously, to describe the low-energy phonons, it is convenient to focus on the collective excitations of the moir\'e lattice itself \cite{Koshino_phonons,phasons}, rather than on the lattice vibrations of the individual graphene layers. We focus on the in-plane motion as flexural modes are not expected to be very different than in monolayer graphene \cite{Koshino_phonons}. Importantly, the moir\'e pattern is a sixfold symmetric incommensurate superlattice, and not a rigid crystal. As a result, its low-energy elastic excitations are not described by acoustic phonons, but by so-called phasons \cite{phasons,phasons2} -- similarly to quasicrystals (see also \cite{Gaa2021}). The crucial difference is that while the dispersion of acoustic phonons is governed by the conservation of linear momentum of the ions of a rigid lattice, phasons in TBG are related to the relative translation between the layers, which is not a symmetry of the system since the layers are subjected to adhesion forces.

In this paper, we show that the phasons' dispersion is qualitatively altered by disorder and by anharmonic vibrations of the underlying graphene layers. In particular, we show that random forces affecting the relative orientation and displacement between the layers, no matter how small they are, give rise to a characteristic length scale $L_c$ beyond which stacking order is lost. Thus, $L_c$, which depends on the elastic constants of graphene and on the strength of the disorder potentials, is the length scale associated with twist-angle inhomogeneity. Experimentally, inhomogeneous twist angles have been widely observed in TBG devices \cite{Zeldov,Milan2021,nat_mat}, and shown to strongly affect the electronic properties (see also \cite{Pixley2019,Ryu2020}). Within a region of size $L_c$, the distribution of twist angles $\theta$ has a variance:

\begin{align}
\label{eq:variance}
\frac{\overline{\delta\theta^2}}{\overline{\theta}^2}\sim\frac{L_{\textrm{m}}^2}{\pi^2L_c^2}\,\ln\left(\frac{L_c}{L_{\textrm{m}}}\right),
\end{align}
where the bar denotes disorder average, $\overline{\mathcal{O}}\equiv\langle \mathcal{O} \rangle_{\textrm{dis}}$, $\delta \theta \equiv \theta - \overline{\theta}$, and $L_{\textrm{m}}$ is the period of the moir\'e pattern. Thus, while for a twist angle variation of $1 \%$, $L_c \gg L_{\textrm{m}}$, for variations of $10 \%$, $L_c \gtrsim L_{\textrm{m}}$. 

The twist-angle inhomogeneity scale $L_c$ also gaps out the phason dispersion, introducing a new relevant energy scale of the order of $\delta \omega \sim \omega_{\rm{ZB}} L_{\rm{m}}/L_c$, where  $\omega_{\rm{ZB}}$ is the acoustic phonon frequency at the moir\'e Brillouin zone boundary. Combining with Eq. (\ref{eq:variance}), we find an implicit and universal relationship between the twist angle variance and the phason-dispersion gap. Thus, twist angle inhomogeneity is not only manifested in the non-interacting electronic structure, but also in the low-energy elastic properties of TBG. 

A gap opening in the phasons' dispersion is consistent with the fact that they are not protected by an underlying conservation law. For the same reason, the phasons' low-energy propagating dynamics is not robust either. Here, we demonstrate that anharmonic excitations introduce a damping term for the phason modes that does not vanish in the long-wavelength limit. Consequently, the phasons dynamics changes from propagating to diffusive. The typical scattering rate $\tau^{-1}$ increases substantially for decreasing twist angle as $\tau  \sim \overline{\theta}^3$, highlighting the importance of this effect for small twist angles.

\begin{figure}[t!]
\begin{center}
\includegraphics[width=\columnwidth]{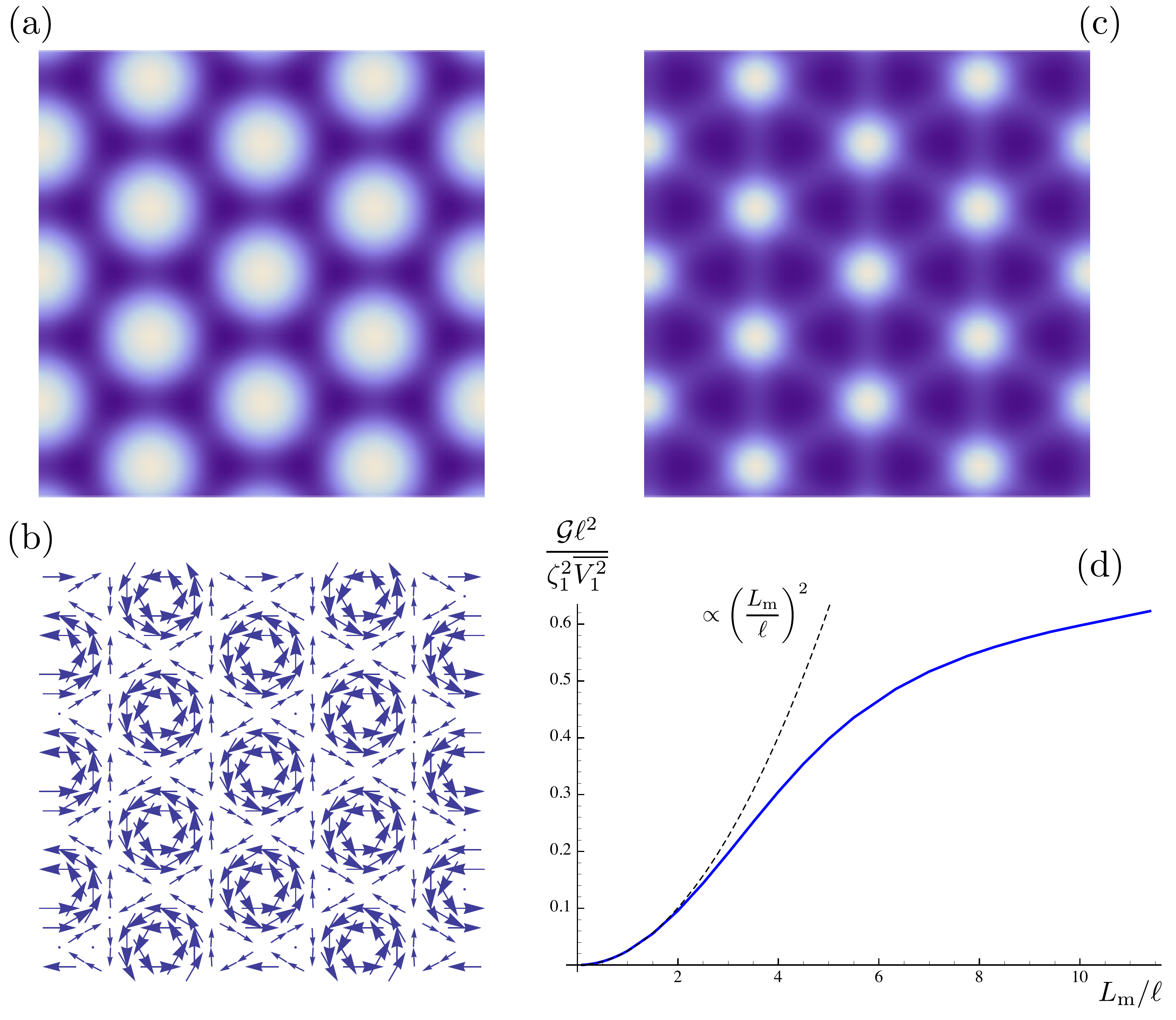}
\caption{\textbf{Stacking texture and disorder strength}. (a) Amplitude of the mass density wave generated by a rigid rotation of the layers. Bright spots correspond to areas of large overlap (AA-like stacking), dark areas are AB/BA minima. (b) Displacement field $\boldsymbol{u}_0(\mathbf{r})$ for $\bar{\theta}=1.05^{\textrm{o}}$.  (c) The same as in (a), but now including the lattice relaxation in (b). (d) Long-wavelength limit of twist-angle disorder correlator $\mathcal{G}$ as a function of $L_{\textrm{m}}/\ell\propto\overline{\theta}^{-1}$.} 
\label{fig:textures}
\end{center}
\end{figure}

\textit{Stacking order and stacking susceptibility}. Starting from a local, continuum approximation, the TBG interlayer tunneling Hamiltonian is given by \cite{dosSantos2007,BM},
\label{eq:tunneling}
\begin{align}
\hat{H}_{\textrm{inter}}=\sum_{\zeta=\pm 1}\int d\mathbf{r}\,\hat{\psi}_{\zeta,\textrm{t}}^{\dagger}\left(\mathbf{r}\right) \hat{T}_{\zeta}\left(\mathbf{r}\right) \hat{\psi}_{\zeta,\textrm{b}}\left(\mathbf{r}\right) + \textrm{h.c.}
\end{align}
Here $\hat{\psi}_{\zeta,\textrm{b(t)}}$ is a Dirac spinor field defined in the sublattice space of bottom (top) layer around valleys $\mathbf{K}_{\zeta}=\zeta\, \mathbf{K}_{+}$ labelled by the chirality index $\zeta=\pm1$. $\mathbf{K}_{+}=(\boldsymbol{g}_1-\boldsymbol{g}_2)/3$ is located at one of the corners of graphene's Brillouin zone, with $\boldsymbol{g}_{1,2}$ denoting the primitive vectors of the reciprocal lattice. The local tunneling matrix is
\begin{align}
 \hat{T}_{\zeta}\left(\mathbf{r}\right) =\sum_{n=0}^2e^{i\zeta\left(\mathbf{q}_n+\mathbf{K}_{+}\right)\cdot\boldsymbol{\phi}\left(\mathbf{r}\right)}\,e^{i\zeta\frac{2\pi n}{3}\hat{\sigma}_z}\,\hat{\rm{T}}_{\zeta}\,e^{-i\zeta\frac{2\pi n}{3}\hat{\sigma}_z},
 \end{align}
where $\mathbf{q}_{0,1,2}=\mathbf{0},\boldsymbol{g}_2,-\boldsymbol{g}_1$ are momentum transfers between equivalent Dirac points \cite{dosSantos2007,BM}, and $\hat{\rm{T}}_{\zeta}=w_{\textrm{AA}}\,\hat{1}+w_{\textrm{AB}}\,\hat{\sigma}_x$ contains the inter-layer tunneling amplitude $w_{\textrm{AA}}$ ($w_{\textrm{AB}}$) involving the same (opposite) sublattices. In this equation, $\boldsymbol{\phi}(\mathbf{r})$ describes the spatial modulation of the pattern resulting from the overlap between the two layers (see Supplemental Material \cite{SM}):
\begin{align}
 \boldsymbol{\phi}\left(\mathbf{r}\right)=2\sin\frac{\theta}{2}\,\mathbf{\hat{z}}\times\mathbf{r}+\boldsymbol{u}_{0}\left(\mathbf{r}\right)+\delta\boldsymbol{\phi}\left(\mathbf{r}\right).
 \label{eq:staking_order}
\end{align}
Hereafter, we dub it the local \textit{stacking order} function. If $\boldsymbol{\phi} = 0$, one would obtain a uniform AA stacking configuration and no moir\'e pattern. It is the twist angle $\theta$, which appears in the first term of the equation above, that yields a sixfold symmetric moir\'e pattern with alternating AA and AB/BA stacking regions, as illustrated in Fig.~\ref{fig:textures}~(a) (light and dark areas, respectively). However, this term corresponds to a rigid rotation between the layers, which in practice is never realized due to the non-negligible adhesion potential between the layers. For a nominal twist angle $\theta\rightarrow\overline{\theta}$, the free energy is minimized by a relative displacement between the layers,  $\boldsymbol{u}_0(\mathbf{r})$. It is shown in Fig.~\ref{fig:textures}~(b) for a relaxed structure around the magic angle, $\overline{\theta}=1.05^{\textrm{o}}$, yielding the moir\'e pattern of Fig.~\ref{fig:textures}~(c) \cite{SM} (see also Refs.~\onlinecite{relax1,relax2,relax3}). Note that by keeping a non-zero $\overline{\theta}$ we are implicitly incorporating the action of lateral forces needed to stabilize the moir\'e pattern over the Bernal stacking. Determining the origin of these lateral forces require first-principles calculations that are beyond the scope of this work \cite{ML_methods}. The main focus of this Letter is the last term in Eq.~\eqref{eq:staking_order}, $\delta\boldsymbol{\phi}\left(\mathbf{r}\right)$, which describes local stacking deviations around the local minimum of the mechanical energy.

Mechanical forces acting on the layers will naturally give rise to a finite stacking deviation $\delta\boldsymbol{\phi}\left(\mathbf{r}\right)$. These forces can be extrinsic (e.g. applied strain) or intrinsic (e.g. due to random strain or thermal fluctuations). They are generically described by a function $\boldsymbol{f}\left(\mathbf{r}\right)$, which in turn can be decomposed into a relative stress and a relative torque between the layers, causing changes in their relative displacement and orientation, respectively. In frequency domain, the stacking deviation caused by such a force is given by \begin{align}
\label{eq:linear_response}
\delta\boldsymbol{\phi}\left(\mathbf{r},\omega\right)=\int d^2\mathbf{r}'\,\,\hat{\chi}_0\left(\mathbf{r},\mathbf{r}',\omega\right)\cdot\boldsymbol{f}\left(\mathbf{r}',\omega\right).
\end{align}
where we introduced the dynamic stacking susceptibility tensor $\hat{\chi}_0(\mathbf{r},\mathbf{r}',\omega)$. 

We first focus on the static case. Due to the (approximate) translational symmetry of the moir\'e superlattice, the stacking susceptibility can be parametrized in momentum space as $\chi_{0,(i,j);(\mathbf{G}_1,\mathbf{G}_2)}\left(\mathbf{q}\right)$, with $\mathbf{q}$ restricted to the moir\'e Brillouin zone and $\mathbf{G}_i=-2\sin\frac{\overline{\theta}}{2}\,\mathbf{\hat{z}}\times\boldsymbol{g}_i$ the vectors of the moir\'e reciprocal lattice. One can directly compute it from the ``mechanical'' free-energy functional of TBG, which includes the intrinsic elastic contributions from the individual layers and the adhesion potential $\mathcal{V}_{\textrm{ad}}$ between them. We obtain \cite{SM}:
\begin{widetext}
\begin{align}
\label{eq:harmonic_spectrum}
\left[\hat{\chi}_0^{-1}\left(\mathbf{q}\right)\right]_{i,j;\mathbf{G}_1,\mathbf{G}_2}=\frac{\lambda+\mu}{2}\left(\mathbf{q}+\mathbf{G}_1\right)_i\left(\mathbf{q}+\mathbf{G}_2\right)_j\delta_{\mathbf{G}_1,\mathbf{G}_2}+\frac{\mu}{2}\left(\mathbf{q}+\mathbf{G}_1\right)\cdot\left(\mathbf{q}+\mathbf{G}_2\right)\delta_{ij}\delta_{\mathbf{G}_1,\mathbf{G}_2}+\mathcal{V}_{ij}^{(2)}\left(\mathbf{G}_2-\mathbf{G}_1\right),
\end{align}
\end{widetext}
where $\mu=9.57$ eV/\AA$^{2}$ and $\lambda\approx 3.25$ eV/\AA$^2$ are graphene's Lam\'e coefficients \cite{elastic_constants} and $\mathcal{V}_{ij}^{(2)}(\mathbf{G})$ are the Fourier components of the harmonic expansion of the adhesion potential $\mathcal{V}_{\textrm{ad}}$,\begin{align} 
\label{eq:V_adhesion}
\mathcal{V}_{\textrm{ad}}\left(\boldsymbol{\phi}\right)=\frac{V_{\textrm{AA}}}{3}+\frac{2V_{\textrm{AA}}}{9}\sum_{i=1}^{3}\cos\left(\boldsymbol{g}_i\cdot\boldsymbol{\phi}\right),
\end{align}
where $\boldsymbol{g}_3=-\boldsymbol{g}_1-\boldsymbol{g}_2$. The key quantity here is $V_{\textrm{AA}}\approx 4$ meV/\AA$^{2}$ \cite{Carr}, which is the free energy difference between AA and AB/BA stacking configurations.

Diagonalization of Eq.~\eqref{eq:harmonic_spectrum} determines the spectrum of harmonic oscillations around the minimum energy configuration, $\boldsymbol{\phi}_{0}(\mathbf{r})= 2\sin\frac{\overline{\theta}}{2}\,\mathbf{\hat{z}}\times\mathbf{r}+\boldsymbol{u}_{0}(\mathbf{r})$, which is controlled by the ratio of the two length scales in the problem: the moir\'e pitch $L_{\textrm{m}}=a/(2\sin\frac{\overline{\theta}}{2})$ and the characteristic width of stacking domain walls connecting degenerate AB and BA minima (dark regions in Fig. \ref{fig:textures}c),
\begin{align}
\label{eq:width}
\ell=\frac{3a}{4\pi}\sqrt{\frac{\mu}{V_{\textrm{AA}}}}\approx 3\,\text{nm}.
\end{align}
The ratio $L_{\textrm{m}}/\ell\propto\overline{\theta}^{-1}$ characterizes the amount of lattice relaxation, i.e., how sharp $\boldsymbol{\phi}_{0}(\mathbf{r})$ is on the moir\'e scale. The spectral decomposition of the stacking susceptibility then reads\begin{align} \label{eq:chi0}
\hat{\chi}_0\left(\mathbf{q}\right)=2\sum_{n}\frac{\hat{\boldsymbol{e}}_{n}\left(\mathbf{q}\right)\otimes\hat{\boldsymbol{e}}_{n}^{\dagger}\left(\mathbf{q}\right)}{\varrho\,\omega_n^2\left(\mathbf{q}\right)},
\end{align}
where $\varrho\approx 7.6\times 10^{-7}$ kg/m$^2$ is graphene's mass density and $\omega_n(\mathbf{q})$, $\hat{\boldsymbol{e}}_{n}(\mathbf{q})=\sum_{i,\mathbf{G}}c_{i,\mathbf{G}}^{n}(\mathbf{q})\,\hat{\boldsymbol{e}}_{i,\mathbf{G}}$ are the dispersion and the polarization vector of vibrational mode $n$ of the moir\'e pattern, respectively. In the long-wavelength and low-energy limit, one obtains two acoustic-like longitudinal and transverse modes, $\omega_{L,T}=c_{L,T}|\mathbf{q}|$ \cite{phasons,Koshino_phonons,phasons2}. These so-called phasons are associated with the invariance of the \textit{equilibrium} free energy with respect to a uniform translation $\boldsymbol{\tilde{u}}$ of the center of the stacking texture, $\boldsymbol{\phi}_{0}(\mathbf{r})\rightarrow\boldsymbol{\phi}_{0}(\mathbf{r}-\boldsymbol{\tilde{u}})$. In the limit of vanishing adhesion forces and small twist angles, the latter can be written in terms of the relative displacement between the layers as $\boldsymbol{\tilde{u}} = \overline{\theta}^{-1}\mathbf{\hat{z}}\times \left( \boldsymbol{u}_{\rm t} - \boldsymbol{u}_{\rm b} \right)$; however, the general relationship between them is more complicated. Note also that longitudinal phason fluctuations (of the collective coordinate $\boldsymbol{\tilde{u}}$) involve transverse stacking fluctuations, and vice-versa; hereafter the indices $L,T$ refer to the latter. The sound velocities $c_{L,T}$ are only slightly smaller than for the corresponding acoustic phonons of monolayer graphene, as the lower stiffness of the stacking domains walls is compensated by the smaller inertia of the sliding motion \cite{phasons,Koshino_phonons}.

\textit{Static response: disorder effects}. We first investigate the impact of quenched disorder on the phason modes. The most relevant types of disorder that affect the moir\'e pattern (i.e. the stacking order) are those that locally change the relative orientation (twist angle) as well as the relative displacement (heterostrain) between the layers. The corresponding disorder potentials can be parametrized, respectively, in terms of a random layer-symmetric potential $V_1(\mathbf{r})$ and a random layer asymmetric potential $V_2(\mathbf{r})$, which give rise to forces of the form:
\begin{align}
\boldsymbol{f}\left(\mathbf{r}\right)=-\left(\boldsymbol{\hat{z}}\times\boldsymbol{\nabla}\right) V_1\left(\mathbf{r}\right)-\boldsymbol{\nabla}V_2\left(\mathbf{r}\right).
\end{align}

The disorder-averaged correlation function between local stacking configurations, $C_{ij}=\langle \delta\phi_i(\mathbf{r})\delta\phi_j(\mathbf{r}')\rangle_{\textrm{dis}}$, is given in linear response by
\begin{align}
\label{eq:disorder_C}
\hat{C}\left(\mathbf{r},\mathbf{r}'\right)=\int d^2\mathbf{r}_1\int d^2\mathbf{r}_2\,\, \hat{\chi}_0\left(\mathbf{r},\mathbf{r}_1\right)\cdot \hat{\mathcal{G}}\left(\mathbf{r}_1,\mathbf{r}_2\right) \cdot\hat{\chi}_0\left(\mathbf{r}_2,\mathbf{r}'\right),
\end{align}
where \begin{align}
\mathcal{G}_{ij}\left(\mathbf{r}_1,\mathbf{r}_2\right)\equiv\left\langle f_i\left(\mathbf{r}_1\right) f_j\left(\mathbf{r}_2\right) \right\rangle_{\textrm{dis}}.
\end{align}
Note that the correlation function $\hat{C}$ is related to the (renormalized) susceptibility $\hat{\chi}$ via the fluctuation-dissipation theorem. To proceed, we assume that the disorder potentials $V_{\alpha}$ are smooth on the inter-atomic scale and decay on a characteristic length scale $\zeta_{\alpha}\ll L_{\textrm{m}}$. Moreover, we consider them to follow random Gaussian distributions with $\overline{V_{\alpha}} = 0$ and finite $\overline{V_{\alpha}^2}$.

To determine how disorder affects the stacking order and the vibrational modes, Eq. (\ref{eq:chi0}), we project the averaged force on the subspace of the low-energy phason modes $n_{1,2}=L,T$, $\mathcal{G}_{n_1,n_2}\left(\mathbf{q}\right)=\hat{\boldsymbol{e}}_{n_1}^{\dagger}(\mathbf{q})\cdot\hat{\mathcal{G}}\left(\mathbf{q}\right)\cdot \hat{\boldsymbol{e}}_{n_2}(\mathbf{q})$. Its long-wavelength behavior depends crucially on the adhesion potential $\mathcal{V}_{\textrm{ad}}$ that couples the two layers, Eq. (\ref{eq:V_adhesion}). If $V_{\textrm{AA}}=0$, the system would be invariant under relative translations of the layers, leading to $\mathcal{G}_{n_1,n_2}(\mathbf{q})\sim |\mathbf{q}|^2$. As a result, from Eq. (\ref{eq:disorder_C}), since $\hat{\chi}_0^{-1} \sim |\mathbf{q}|^2$, $\hat{C}(\mathbf{q})$ would diverge only as $\sim |\mathbf{q}|^{-2}$, implying quasi-long-range stacking order \cite{SM}. However, once $V_{\textrm{AA}}>0$, as in TBG, relative translations of the layers are no longer a symmetry, and we obtain $\mathcal{G}_{n_1,n_2}\left(\mathbf{q}\rightarrow\mathbf{0}\right)=\mathcal{G}\,\delta_{n_1,n_2}$ \cite{SM}. Consequently, $\hat{C}(\mathbf{q})$ now diverges as $\sim |\mathbf{q}|^{-4}$, which implies loss of stacking order, in agreement with the Imry-Ma criterion \cite{Imry-Ma}. More specifically, stacking correlations in the relaxed moir\'e structure decay exponentially \cite{SM},
\begin{align}
\label{eq:scaling_pinning}
\left\langle e^{i\boldsymbol{g}\cdot\left(\delta\boldsymbol{\phi}(\mathbf{r})-\delta\boldsymbol{\phi}(\mathbf{r}')\right)} \right\rangle_{\textrm{dis}}\sim \left(\frac{L}{\left|\mathbf{r}-\mathbf{r}'\right|}\right)^{-\frac{\left|\mathbf{r}-\mathbf{r}'\right|^2}{2L_c^2}},
\end{align}
where $L$ is the size of the system and $L_c$ is the characteristic length scale beyond which the moir\'e pattern ceases to respond elastically to external forces,
\begin{align}
\label{eq:pinning}
L_c=\frac{a\varrho}{\sqrt{4\pi\mathcal{G}}}\frac{c_L^2c_T^2}{\sqrt{c_L^4+c_T^4}}.
\end{align}
This result follows from the first cumulant approximation for the correlator in Eq.~\eqref{eq:scaling_pinning}, where the fast growth of $C(\mathbf{x},\mathbf{y})$ with the relative distance gives rise to an exponential suppression of stacking order. The characteristic scale of this decay reflects the competition between the stiffness of the moir\'e pattern and the accumulated action of disorder forces on the stacking texture. Equation~\eqref{eq:variance} for the twist-angle variation inside a region of length $L_c$ is obtained by computing the fluctuations of $\delta\theta=(\boldsymbol{\nabla}\times\delta\boldsymbol{\phi})_z$ cut-off by $L_c$ \cite{SM}. We emphasize that this last result does not depend on the form of the disorder potentials, but rather on the presence of interlayer adhesion forces.

\begin{figure}[t!]
\begin{center}
\includegraphics[width=\columnwidth]{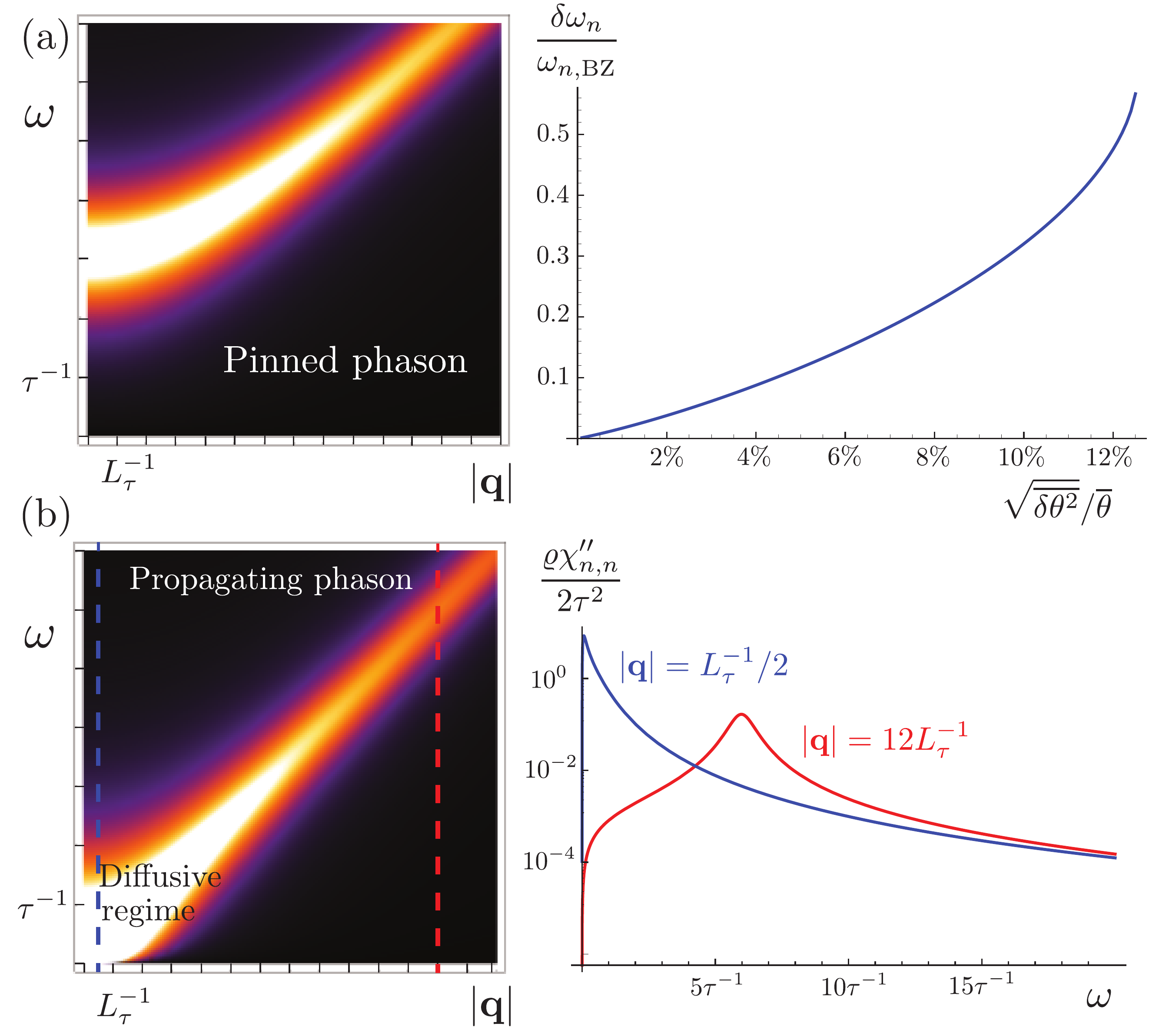}
\caption{\textbf{Phason dispersion relation}. (a) Density plot of the imaginary part of the stacking susceptibility, $\chi_{n,n}''$ , in the presence of disorder ($L_c=L_{\tau}$). The right panel shows the evolution of the phason gap as a function of twist-angle disorder. (b) The same as in (a) in the absence of disorder. At long wavelengths, $|\mathbf{q}|<1/L_{\tau}$, phasons are overdamped. The right panel shows constant momentum cuts in the diffusive (blue) and propagating (red) regimes, the latter characterized by a sharp quasiparticle peak (note that the vertical axis is in logarithmic scale).}
\label{fig:phason}
\end{center}
\end{figure}

Upon computing the ``disorder strength'' $\mathcal{G}$, we find that the contribution from the potential $V_1$ (associated with twist angle disorder) is several orders of magnitude larger than that from $V_2$ (associated with heterostrain disorder). The reason is that the relaxed structure generates very little longitudinal strain in order to preserve the symmetry of the moir\'e pattern. Thus, we associate $L_c$ with the length-scale of twist-angle inhomogeneity. Fig.~\ref{fig:textures}~(d) shows a numerical evaluation of $\mathcal{G}$ as a function of $L_{\textrm{m}}/\ell\propto\overline{\theta}^{-1}$ \cite{SM}. While $\mathcal{G}$ increases quadratically with $L_{\textrm{m}}/\ell$ in the region of large twist angles, it seems to saturate to an angle-independent value for large $L_{\textrm{m}}/\ell$, $L_c\sim a\mu\ell/\zeta_1(\overline{V_1^2})^{1/2}$, which can be interpreted as a collective pinning length of the stacking domain wall system, akin to the case of an incommensurate charge density wave \cite{collective_pinning}.

Because stacking order is lost at the length scale $L_c$, the phasons acquire a gap at a momentum scale $q_c = 2\pi/L_c$ [Fig. \ref{fig:phason}(a)]. We can use the long-wavelength dispersion $\omega_{n}=c_{n}|\mathbf{q}|$ to estimate the gap $\delta \omega_{n} = \omega_{n, \rm ZB} L_{\rm m}/L_c$, where $\omega_{n, \rm ZB}$ is the extrapolated phonon frequency at the zone boundary $q_{\rm ZB} = 2\pi/L_{\rm m}$. Since the ratio $L_{\rm m}/L_c$ is an implicit function of the twist angle variance, Eq. (\ref{eq:variance}), the relative phason gap is a universal function of $\overline{\delta\theta^2} / \overline{\theta}^2$, as illustrated in Fig. \ref{fig:phason}(a).
 
\textit{Dynamical response: anharmonic effects}. While quenched disorder impacts the static properties of the stacking susceptibility, thermal fluctuations affect its dynamics. Indeed, the lack of a conservation law that protects the gapless dispersion of the phasons also leaves its propagating dynamics unprotected. Quite generally, the renormalized stacking susceptibility $\hat{\chi}$ can be parametrized as
\cite{Forster},\begin{align} \label{eq:sigma_dynamics}
\hat{\chi}\left(\mathbf{q},\omega\right)=\left[\hat{\chi}_0^{-1}\left(\mathbf{q}\right)-\frac{\varrho}{2}\omega^2\hat{1}-i\omega\hat{\sigma}\left(\mathbf{q},\omega\right)\right]^{-1},
\end{align}
where $\hat{\chi}_0^{-1}$ is given by Eq. (\ref{eq:harmonic_spectrum}) and the memory matrix function $\hat{\sigma}(\mathbf{q},\omega)$ can be extracted from the imaginary part of the phason self-energy. Specifically, $\sigma_{n_1,n_2}(\mathbf{q},\omega)$ can be expressed as a thermal correlator of forces on the specific phason modes $n_{1,2}$ exerted by the other stacking degrees of freedom \cite{SM}. The most relevant type of thermal fluctuations in our model are those arising from anharmonic contributions in the adhesion potential of Eq.~\eqref{eq:V_adhesion}. Formally, these anharmonic terms induce a self-interaction for the stacking fluctuations and, consequently, dissipation.

The detailed computation of $\hat{\sigma}(\mathbf{q},\omega)$ projected onto the $n_1, \, n_2$ phason modes, $\sigma_{n_1,n_2}(\mathbf{q},\omega)$, is shown in the Supplemental Material \cite{SM}. In the absence of an adhesion potential (i.e. $V_{\textrm{AA}}=0$), the linear momentum of each layer is locally conserved, and we find $\sigma_{n_1,n_2}(\mathbf{q},\omega)\propto |\mathbf{q}|^2$. In this case, the low-frequency phason modes display propagating dynamics, akin to the case of regular acoustic phonons. However, in the realistic case of a non-zero adhesion potential, $\sigma_{n_1,n_2}(\mathbf{q},\omega)$ remains finite in the long-wavelength limit $\mathbf{q} \rightarrow 0$, where $\tau^{-1}\equiv 2 \sigma_{n,n}(0,0)/\varrho$ can then be identified as the relaxation rate of the relative momentum between the layers. This is determined by resonant processes in which thermally populated amplitude vibrations in mode $n_i$ are converted into mode $n_j$ via phason scattering conserving energy and quasi-momentum. At large angles, these processes are dominated by interlayer phonon umklapp in the moir\'e superlattice, leading to a quick growth with decreasing twist angle, $\tau^{-1} \propto(\overline{\theta})^{-3}$ \cite{SM}. The main consequence of a finite $\tau$ in Eq. (\ref{eq:sigma_dynamics}) is that, for small momenta, $|\mathbf{q}|< 1/L_\tau$, with $L_\tau = 2\tau c_n$, the dynamics of the phason modes becomes diffusive,  $\omega_n(\mathbf{q})\approx-i\tau c_n^2|\mathbf{q}|^2$. This behavior is illustrated in Fig. \ref{fig:phason}(b).

\textit{Discussion}. In this paper, we showed how deviations in the stacking order of the graphene layers in TBG fundamentally alter the elastic properties of the resulting moir\'e pattern. These stacking deviations are promoted either by disorder in the relative orientation (twist angle) and relative displacement (heterostrain) between the layers, or by thermally-excited anharmonic fluctuations of the lattice. In the presence of adhesion forces between the layers, the former introduce a length scale $L_c$ beyond which the moir\'e pattern lacks positional order, while the latter generates a time scale $\tau$ beyond which the relative momentum between the layers relaxes. Whereas $L_c$ has a monotonic dependence on the twist-angle variance, $\overline{\delta\theta^2} / \overline{\theta}^2$ [Eq.~\ref{eq:variance}], $\tau$ is strongly suppressed for decreasing twist angles, $\tau \propto \overline{\theta}^3$. Both quantities qualitatively change the low-energy, long-wavelength 
excitations of the moir\'e superlattice. Instead of behaving like gapless propagating acoustic phonons, these phasons become gapped (due to finite $L_c$) and diffusive (due to finite $\tau$). These are the consequences of the absence of a conservation law protecting these soft modes. Atomistic models for interfacial forces \cite{ML_methods} can provide better numerical estimates and shed more light on the stability of moir\'e patterns. 
 
Twist angle variations are manifested as local changes in the electronic density. Our work reveals a hitherto unexplored facet of this ubiquitous property of TBG devices, showing its crucial role in shaping the phason modes of the moir\'e superlattice. Indeed, the typical experimental values $\sqrt{\overline{\delta\theta^2}}\sim 0.02^{\textrm{o}}-0.04^{\circ}$ around the magic angle $\overline{\theta} \sim 1^{\circ}$ \cite{Zeldov,nat_mat} give $L_c\sim400-200$ nm, which imply random pinning forces $\zeta_1(\overline{V_1^2})^{1/2}$ comparable to the surface tension of stacking domain walls ($\ell\, V_{\textrm{AA}}\approx 1$ eV/nm) and a phason gap of $4\%-9\%$ of $\omega_{n,\rm ZB}\sim 40$ K. The phason dynamics should be manifested in thermodynamic properties at low temperatures, such as in the specific heat capacity \cite{Cv,BP}. Importantly, via the electron-phonon coupling, the changes in the phason modes promoted by twist-angle variations -- and anharmonic forces -- will inevitably impact the electronic properties in different ways. This includes the renormalized dispersion of the remote and narrow bands \cite{Fu_heterostrain,Vafek20_RG}, the contribution to the resistivity arising from electron-phonon scattering \cite{Wu_etal2,phasons2,scattering1,scattering2}, and the ordered states that can be either promoted or strongly affected by electron-phonon interactions, such as superconductivity \cite{Wu_etal,Bernevig_phonons,Wu_etal2,Fabrizio19,Lewandoski_etal,Cea_Guinea} and nematicity \cite{moire_nematicity,Pablo_nematics,Carmen_etal}.

\begin{acknowledgments}
H.O. acknowledges NSF MRSEC program Grant No. DMR-1420634. R.M.F. was supported by the U.S. Department of Energy, Office of Science, Basic Energy Sciences, Materials Science and Engineering Division, under Award No. DE-SC0020045.
\end{acknowledgments}

\bibliography{references}

\begin{thebibliography}{100}%
\makeatletter
\providecommand \@ifxundefined [1]{%
 \@ifx{#1\undefined}
}%
\providecommand \@ifnum [1]{%
 \ifnum #1\expandafter \@firstoftwo
 \else \expandafter \@secondoftwo
 \fi
}%
\providecommand \@ifx [1]{%
 \ifx #1\expandafter \@firstoftwo
 \else \expandafter \@secondoftwo
 \fi
}%
\providecommand \natexlab [1]{#1}%
\providecommand \enquote  [1]{``#1''}%
\providecommand \bibnamefont  [1]{#1}%
\providecommand \bibfnamefont [1]{#1}%
\providecommand \citenamefont [1]{#1}%
\providecommand \href@noop [0]{\@secondoftwo}%
\providecommand \href [0]{\begingroup \@sanitize@url \@href}%
\providecommand \@href[1]{\@@startlink{#1}\@@href}%
\providecommand \@@href[1]{\endgroup#1\@@endlink}%
\providecommand \@sanitize@url [0]{\catcode `\\12\catcode `\$12\catcode
  `\&12\catcode `\#12\catcode `\^12\catcode `\_12\catcode `\%12\relax}%
\providecommand \@@startlink[1]{}%
\providecommand \@@endlink[0]{}%
\providecommand \url  [0]{\begingroup\@sanitize@url \@url }%
\providecommand \@url [1]{\endgroup\@href {#1}{\urlprefix }}%
\providecommand \urlprefix  [0]{URL }%
\providecommand \Eprint [0]{\href }%
\providecommand \doibase [0]{http://dx.doi.org/}%
\providecommand \selectlanguage [0]{\@gobble}%
\providecommand \bibinfo  [0]{\@secondoftwo}%
\providecommand \bibfield  [0]{\@secondoftwo}%
\providecommand \translation [1]{[#1]}%
\providecommand \BibitemOpen [0]{}%
\providecommand \bibitemStop [0]{}%
\providecommand \bibitemNoStop [0]{.\EOS\space}%
\providecommand \EOS [0]{\spacefactor3000\relax}%
\providecommand \BibitemShut  [1]{\csname bibitem#1\endcsname}%
\let\auto@bib@innerbib\@empty
\bibitem [{\citenamefont {Balents}\ \emph {et~al.}(2020)\citenamefont
  {Balents}, \citenamefont {Dean}, \citenamefont {Efetov},\ and\ \citenamefont
  {Young}}]{Balents_review}%
  \BibitemOpen
  \bibfield  {author} {\bibinfo {author} {\bibfnamefont {L.}~\bibnamefont
  {Balents}}, \bibinfo {author} {\bibfnamefont {C.~R.}\ \bibnamefont {Dean}},
  \bibinfo {author} {\bibfnamefont {D.~K.}\ \bibnamefont {Efetov}}, \ and\
  \bibinfo {author} {\bibfnamefont {A.~F.}\ \bibnamefont {Young}},\ }\href@noop
  {} {\bibfield  {journal} {\bibinfo  {journal} {Nat. Phys.}\ }\textbf
  {\bibinfo {volume} {16}},\ \bibinfo {pages} {725} (\bibinfo {year}
  {2020})}\BibitemShut {NoStop}%
\bibitem [{\citenamefont {Cao}\ \emph {et~al.}(2018{\natexlab{a}})\citenamefont
  {Cao}, \citenamefont {Fatemi}, \citenamefont {Fang}, \citenamefont
  {Watanabe}, \citenamefont {Taniguchi}, \citenamefont {Kaxiras},\ and\
  \citenamefont {Jarillo-Herrero}}]{Cao2018a}%
  \BibitemOpen
  \bibfield  {author} {\bibinfo {author} {\bibfnamefont {Y.}~\bibnamefont
  {Cao}}, \bibinfo {author} {\bibfnamefont {V.}~\bibnamefont {Fatemi}},
  \bibinfo {author} {\bibfnamefont {S.}~\bibnamefont {Fang}}, \bibinfo {author}
  {\bibfnamefont {K.}~\bibnamefont {Watanabe}}, \bibinfo {author}
  {\bibfnamefont {T.}~\bibnamefont {Taniguchi}}, \bibinfo {author}
  {\bibfnamefont {E.}~\bibnamefont {Kaxiras}}, \ and\ \bibinfo {author}
  {\bibfnamefont {P.}~\bibnamefont {Jarillo-Herrero}},\ }\href {\doibase
  10.1038/nature26160} {\bibfield  {journal} {\bibinfo  {journal} {Nature}\
  }\textbf {\bibinfo {volume} {556}},\ \bibinfo {pages} {43} (\bibinfo {year}
  {2018}{\natexlab{a}})}\BibitemShut {NoStop}%
\bibitem [{\citenamefont {Cao}\ \emph {et~al.}(2018{\natexlab{b}})\citenamefont
  {Cao}, \citenamefont {Fatemi}, \citenamefont {Demir}, \citenamefont {Fang},
  \citenamefont {Tomarken}, \citenamefont {Luo}, \citenamefont
  {Sanchez-Yamagishi}, \citenamefont {Watanabe}, \citenamefont {Taniguchi},
  \citenamefont {Kaxiras}, \citenamefont {Ashoori},\ and\ \citenamefont
  {Jarillo-Herrero}}]{Cao2018b}%
  \BibitemOpen
  \bibfield  {author} {\bibinfo {author} {\bibfnamefont {Y.}~\bibnamefont
  {Cao}}, \bibinfo {author} {\bibfnamefont {V.}~\bibnamefont {Fatemi}},
  \bibinfo {author} {\bibfnamefont {A.}~\bibnamefont {Demir}}, \bibinfo
  {author} {\bibfnamefont {S.}~\bibnamefont {Fang}}, \bibinfo {author}
  {\bibfnamefont {S.~L.}\ \bibnamefont {Tomarken}}, \bibinfo {author}
  {\bibfnamefont {J.~Y.}\ \bibnamefont {Luo}}, \bibinfo {author} {\bibfnamefont
  {J.~D.}\ \bibnamefont {Sanchez-Yamagishi}}, \bibinfo {author} {\bibfnamefont
  {K.}~\bibnamefont {Watanabe}}, \bibinfo {author} {\bibfnamefont
  {T.}~\bibnamefont {Taniguchi}}, \bibinfo {author} {\bibfnamefont
  {E.}~\bibnamefont {Kaxiras}}, \bibinfo {author} {\bibfnamefont {R.~C.}\
  \bibnamefont {Ashoori}}, \ and\ \bibinfo {author} {\bibfnamefont
  {P.}~\bibnamefont {Jarillo-Herrero}},\ }\href {\doibase 10.1038/nature26154}
  {\bibfield  {journal} {\bibinfo  {journal} {Nature}\ }\textbf {\bibinfo
  {volume} {556}},\ \bibinfo {pages} {80} (\bibinfo {year}
  {2018}{\natexlab{b}})}\BibitemShut {NoStop}%
\bibitem [{\citenamefont {Yankowitz}\ \emph {et~al.}(2019)\citenamefont
  {Yankowitz}, \citenamefont {Chen}, \citenamefont {Polshyn}, \citenamefont
  {Zhang}, \citenamefont {Watanabe}, \citenamefont {Taniguchi}, \citenamefont
  {Graf}, \citenamefont {Young},\ and\ \citenamefont {Dean}}]{Yankowitz1059}%
  \BibitemOpen
  \bibfield  {author} {\bibinfo {author} {\bibfnamefont {M.}~\bibnamefont
  {Yankowitz}}, \bibinfo {author} {\bibfnamefont {S.}~\bibnamefont {Chen}},
  \bibinfo {author} {\bibfnamefont {H.}~\bibnamefont {Polshyn}}, \bibinfo
  {author} {\bibfnamefont {Y.}~\bibnamefont {Zhang}}, \bibinfo {author}
  {\bibfnamefont {K.}~\bibnamefont {Watanabe}}, \bibinfo {author}
  {\bibfnamefont {T.}~\bibnamefont {Taniguchi}}, \bibinfo {author}
  {\bibfnamefont {D.}~\bibnamefont {Graf}}, \bibinfo {author} {\bibfnamefont
  {A.~F.}\ \bibnamefont {Young}}, \ and\ \bibinfo {author} {\bibfnamefont
  {C.~R.}\ \bibnamefont {Dean}},\ }\href {\doibase 10.1126/science.aav1910}
  {\bibfield  {journal} {\bibinfo  {journal} {Science}\ }\textbf {\bibinfo
  {volume} {363}},\ \bibinfo {pages} {1059} (\bibinfo {year}
  {2019})}\BibitemShut {NoStop}%
\bibitem [{\citenamefont {Sharpe}\ \emph {et~al.}(2019)\citenamefont {Sharpe},
  \citenamefont {Fox}, \citenamefont {Barnard}, \citenamefont {Finney},
  \citenamefont {Watanabe}, \citenamefont {Taniguchi}, \citenamefont
  {Kastner},\ and\ \citenamefont {Goldhaber-Gordon}}]{Sharpe19}%
  \BibitemOpen
  \bibfield  {author} {\bibinfo {author} {\bibfnamefont {A.~L.}\ \bibnamefont
  {Sharpe}}, \bibinfo {author} {\bibfnamefont {E.~J.}\ \bibnamefont {Fox}},
  \bibinfo {author} {\bibfnamefont {A.~W.}\ \bibnamefont {Barnard}}, \bibinfo
  {author} {\bibfnamefont {J.}~\bibnamefont {Finney}}, \bibinfo {author}
  {\bibfnamefont {K.}~\bibnamefont {Watanabe}}, \bibinfo {author}
  {\bibfnamefont {T.}~\bibnamefont {Taniguchi}}, \bibinfo {author}
  {\bibfnamefont {M.~A.}\ \bibnamefont {Kastner}}, \ and\ \bibinfo {author}
  {\bibfnamefont {D.}~\bibnamefont {Goldhaber-Gordon}},\ }\href {\doibase
  10.1126/science.aaw3780} {\bibfield  {journal} {\bibinfo  {journal}
  {Science}\ }\textbf {\bibinfo {volume} {365}},\ \bibinfo {pages} {605}
  (\bibinfo {year} {2019})}\BibitemShut {NoStop}%
\bibitem [{\citenamefont {Jiang}\ \emph {et~al.}(2019)\citenamefont {Jiang},
  \citenamefont {Lai}, \citenamefont {Watanabe}, \citenamefont {Taniguchi},
  \citenamefont {Haule}, \citenamefont {Mao},\ and\ \citenamefont
  {Andrei}}]{STM_Andrei19}%
  \BibitemOpen
  \bibfield  {author} {\bibinfo {author} {\bibfnamefont {Y.}~\bibnamefont
  {Jiang}}, \bibinfo {author} {\bibfnamefont {X.}~\bibnamefont {Lai}}, \bibinfo
  {author} {\bibfnamefont {K.}~\bibnamefont {Watanabe}}, \bibinfo {author}
  {\bibfnamefont {T.}~\bibnamefont {Taniguchi}}, \bibinfo {author}
  {\bibfnamefont {K.}~\bibnamefont {Haule}}, \bibinfo {author} {\bibfnamefont
  {J.}~\bibnamefont {Mao}}, \ and\ \bibinfo {author} {\bibfnamefont {E.~Y.}\
  \bibnamefont {Andrei}},\ }\href@noop {} {\bibfield  {journal} {\bibinfo
  {journal} {Nature}\ }\textbf {\bibinfo {volume} {573}},\ \bibinfo {pages}
  {91} (\bibinfo {year} {2019})}\BibitemShut {NoStop}%
\bibitem [{\citenamefont {Kerelsky}\ \emph {et~al.}(2019)\citenamefont
  {Kerelsky}, \citenamefont {McGilly}, \citenamefont {Kennes}, \citenamefont
  {Xian}, \citenamefont {Yankowitz}, \citenamefont {Chen}, \citenamefont
  {Watanabe}, \citenamefont {Taniguchi}, \citenamefont {Hone}, \citenamefont
  {Dean}, \citenamefont {Rubio},\ and\ \citenamefont
  {Pasupathy}}]{STM_Pasupathy19}%
  \BibitemOpen
  \bibfield  {author} {\bibinfo {author} {\bibfnamefont {A.}~\bibnamefont
  {Kerelsky}}, \bibinfo {author} {\bibfnamefont {L.~J.}\ \bibnamefont
  {McGilly}}, \bibinfo {author} {\bibfnamefont {D.~M.}\ \bibnamefont {Kennes}},
  \bibinfo {author} {\bibfnamefont {L.}~\bibnamefont {Xian}}, \bibinfo {author}
  {\bibfnamefont {M.}~\bibnamefont {Yankowitz}}, \bibinfo {author}
  {\bibfnamefont {S.}~\bibnamefont {Chen}}, \bibinfo {author} {\bibfnamefont
  {K.}~\bibnamefont {Watanabe}}, \bibinfo {author} {\bibfnamefont
  {T.}~\bibnamefont {Taniguchi}}, \bibinfo {author} {\bibfnamefont
  {J.}~\bibnamefont {Hone}}, \bibinfo {author} {\bibfnamefont {C.}~\bibnamefont
  {Dean}}, \bibinfo {author} {\bibfnamefont {A.}~\bibnamefont {Rubio}}, \ and\
  \bibinfo {author} {\bibfnamefont {A.~N.}\ \bibnamefont {Pasupathy}},\
  }\href@noop {} {\bibfield  {journal} {\bibinfo  {journal} {Nature}\ }\textbf
  {\bibinfo {volume} {572}},\ \bibinfo {pages} {95} (\bibinfo {year}
  {2019})}\BibitemShut {NoStop}%
\bibitem [{\citenamefont {Lu}\ \emph {et~al.}(2019)\citenamefont {Lu},
  \citenamefont {Stepanov}, \citenamefont {Yang}, \citenamefont {Xie},
  \citenamefont {Aamir}, \citenamefont {Das}, \citenamefont {Urgell},
  \citenamefont {Watanabe}, \citenamefont {Taniguchi}, \citenamefont {Zhang},
  \citenamefont {Bachtold}, \citenamefont {MacDonald},\ and\ \citenamefont
  {Efetov}}]{Efetov19}%
  \BibitemOpen
  \bibfield  {author} {\bibinfo {author} {\bibfnamefont {X.}~\bibnamefont
  {Lu}}, \bibinfo {author} {\bibfnamefont {P.}~\bibnamefont {Stepanov}},
  \bibinfo {author} {\bibfnamefont {W.}~\bibnamefont {Yang}}, \bibinfo {author}
  {\bibfnamefont {M.}~\bibnamefont {Xie}}, \bibinfo {author} {\bibfnamefont
  {M.~A.}\ \bibnamefont {Aamir}}, \bibinfo {author} {\bibfnamefont
  {I.}~\bibnamefont {Das}}, \bibinfo {author} {\bibfnamefont {C.}~\bibnamefont
  {Urgell}}, \bibinfo {author} {\bibfnamefont {K.}~\bibnamefont {Watanabe}},
  \bibinfo {author} {\bibfnamefont {T.}~\bibnamefont {Taniguchi}}, \bibinfo
  {author} {\bibfnamefont {G.}~\bibnamefont {Zhang}}, \bibinfo {author}
  {\bibfnamefont {A.}~\bibnamefont {Bachtold}}, \bibinfo {author}
  {\bibfnamefont {A.~H.}\ \bibnamefont {MacDonald}}, \ and\ \bibinfo {author}
  {\bibfnamefont {D.~K.}\ \bibnamefont {Efetov}},\ }\href@noop {} {\bibfield
  {journal} {\bibinfo  {journal} {Nature}\ }\textbf {\bibinfo {volume} {574}},\
  \bibinfo {pages} {653} (\bibinfo {year} {2019})}\BibitemShut {NoStop}%
\bibitem [{\citenamefont {Xie}\ \emph {et~al.}(2019)\citenamefont {Xie},
  \citenamefont {Lian}, \citenamefont {J{\"a}ck}, \citenamefont {Liu},
  \citenamefont {Chiu}, \citenamefont {Watanabe}, \citenamefont {Taniguchi},
  \citenamefont {Bernevig},\ and\ \citenamefont {Yazdani}}]{STM_Yazdani19}%
  \BibitemOpen
  \bibfield  {author} {\bibinfo {author} {\bibfnamefont {Y.}~\bibnamefont
  {Xie}}, \bibinfo {author} {\bibfnamefont {B.}~\bibnamefont {Lian}}, \bibinfo
  {author} {\bibfnamefont {B.}~\bibnamefont {J{\"a}ck}}, \bibinfo {author}
  {\bibfnamefont {X.}~\bibnamefont {Liu}}, \bibinfo {author} {\bibfnamefont
  {C.-L.}\ \bibnamefont {Chiu}}, \bibinfo {author} {\bibfnamefont
  {K.}~\bibnamefont {Watanabe}}, \bibinfo {author} {\bibfnamefont
  {T.}~\bibnamefont {Taniguchi}}, \bibinfo {author} {\bibfnamefont {B.~A.}\
  \bibnamefont {Bernevig}}, \ and\ \bibinfo {author} {\bibfnamefont
  {A.}~\bibnamefont {Yazdani}},\ }\href@noop {} {\bibfield  {journal} {\bibinfo
   {journal} {Nature}\ }\textbf {\bibinfo {volume} {572}},\ \bibinfo {pages}
  {101} (\bibinfo {year} {2019})}\BibitemShut {NoStop}%
\bibitem [{\citenamefont {Choi}\ \emph {et~al.}(2019)\citenamefont {Choi},
  \citenamefont {Kemmer}, \citenamefont {Peng}, \citenamefont {Thomson},
  \citenamefont {Arora}, \citenamefont {Polski}, \citenamefont {Zhang},
  \citenamefont {Ren}, \citenamefont {Alicea}, \citenamefont {Refael},
  \citenamefont {Oppen}, \citenamefont {Watanabe}, \citenamefont {Taniguchi},\
  and\ \citenamefont {Nadj-Perge}}]{STM_Perge19}%
  \BibitemOpen
  \bibfield  {author} {\bibinfo {author} {\bibfnamefont {Y.}~\bibnamefont
  {Choi}}, \bibinfo {author} {\bibfnamefont {J.}~\bibnamefont {Kemmer}},
  \bibinfo {author} {\bibfnamefont {Y.}~\bibnamefont {Peng}}, \bibinfo {author}
  {\bibfnamefont {A.}~\bibnamefont {Thomson}}, \bibinfo {author} {\bibfnamefont
  {H.}~\bibnamefont {Arora}}, \bibinfo {author} {\bibfnamefont
  {R.}~\bibnamefont {Polski}}, \bibinfo {author} {\bibfnamefont
  {Y.}~\bibnamefont {Zhang}}, \bibinfo {author} {\bibfnamefont
  {H.}~\bibnamefont {Ren}}, \bibinfo {author} {\bibfnamefont {J.}~\bibnamefont
  {Alicea}}, \bibinfo {author} {\bibfnamefont {G.}~\bibnamefont {Refael}},
  \bibinfo {author} {\bibfnamefont {F.~v.}\ \bibnamefont {Oppen}}, \bibinfo
  {author} {\bibfnamefont {K.}~\bibnamefont {Watanabe}}, \bibinfo {author}
  {\bibfnamefont {T.}~\bibnamefont {Taniguchi}}, \ and\ \bibinfo {author}
  {\bibfnamefont {S.}~\bibnamefont {Nadj-Perge}},\ }\href@noop {} {\bibfield
  {journal} {\bibinfo  {journal} {Nat. Phys.}\ }\textbf {\bibinfo {volume}
  {15}},\ \bibinfo {pages} {1174} (\bibinfo {year} {2019})}\BibitemShut
  {NoStop}%
\bibitem [{\citenamefont {Serlin}\ \emph {et~al.}(2020)\citenamefont {Serlin},
  \citenamefont {Tschirhart}, \citenamefont {Polshyn}, \citenamefont {Zhang},
  \citenamefont {Zhu}, \citenamefont {Watanabe}, \citenamefont {Taniguchi},
  \citenamefont {Balents},\ and\ \citenamefont {Young}}]{Young19}%
  \BibitemOpen
  \bibfield  {author} {\bibinfo {author} {\bibfnamefont {M.}~\bibnamefont
  {Serlin}}, \bibinfo {author} {\bibfnamefont {C.}~\bibnamefont {Tschirhart}},
  \bibinfo {author} {\bibfnamefont {H.}~\bibnamefont {Polshyn}}, \bibinfo
  {author} {\bibfnamefont {Y.}~\bibnamefont {Zhang}}, \bibinfo {author}
  {\bibfnamefont {J.}~\bibnamefont {Zhu}}, \bibinfo {author} {\bibfnamefont
  {K.}~\bibnamefont {Watanabe}}, \bibinfo {author} {\bibfnamefont
  {T.}~\bibnamefont {Taniguchi}}, \bibinfo {author} {\bibfnamefont
  {L.}~\bibnamefont {Balents}}, \ and\ \bibinfo {author} {\bibfnamefont
  {A.}~\bibnamefont {Young}},\ }\href@noop {} {\bibfield  {journal} {\bibinfo
  {journal} {Science}\ }\textbf {\bibinfo {volume} {367}},\ \bibinfo {pages}
  {900} (\bibinfo {year} {2020})}\BibitemShut {NoStop}%
\bibitem [{\citenamefont {Cao}\ \emph {et~al.}(2021)\citenamefont {Cao},
  \citenamefont {Rodan-Legrain}, \citenamefont {Park}, \citenamefont {Yuan},
  \citenamefont {Watanabe}, \citenamefont {Taniguchi}, \citenamefont
  {Fernandes}, \citenamefont {Fu},\ and\ \citenamefont
  {Jarillo-Herrero}}]{Pablo_nematics}%
  \BibitemOpen
  \bibfield  {author} {\bibinfo {author} {\bibfnamefont {Y.}~\bibnamefont
  {Cao}}, \bibinfo {author} {\bibfnamefont {D.}~\bibnamefont {Rodan-Legrain}},
  \bibinfo {author} {\bibfnamefont {J.~M.}\ \bibnamefont {Park}}, \bibinfo
  {author} {\bibfnamefont {N.~F.~Q.}\ \bibnamefont {Yuan}}, \bibinfo {author}
  {\bibfnamefont {K.}~\bibnamefont {Watanabe}}, \bibinfo {author}
  {\bibfnamefont {T.}~\bibnamefont {Taniguchi}}, \bibinfo {author}
  {\bibfnamefont {R.~M.}\ \bibnamefont {Fernandes}}, \bibinfo {author}
  {\bibfnamefont {L.}~\bibnamefont {Fu}}, \ and\ \bibinfo {author}
  {\bibfnamefont {P.}~\bibnamefont {Jarillo-Herrero}},\ }\href {\doibase
  10.1126/science.abc2836} {\bibfield  {journal} {\bibinfo  {journal}
  {Science}\ }\textbf {\bibinfo {volume} {372}},\ \bibinfo {pages} {264}
  (\bibinfo {year} {2021})}\BibitemShut {NoStop}%
\bibitem [{\citenamefont {Rozen}\ \emph {et~al.}(2021)\citenamefont {Rozen},
  \citenamefont {Park}, \citenamefont {Zondiner}, \citenamefont {Cao},
  \citenamefont {Rodan-Legrain}, \citenamefont {Taniguchi}, \citenamefont
  {Watanabe}, \citenamefont {Oreg}, \citenamefont {Stern}, \citenamefont {Berg}
  \emph {et~al.}}]{Pomeranchuk1}%
  \BibitemOpen
  \bibfield  {author} {\bibinfo {author} {\bibfnamefont {A.}~\bibnamefont
  {Rozen}}, \bibinfo {author} {\bibfnamefont {J.~M.}\ \bibnamefont {Park}},
  \bibinfo {author} {\bibfnamefont {U.}~\bibnamefont {Zondiner}}, \bibinfo
  {author} {\bibfnamefont {Y.}~\bibnamefont {Cao}}, \bibinfo {author}
  {\bibfnamefont {D.}~\bibnamefont {Rodan-Legrain}}, \bibinfo {author}
  {\bibfnamefont {T.}~\bibnamefont {Taniguchi}}, \bibinfo {author}
  {\bibfnamefont {K.}~\bibnamefont {Watanabe}}, \bibinfo {author}
  {\bibfnamefont {Y.}~\bibnamefont {Oreg}}, \bibinfo {author} {\bibfnamefont
  {A.}~\bibnamefont {Stern}}, \bibinfo {author} {\bibfnamefont
  {E.}~\bibnamefont {Berg}},  \emph {et~al.},\ }\href@noop {} {\bibfield
  {journal} {\bibinfo  {journal} {Nature}\ }\textbf {\bibinfo {volume} {592}},\
  \bibinfo {pages} {214} (\bibinfo {year} {2021})}\BibitemShut {NoStop}%
\bibitem [{\citenamefont {Saito}\ \emph {et~al.}(2021)\citenamefont {Saito},
  \citenamefont {Yang}, \citenamefont {Ge}, \citenamefont {Liu}, \citenamefont
  {Taniguchi}, \citenamefont {Watanabe}, \citenamefont {Li}, \citenamefont
  {Berg},\ and\ \citenamefont {Young}}]{Pomeranchuk2}%
  \BibitemOpen
  \bibfield  {author} {\bibinfo {author} {\bibfnamefont {Y.}~\bibnamefont
  {Saito}}, \bibinfo {author} {\bibfnamefont {F.}~\bibnamefont {Yang}},
  \bibinfo {author} {\bibfnamefont {J.}~\bibnamefont {Ge}}, \bibinfo {author}
  {\bibfnamefont {X.}~\bibnamefont {Liu}}, \bibinfo {author} {\bibfnamefont
  {T.}~\bibnamefont {Taniguchi}}, \bibinfo {author} {\bibfnamefont
  {K.}~\bibnamefont {Watanabe}}, \bibinfo {author} {\bibfnamefont
  {J.}~\bibnamefont {Li}}, \bibinfo {author} {\bibfnamefont {E.}~\bibnamefont
  {Berg}}, \ and\ \bibinfo {author} {\bibfnamefont {A.~F.}\ \bibnamefont
  {Young}},\ }\href@noop {} {\bibfield  {journal} {\bibinfo  {journal}
  {Nature}\ }\textbf {\bibinfo {volume} {592}},\ \bibinfo {pages} {220}
  (\bibinfo {year} {2021})}\BibitemShut {NoStop}%
\bibitem [{\citenamefont {Xu}\ and\ \citenamefont {Balents}(2018)}]{Xu2018a}%
  \BibitemOpen
  \bibfield  {author} {\bibinfo {author} {\bibfnamefont {C.}~\bibnamefont
  {Xu}}\ and\ \bibinfo {author} {\bibfnamefont {L.}~\bibnamefont {Balents}},\
  }\href {\doibase 10.1103/PhysRevLett.121.087001} {\bibfield  {journal}
  {\bibinfo  {journal} {Phys. Rev. Lett.}\ }\textbf {\bibinfo {volume} {121}},\
  \bibinfo {pages} {087001} (\bibinfo {year} {2018})}\BibitemShut {NoStop}%
\bibitem [{\citenamefont {Po}\ \emph {et~al.}(2018)\citenamefont {Po},
  \citenamefont {Zou}, \citenamefont {Vishwanath},\ and\ \citenamefont
  {Senthil}}]{Po2018}%
  \BibitemOpen
  \bibfield  {author} {\bibinfo {author} {\bibfnamefont {H.~C.}\ \bibnamefont
  {Po}}, \bibinfo {author} {\bibfnamefont {L.}~\bibnamefont {Zou}}, \bibinfo
  {author} {\bibfnamefont {A.}~\bibnamefont {Vishwanath}}, \ and\ \bibinfo
  {author} {\bibfnamefont {T.}~\bibnamefont {Senthil}},\ }\href {\doibase
  10.1103/PhysRevX.8.031089} {\bibfield  {journal} {\bibinfo  {journal} {Phys.
  Rev. X}\ }\textbf {\bibinfo {volume} {8}},\ \bibinfo {pages} {031089}
  (\bibinfo {year} {2018})}\BibitemShut {NoStop}%
\bibitem [{\citenamefont {Isobe}\ \emph {et~al.}(2018)\citenamefont {Isobe},
  \citenamefont {Yuan},\ and\ \citenamefont {Fu}}]{Isobe2018}%
  \BibitemOpen
  \bibfield  {author} {\bibinfo {author} {\bibfnamefont {H.}~\bibnamefont
  {Isobe}}, \bibinfo {author} {\bibfnamefont {N.~F.~Q.}\ \bibnamefont {Yuan}},
  \ and\ \bibinfo {author} {\bibfnamefont {L.}~\bibnamefont {Fu}},\ }\href
  {\doibase 10.1103/PhysRevX.8.041041} {\bibfield  {journal} {\bibinfo
  {journal} {Phys. Rev. X}\ }\textbf {\bibinfo {volume} {8}},\ \bibinfo {pages}
  {041041} (\bibinfo {year} {2018})}\BibitemShut {NoStop}%
\bibitem [{\citenamefont {Kennes}\ \emph {et~al.}(2018)\citenamefont {Kennes},
  \citenamefont {Lischner},\ and\ \citenamefont {Karrasch}}]{Kennes2018}%
  \BibitemOpen
  \bibfield  {author} {\bibinfo {author} {\bibfnamefont {D.~M.}\ \bibnamefont
  {Kennes}}, \bibinfo {author} {\bibfnamefont {J.}~\bibnamefont {Lischner}}, \
  and\ \bibinfo {author} {\bibfnamefont {C.}~\bibnamefont {Karrasch}},\ }\href
  {\doibase 10.1103/PhysRevB.98.241407} {\bibfield  {journal} {\bibinfo
  {journal} {Phys. Rev. B}\ }\textbf {\bibinfo {volume} {98}},\ \bibinfo
  {pages} {241407} (\bibinfo {year} {2018})}\BibitemShut {NoStop}%
\bibitem [{\citenamefont {Rademaker}\ and\ \citenamefont
  {Mellado}(2018)}]{Rademaker2018}%
  \BibitemOpen
  \bibfield  {author} {\bibinfo {author} {\bibfnamefont {L.}~\bibnamefont
  {Rademaker}}\ and\ \bibinfo {author} {\bibfnamefont {P.}~\bibnamefont
  {Mellado}},\ }\href {\doibase 10.1103/PhysRevB.98.235158} {\bibfield
  {journal} {\bibinfo  {journal} {Phys. Rev. B}\ }\textbf {\bibinfo {volume}
  {98}},\ \bibinfo {pages} {235158} (\bibinfo {year} {2018})}\BibitemShut
  {NoStop}%
\bibitem [{\citenamefont {Dodaro}\ \emph {et~al.}(2018)\citenamefont {Dodaro},
  \citenamefont {Kivelson}, \citenamefont {Schattner}, \citenamefont {Sun},\
  and\ \citenamefont {Wang}}]{Dodaro2018}%
  \BibitemOpen
  \bibfield  {author} {\bibinfo {author} {\bibfnamefont {J.~F.}\ \bibnamefont
  {Dodaro}}, \bibinfo {author} {\bibfnamefont {S.~A.}\ \bibnamefont
  {Kivelson}}, \bibinfo {author} {\bibfnamefont {Y.}~\bibnamefont {Schattner}},
  \bibinfo {author} {\bibfnamefont {X.~Q.}\ \bibnamefont {Sun}}, \ and\
  \bibinfo {author} {\bibfnamefont {C.}~\bibnamefont {Wang}},\ }\href {\doibase
  10.1103/PhysRevB.98.075154} {\bibfield  {journal} {\bibinfo  {journal} {Phys.
  Rev. B}\ }\textbf {\bibinfo {volume} {98}},\ \bibinfo {pages} {075154}
  (\bibinfo {year} {2018})}\BibitemShut {NoStop}%
\bibitem [{\citenamefont {Thomson}\ \emph {et~al.}(2018)\citenamefont
  {Thomson}, \citenamefont {Chatterjee}, \citenamefont {Sachdev},\ and\
  \citenamefont {Scheurer}}]{Thomson2018}%
  \BibitemOpen
  \bibfield  {author} {\bibinfo {author} {\bibfnamefont {A.}~\bibnamefont
  {Thomson}}, \bibinfo {author} {\bibfnamefont {S.}~\bibnamefont {Chatterjee}},
  \bibinfo {author} {\bibfnamefont {S.}~\bibnamefont {Sachdev}}, \ and\
  \bibinfo {author} {\bibfnamefont {M.~S.}\ \bibnamefont {Scheurer}},\ }\href
  {\doibase 10.1103/PhysRevB.98.075109} {\bibfield  {journal} {\bibinfo
  {journal} {Phys. Rev. B}\ }\textbf {\bibinfo {volume} {98}},\ \bibinfo
  {pages} {075109} (\bibinfo {year} {2018})}\BibitemShut {NoStop}%
\bibitem [{\citenamefont {Lin}\ and\ \citenamefont
  {Nandkishore}(2018)}]{Lin2018}%
  \BibitemOpen
  \bibfield  {author} {\bibinfo {author} {\bibfnamefont {Y.-P.}\ \bibnamefont
  {Lin}}\ and\ \bibinfo {author} {\bibfnamefont {R.~M.}\ \bibnamefont
  {Nandkishore}},\ }\href {\doibase 10.1103/PhysRevB.98.214521} {\bibfield
  {journal} {\bibinfo  {journal} {Phys. Rev. B}\ }\textbf {\bibinfo {volume}
  {98}},\ \bibinfo {pages} {214521} (\bibinfo {year} {2018})}\BibitemShut
  {NoStop}%
\bibitem [{\citenamefont {Guinea}\ and\ \citenamefont
  {Walet}(2018)}]{Guinea2018}%
  \BibitemOpen
  \bibfield  {author} {\bibinfo {author} {\bibfnamefont {F.}~\bibnamefont
  {Guinea}}\ and\ \bibinfo {author} {\bibfnamefont {N.~R.}\ \bibnamefont
  {Walet}},\ }\href {\doibase 10.1073/pnas.1810947115} {\bibfield  {journal}
  {\bibinfo  {journal} {Proc. Natl. Acad. Sci. U.S.A.}\ }\textbf {\bibinfo
  {volume} {115}},\ \bibinfo {pages} {13174} (\bibinfo {year}
  {2018})}\BibitemShut {NoStop}%
\bibitem [{\citenamefont {Sherkunov}\ and\ \citenamefont
  {Betouras}(2018)}]{Sherkunov2018}%
  \BibitemOpen
  \bibfield  {author} {\bibinfo {author} {\bibfnamefont {Y.}~\bibnamefont
  {Sherkunov}}\ and\ \bibinfo {author} {\bibfnamefont {J.~J.}\ \bibnamefont
  {Betouras}},\ }\href {\doibase 10.1103/PhysRevB.98.205151} {\bibfield
  {journal} {\bibinfo  {journal} {Phys. Rev. B}\ }\textbf {\bibinfo {volume}
  {98}},\ \bibinfo {pages} {205151} (\bibinfo {year} {2018})}\BibitemShut
  {NoStop}%
\bibitem [{\citenamefont {Liu}\ \emph {et~al.}(2018)\citenamefont {Liu},
  \citenamefont {Zhang}, \citenamefont {Chen},\ and\ \citenamefont
  {Yang}}]{Liu2018}%
  \BibitemOpen
  \bibfield  {author} {\bibinfo {author} {\bibfnamefont {C.-C.}\ \bibnamefont
  {Liu}}, \bibinfo {author} {\bibfnamefont {L.-D.}\ \bibnamefont {Zhang}},
  \bibinfo {author} {\bibfnamefont {W.-Q.}\ \bibnamefont {Chen}}, \ and\
  \bibinfo {author} {\bibfnamefont {F.}~\bibnamefont {Yang}},\ }\href {\doibase
  10.1103/PhysRevLett.121.217001} {\bibfield  {journal} {\bibinfo  {journal}
  {Phys. Rev. Lett.}\ }\textbf {\bibinfo {volume} {121}},\ \bibinfo {pages}
  {217001} (\bibinfo {year} {2018})}\BibitemShut {NoStop}%
\bibitem [{\citenamefont {Venderbos}\ and\ \citenamefont
  {Fernandes}(2018)}]{Venderbos18}%
  \BibitemOpen
  \bibfield  {author} {\bibinfo {author} {\bibfnamefont {J.~W.~F.}\
  \bibnamefont {Venderbos}}\ and\ \bibinfo {author} {\bibfnamefont {R.~M.}\
  \bibnamefont {Fernandes}},\ }\href {\doibase 10.1103/PhysRevB.98.245103}
  {\bibfield  {journal} {\bibinfo  {journal} {Phys. Rev. B}\ }\textbf {\bibinfo
  {volume} {98}},\ \bibinfo {pages} {245103} (\bibinfo {year}
  {2018})}\BibitemShut {NoStop}%
\bibitem [{\citenamefont {Song}\ \emph {et~al.}(2019)\citenamefont {Song},
  \citenamefont {Wang}, \citenamefont {Shi}, \citenamefont {Li}, \citenamefont
  {Fang},\ and\ \citenamefont {Bernevig}}]{Song2019}%
  \BibitemOpen
  \bibfield  {author} {\bibinfo {author} {\bibfnamefont {Z.}~\bibnamefont
  {Song}}, \bibinfo {author} {\bibfnamefont {Z.}~\bibnamefont {Wang}}, \bibinfo
  {author} {\bibfnamefont {W.}~\bibnamefont {Shi}}, \bibinfo {author}
  {\bibfnamefont {G.}~\bibnamefont {Li}}, \bibinfo {author} {\bibfnamefont
  {C.}~\bibnamefont {Fang}}, \ and\ \bibinfo {author} {\bibfnamefont {B.~A.}\
  \bibnamefont {Bernevig}},\ }\href {\doibase 10.1103/PhysRevLett.123.036401}
  {\bibfield  {journal} {\bibinfo  {journal} {Phys. Rev. Lett.}\ }\textbf
  {\bibinfo {volume} {123}},\ \bibinfo {pages} {036401} (\bibinfo {year}
  {2019})}\BibitemShut {NoStop}%
\bibitem [{\citenamefont {Kang}\ and\ \citenamefont {Vafek}(2019)}]{Kang2019}%
  \BibitemOpen
  \bibfield  {author} {\bibinfo {author} {\bibfnamefont {J.}~\bibnamefont
  {Kang}}\ and\ \bibinfo {author} {\bibfnamefont {O.}~\bibnamefont {Vafek}},\
  }\href {\doibase 10.1103/PhysRevLett.122.246401} {\bibfield  {journal}
  {\bibinfo  {journal} {Phys. Rev. Lett.}\ }\textbf {\bibinfo {volume} {122}},\
  \bibinfo {pages} {246401} (\bibinfo {year} {2019})}\BibitemShut {NoStop}%
\bibitem [{\citenamefont {Pizarro}\ \emph {et~al.}(2019)\citenamefont
  {Pizarro}, \citenamefont {Calder{\'o}n},\ and\ \citenamefont
  {Bascones}}]{Bascones19}%
  \BibitemOpen
  \bibfield  {author} {\bibinfo {author} {\bibfnamefont {J.~M.}\ \bibnamefont
  {Pizarro}}, \bibinfo {author} {\bibfnamefont {M.~J.}\ \bibnamefont
  {Calder{\'o}n}}, \ and\ \bibinfo {author} {\bibfnamefont {E.}~\bibnamefont
  {Bascones}},\ }\href@noop {} {\bibfield  {journal} {\bibinfo  {journal} {J.
  Phys. Commun.}\ }\textbf {\bibinfo {volume} {3}},\ \bibinfo {pages} {035024}
  (\bibinfo {year} {2019})}\BibitemShut {NoStop}%
\bibitem [{\citenamefont {Tarnopolsky}\ \emph {et~al.}(2019)\citenamefont
  {Tarnopolsky}, \citenamefont {Kruchkov},\ and\ \citenamefont
  {Vishwanath}}]{Tarnopolsky2019}%
  \BibitemOpen
  \bibfield  {author} {\bibinfo {author} {\bibfnamefont {G.}~\bibnamefont
  {Tarnopolsky}}, \bibinfo {author} {\bibfnamefont {A.~J.}\ \bibnamefont
  {Kruchkov}}, \ and\ \bibinfo {author} {\bibfnamefont {A.}~\bibnamefont
  {Vishwanath}},\ }\href {\doibase 10.1103/PhysRevLett.122.106405} {\bibfield
  {journal} {\bibinfo  {journal} {Phys. Rev. Lett.}\ }\textbf {\bibinfo
  {volume} {122}},\ \bibinfo {pages} {106405} (\bibinfo {year}
  {2019})}\BibitemShut {NoStop}%
\bibitem [{\citenamefont {Roy}\ and\ \citenamefont {Juricic}(2019)}]{Roy2019}%
  \BibitemOpen
  \bibfield  {author} {\bibinfo {author} {\bibfnamefont {B.}~\bibnamefont
  {Roy}}\ and\ \bibinfo {author} {\bibfnamefont {V.}~\bibnamefont {Juricic}},\
  }\href {\doibase 10.1103/PhysRevB.99.121407} {\bibfield  {journal} {\bibinfo
  {journal} {Phys. Rev. B}\ }\textbf {\bibinfo {volume} {99}},\ \bibinfo
  {pages} {121407} (\bibinfo {year} {2019})}\BibitemShut {NoStop}%
\bibitem [{\citenamefont {Lin}\ and\ \citenamefont
  {Nandkishore}(2019)}]{Nandkishore2019}%
  \BibitemOpen
  \bibfield  {author} {\bibinfo {author} {\bibfnamefont {Y.-P.}\ \bibnamefont
  {Lin}}\ and\ \bibinfo {author} {\bibfnamefont {R.~M.}\ \bibnamefont
  {Nandkishore}},\ }\href {\doibase 10.1103/PhysRevB.100.085136} {\bibfield
  {journal} {\bibinfo  {journal} {Phys. Rev. B}\ }\textbf {\bibinfo {volume}
  {100}},\ \bibinfo {pages} {085136} (\bibinfo {year} {2019})}\BibitemShut
  {NoStop}%
\bibitem [{\citenamefont {Hejazi}\ \emph {et~al.}(2019)\citenamefont {Hejazi},
  \citenamefont {Liu}, \citenamefont {Shapourian}, \citenamefont {Chen},\ and\
  \citenamefont {Balents}}]{Balents19}%
  \BibitemOpen
  \bibfield  {author} {\bibinfo {author} {\bibfnamefont {K.}~\bibnamefont
  {Hejazi}}, \bibinfo {author} {\bibfnamefont {C.}~\bibnamefont {Liu}},
  \bibinfo {author} {\bibfnamefont {H.}~\bibnamefont {Shapourian}}, \bibinfo
  {author} {\bibfnamefont {X.}~\bibnamefont {Chen}}, \ and\ \bibinfo {author}
  {\bibfnamefont {L.}~\bibnamefont {Balents}},\ }\href {\doibase
  10.1103/PhysRevB.99.035111} {\bibfield  {journal} {\bibinfo  {journal} {Phys.
  Rev. B}\ }\textbf {\bibinfo {volume} {99}},\ \bibinfo {pages} {035111}
  (\bibinfo {year} {2019})}\BibitemShut {NoStop}%
\bibitem [{\citenamefont {Huang}\ \emph {et~al.}(2019)\citenamefont {Huang},
  \citenamefont {Zhang},\ and\ \citenamefont {Ma}}]{Huang2019}%
  \BibitemOpen
  \bibfield  {author} {\bibinfo {author} {\bibfnamefont {T.}~\bibnamefont
  {Huang}}, \bibinfo {author} {\bibfnamefont {L.}~\bibnamefont {Zhang}}, \ and\
  \bibinfo {author} {\bibfnamefont {T.}~\bibnamefont {Ma}},\ }\href {\doibase
  10.1016/j.scib.2019.01.026} {\bibfield  {journal} {\bibinfo  {journal} {Sci.
  Bull.}\ }\textbf {\bibinfo {volume} {64}},\ \bibinfo {pages} {310} (\bibinfo
  {year} {2019})}\BibitemShut {NoStop}%
\bibitem [{\citenamefont {Zhang}\ \emph {et~al.}(2019)\citenamefont {Zhang},
  \citenamefont {Mao}, \citenamefont {Cao}, \citenamefont {Jarillo-Herrero},\
  and\ \citenamefont {Senthil}}]{Zhang2019a}%
  \BibitemOpen
  \bibfield  {author} {\bibinfo {author} {\bibfnamefont {Y.-H.}\ \bibnamefont
  {Zhang}}, \bibinfo {author} {\bibfnamefont {D.}~\bibnamefont {Mao}}, \bibinfo
  {author} {\bibfnamefont {Y.}~\bibnamefont {Cao}}, \bibinfo {author}
  {\bibfnamefont {P.}~\bibnamefont {Jarillo-Herrero}}, \ and\ \bibinfo {author}
  {\bibfnamefont {T.}~\bibnamefont {Senthil}},\ }\href {\doibase
  10.1103/PhysRevB.99.075127} {\bibfield  {journal} {\bibinfo  {journal} {Phys.
  Rev. B}\ }\textbf {\bibinfo {volume} {99}},\ \bibinfo {pages} {075127}
  (\bibinfo {year} {2019})}\BibitemShut {NoStop}%
\bibitem [{\citenamefont {Gonz{\'a}lez}\ and\ \citenamefont
  {Stauber}(2019)}]{Gonzalez2019}%
  \BibitemOpen
  \bibfield  {author} {\bibinfo {author} {\bibfnamefont {J.}~\bibnamefont
  {Gonz{\'a}lez}}\ and\ \bibinfo {author} {\bibfnamefont {T.}~\bibnamefont
  {Stauber}},\ }\href {\doibase 10.1103/PhysRevLett.122.026801} {\bibfield
  {journal} {\bibinfo  {journal} {Phys. Rev. Lett.}\ }\textbf {\bibinfo
  {volume} {122}},\ \bibinfo {pages} {026801} (\bibinfo {year}
  {2019})}\BibitemShut {NoStop}%
\bibitem [{\citenamefont {Classen}\ \emph {et~al.}(2019)\citenamefont
  {Classen}, \citenamefont {Honerkamp},\ and\ \citenamefont
  {Scherer}}]{Classen19}%
  \BibitemOpen
  \bibfield  {author} {\bibinfo {author} {\bibfnamefont {L.}~\bibnamefont
  {Classen}}, \bibinfo {author} {\bibfnamefont {C.}~\bibnamefont {Honerkamp}},
  \ and\ \bibinfo {author} {\bibfnamefont {M.~M.}\ \bibnamefont {Scherer}},\
  }\href {\doibase 10.1103/PhysRevB.99.195120} {\bibfield  {journal} {\bibinfo
  {journal} {Phys. Rev. B}\ }\textbf {\bibinfo {volume} {99}},\ \bibinfo
  {pages} {195120} (\bibinfo {year} {2019})}\BibitemShut {NoStop}%
\bibitem [{\citenamefont {Seo}\ \emph {et~al.}(2019)\citenamefont {Seo},
  \citenamefont {Kotov},\ and\ \citenamefont {Uchoa}}]{Uchoa19}%
  \BibitemOpen
  \bibfield  {author} {\bibinfo {author} {\bibfnamefont {K.}~\bibnamefont
  {Seo}}, \bibinfo {author} {\bibfnamefont {V.~N.}\ \bibnamefont {Kotov}}, \
  and\ \bibinfo {author} {\bibfnamefont {B.}~\bibnamefont {Uchoa}},\ }\href
  {\doibase 10.1103/PhysRevLett.122.246402} {\bibfield  {journal} {\bibinfo
  {journal} {Phys. Rev. Lett.}\ }\textbf {\bibinfo {volume} {122}},\ \bibinfo
  {pages} {246402} (\bibinfo {year} {2019})}\BibitemShut {NoStop}%
\bibitem [{\citenamefont {Yuan}\ \emph {et~al.}(2019)\citenamefont {Yuan},
  \citenamefont {Isobe},\ and\ \citenamefont {Fu}}]{Yuan2019magic}%
  \BibitemOpen
  \bibfield  {author} {\bibinfo {author} {\bibfnamefont {N.~F.}\ \bibnamefont
  {Yuan}}, \bibinfo {author} {\bibfnamefont {H.}~\bibnamefont {Isobe}}, \ and\
  \bibinfo {author} {\bibfnamefont {L.}~\bibnamefont {Fu}},\ }\href@noop {}
  {\bibfield  {journal} {\bibinfo  {journal} {Nature Communications}\ }\textbf
  {\bibinfo {volume} {10}},\ \bibinfo {pages} {5769} (\bibinfo {year}
  {2019})}\BibitemShut {NoStop}%
\bibitem [{\citenamefont {Kang}\ and\ \citenamefont {Vafek}(2020)}]{Vafek20}%
  \BibitemOpen
  \bibfield  {author} {\bibinfo {author} {\bibfnamefont {J.}~\bibnamefont
  {Kang}}\ and\ \bibinfo {author} {\bibfnamefont {O.}~\bibnamefont {Vafek}},\
  }\href {\doibase 10.1103/PhysRevB.102.035161} {\bibfield  {journal} {\bibinfo
   {journal} {Phys. Rev. B}\ }\textbf {\bibinfo {volume} {102}},\ \bibinfo
  {pages} {035161} (\bibinfo {year} {2020})}\BibitemShut {NoStop}%
\bibitem [{\citenamefont {Xie}\ and\ \citenamefont
  {MacDonald}(2020)}]{MacDonald20}%
  \BibitemOpen
  \bibfield  {author} {\bibinfo {author} {\bibfnamefont {M.}~\bibnamefont
  {Xie}}\ and\ \bibinfo {author} {\bibfnamefont {A.~H.}\ \bibnamefont
  {MacDonald}},\ }\href {\doibase 10.1103/PhysRevLett.124.097601} {\bibfield
  {journal} {\bibinfo  {journal} {Phys. Rev. Lett.}\ }\textbf {\bibinfo
  {volume} {124}},\ \bibinfo {pages} {097601} (\bibinfo {year}
  {2020})}\BibitemShut {NoStop}%
\bibitem [{\citenamefont {Repellin}\ \emph {et~al.}(2020)\citenamefont
  {Repellin}, \citenamefont {Dong}, \citenamefont {Zhang},\ and\ \citenamefont
  {Senthil}}]{Repellin20}%
  \BibitemOpen
  \bibfield  {author} {\bibinfo {author} {\bibfnamefont {C.}~\bibnamefont
  {Repellin}}, \bibinfo {author} {\bibfnamefont {Z.}~\bibnamefont {Dong}},
  \bibinfo {author} {\bibfnamefont {Y.-H.}\ \bibnamefont {Zhang}}, \ and\
  \bibinfo {author} {\bibfnamefont {T.}~\bibnamefont {Senthil}},\ }\href
  {\doibase 10.1103/PhysRevLett.124.187601} {\bibfield  {journal} {\bibinfo
  {journal} {Phys. Rev. Lett.}\ }\textbf {\bibinfo {volume} {124}},\ \bibinfo
  {pages} {187601} (\bibinfo {year} {2020})}\BibitemShut {NoStop}%
\bibitem [{\citenamefont {Xu}\ \emph {et~al.}(2020)\citenamefont {Xu},
  \citenamefont {Wu}, \citenamefont {Jian},\ and\ \citenamefont
  {Xu}}]{Cenke_nematics}%
  \BibitemOpen
  \bibfield  {author} {\bibinfo {author} {\bibfnamefont {Y.}~\bibnamefont
  {Xu}}, \bibinfo {author} {\bibfnamefont {X.-C.}\ \bibnamefont {Wu}}, \bibinfo
  {author} {\bibfnamefont {C.-M.}\ \bibnamefont {Jian}}, \ and\ \bibinfo
  {author} {\bibfnamefont {C.}~\bibnamefont {Xu}},\ }\href {\doibase
  10.1103/PhysRevB.101.205426} {\bibfield  {journal} {\bibinfo  {journal}
  {Phys. Rev. B}\ }\textbf {\bibinfo {volume} {101}},\ \bibinfo {pages}
  {205426} (\bibinfo {year} {2020})}\BibitemShut {NoStop}%
\bibitem [{\citenamefont {Christos}\ \emph {et~al.}(2020)\citenamefont
  {Christos}, \citenamefont {Sachdev},\ and\ \citenamefont
  {Scheurer}}]{Christos20}%
  \BibitemOpen
  \bibfield  {author} {\bibinfo {author} {\bibfnamefont {M.}~\bibnamefont
  {Christos}}, \bibinfo {author} {\bibfnamefont {S.}~\bibnamefont {Sachdev}}, \
  and\ \bibinfo {author} {\bibfnamefont {M.~S.}\ \bibnamefont {Scheurer}},\
  }\href {\doibase 10.1073/pnas.2014691117} {\bibfield  {journal} {\bibinfo
  {journal} {Proc. Natl. Acad. Sci. U.S.A.}\ }\textbf {\bibinfo {volume}
  {117}},\ \bibinfo {pages} {29543} (\bibinfo {year} {2020})}\BibitemShut
  {NoStop}%
\bibitem [{\citenamefont {Bultinck}\ \emph {et~al.}(2020)\citenamefont
  {Bultinck}, \citenamefont {Khalaf}, \citenamefont {Liu}, \citenamefont
  {Chatterjee}, \citenamefont {Vishwanath},\ and\ \citenamefont
  {Zaletel}}]{Bultinck20}%
  \BibitemOpen
  \bibfield  {author} {\bibinfo {author} {\bibfnamefont {N.}~\bibnamefont
  {Bultinck}}, \bibinfo {author} {\bibfnamefont {E.}~\bibnamefont {Khalaf}},
  \bibinfo {author} {\bibfnamefont {S.}~\bibnamefont {Liu}}, \bibinfo {author}
  {\bibfnamefont {S.}~\bibnamefont {Chatterjee}}, \bibinfo {author}
  {\bibfnamefont {A.}~\bibnamefont {Vishwanath}}, \ and\ \bibinfo {author}
  {\bibfnamefont {M.~P.}\ \bibnamefont {Zaletel}},\ }\href {\doibase
  10.1103/PhysRevX.10.031034} {\bibfield  {journal} {\bibinfo  {journal} {Phys.
  Rev. X}\ }\textbf {\bibinfo {volume} {10}},\ \bibinfo {pages} {031034}
  (\bibinfo {year} {2020})}\BibitemShut {NoStop}%
\bibitem [{\citenamefont {Vafek}\ and\ \citenamefont
  {Kang}(2020)}]{Vafek20_RG}%
  \BibitemOpen
  \bibfield  {author} {\bibinfo {author} {\bibfnamefont {O.}~\bibnamefont
  {Vafek}}\ and\ \bibinfo {author} {\bibfnamefont {J.}~\bibnamefont {Kang}},\
  }\href {\doibase 10.1103/PhysRevLett.125.257602} {\bibfield  {journal}
  {\bibinfo  {journal} {Phys. Rev. Lett.}\ }\textbf {\bibinfo {volume} {125}},\
  \bibinfo {pages} {257602} (\bibinfo {year} {2020})}\BibitemShut {NoStop}%
\bibitem [{\citenamefont {Zhang}\ \emph {et~al.}(2020)\citenamefont {Zhang},
  \citenamefont {Jiang}, \citenamefont {Wang},\ and\ \citenamefont
  {Zhang}}]{FCZhang20}%
  \BibitemOpen
  \bibfield  {author} {\bibinfo {author} {\bibfnamefont {Y.}~\bibnamefont
  {Zhang}}, \bibinfo {author} {\bibfnamefont {K.}~\bibnamefont {Jiang}},
  \bibinfo {author} {\bibfnamefont {Z.}~\bibnamefont {Wang}}, \ and\ \bibinfo
  {author} {\bibfnamefont {F.}~\bibnamefont {Zhang}},\ }\href {\doibase
  10.1103/PhysRevB.102.035136} {\bibfield  {journal} {\bibinfo  {journal}
  {Phys. Rev. B}\ }\textbf {\bibinfo {volume} {102}},\ \bibinfo {pages}
  {035136} (\bibinfo {year} {2020})}\BibitemShut {NoStop}%
\bibitem [{\citenamefont {Cea}\ and\ \citenamefont {Guinea}(2020)}]{Cea20}%
  \BibitemOpen
  \bibfield  {author} {\bibinfo {author} {\bibfnamefont {T.}~\bibnamefont
  {Cea}}\ and\ \bibinfo {author} {\bibfnamefont {F.}~\bibnamefont {Guinea}},\
  }\href {\doibase 10.1103/PhysRevB.102.045107} {\bibfield  {journal} {\bibinfo
   {journal} {Phys. Rev. B}\ }\textbf {\bibinfo {volume} {102}},\ \bibinfo
  {pages} {045107} (\bibinfo {year} {2020})}\BibitemShut {NoStop}%
\bibitem [{\citenamefont {Brillaux}\ \emph {et~al.}(2020)\citenamefont
  {Brillaux}, \citenamefont {Carpentier}, \citenamefont {Fedorenko},\ and\
  \citenamefont {Savary}}]{Savary20}%
  \BibitemOpen
  \bibfield  {author} {\bibinfo {author} {\bibfnamefont {E.}~\bibnamefont
  {Brillaux}}, \bibinfo {author} {\bibfnamefont {D.}~\bibnamefont
  {Carpentier}}, \bibinfo {author} {\bibfnamefont {A.~A.}\ \bibnamefont
  {Fedorenko}}, \ and\ \bibinfo {author} {\bibfnamefont {L.}~\bibnamefont
  {Savary}},\ }\href@noop {} {\bibfield  {journal} {\bibinfo  {journal}
  {arXiv:2008.05041}\ } (\bibinfo {year} {2020})}\BibitemShut {NoStop}%
\bibitem [{\citenamefont {Bernevig}\ \emph {et~al.}(2021)\citenamefont
  {Bernevig}, \citenamefont {Song}, \citenamefont {Regnault},\ and\
  \citenamefont {Lian}}]{Bernevig_TBG3}%
  \BibitemOpen
  \bibfield  {author} {\bibinfo {author} {\bibfnamefont {B.~A.}\ \bibnamefont
  {Bernevig}}, \bibinfo {author} {\bibfnamefont {Z.-D.}\ \bibnamefont {Song}},
  \bibinfo {author} {\bibfnamefont {N.}~\bibnamefont {Regnault}}, \ and\
  \bibinfo {author} {\bibfnamefont {B.}~\bibnamefont {Lian}},\ }\href {\doibase
  10.1103/PhysRevB.103.205413} {\bibfield  {journal} {\bibinfo  {journal}
  {Phys. Rev. B}\ }\textbf {\bibinfo {volume} {103}},\ \bibinfo {pages}
  {205413} (\bibinfo {year} {2021})}\BibitemShut {NoStop}%
\bibitem [{\citenamefont {Lian}\ \emph {et~al.}(2021)\citenamefont {Lian},
  \citenamefont {Song}, \citenamefont {Regnault}, \citenamefont {Efetov},
  \citenamefont {Yazdani},\ and\ \citenamefont {Bernevig}}]{Bernevig_TBG4}%
  \BibitemOpen
  \bibfield  {author} {\bibinfo {author} {\bibfnamefont {B.}~\bibnamefont
  {Lian}}, \bibinfo {author} {\bibfnamefont {Z.-D.}\ \bibnamefont {Song}},
  \bibinfo {author} {\bibfnamefont {N.}~\bibnamefont {Regnault}}, \bibinfo
  {author} {\bibfnamefont {D.~K.}\ \bibnamefont {Efetov}}, \bibinfo {author}
  {\bibfnamefont {A.}~\bibnamefont {Yazdani}}, \ and\ \bibinfo {author}
  {\bibfnamefont {B.~A.}\ \bibnamefont {Bernevig}},\ }\href {\doibase
  10.1103/PhysRevB.103.205414} {\bibfield  {journal} {\bibinfo  {journal}
  {Phys. Rev. B}\ }\textbf {\bibinfo {volume} {103}},\ \bibinfo {pages}
  {205414} (\bibinfo {year} {2021})}\BibitemShut {NoStop}%
\bibitem [{\citenamefont {Xie}\ \emph {et~al.}(2021)\citenamefont {Xie},
  \citenamefont {Cowsik}, \citenamefont {Song}, \citenamefont {Lian},
  \citenamefont {Bernevig},\ and\ \citenamefont {Regnault}}]{Bernevig_TBG5}%
  \BibitemOpen
  \bibfield  {author} {\bibinfo {author} {\bibfnamefont {F.}~\bibnamefont
  {Xie}}, \bibinfo {author} {\bibfnamefont {A.}~\bibnamefont {Cowsik}},
  \bibinfo {author} {\bibfnamefont {Z.-D.}\ \bibnamefont {Song}}, \bibinfo
  {author} {\bibfnamefont {B.}~\bibnamefont {Lian}}, \bibinfo {author}
  {\bibfnamefont {B.~A.}\ \bibnamefont {Bernevig}}, \ and\ \bibinfo {author}
  {\bibfnamefont {N.}~\bibnamefont {Regnault}},\ }\href {\doibase
  10.1103/PhysRevB.103.205416} {\bibfield  {journal} {\bibinfo  {journal}
  {Phys. Rev. B}\ }\textbf {\bibinfo {volume} {103}},\ \bibinfo {pages}
  {205416} (\bibinfo {year} {2021})}\BibitemShut {NoStop}%
\bibitem [{\citenamefont {Da~Liao}\ \emph {et~al.}(2021)\citenamefont
  {Da~Liao}, \citenamefont {Kang}, \citenamefont {Brei\o{}}, \citenamefont
  {Xu}, \citenamefont {Wu}, \citenamefont {Andersen}, \citenamefont
  {Fernandes},\ and\ \citenamefont {Meng}}]{ZYMeng21}%
  \BibitemOpen
  \bibfield  {author} {\bibinfo {author} {\bibfnamefont {Y.}~\bibnamefont
  {Da~Liao}}, \bibinfo {author} {\bibfnamefont {J.}~\bibnamefont {Kang}},
  \bibinfo {author} {\bibfnamefont {C.~N.}\ \bibnamefont {Brei\o{}}}, \bibinfo
  {author} {\bibfnamefont {X.~Y.}\ \bibnamefont {Xu}}, \bibinfo {author}
  {\bibfnamefont {H.-Q.}\ \bibnamefont {Wu}}, \bibinfo {author} {\bibfnamefont
  {B.~M.}\ \bibnamefont {Andersen}}, \bibinfo {author} {\bibfnamefont {R.~M.}\
  \bibnamefont {Fernandes}}, \ and\ \bibinfo {author} {\bibfnamefont {Z.~Y.}\
  \bibnamefont {Meng}},\ }\href {\doibase 10.1103/PhysRevX.11.011014}
  {\bibfield  {journal} {\bibinfo  {journal} {Phys. Rev. X}\ }\textbf {\bibinfo
  {volume} {11}},\ \bibinfo {pages} {011014} (\bibinfo {year}
  {2021})}\BibitemShut {NoStop}%
\bibitem [{\citenamefont {Wang}\ \emph {et~al.}(2021)\citenamefont {Wang},
  \citenamefont {Kang},\ and\ \citenamefont {Fernandes}}]{Wang2021}%
  \BibitemOpen
  \bibfield  {author} {\bibinfo {author} {\bibfnamefont {Y.}~\bibnamefont
  {Wang}}, \bibinfo {author} {\bibfnamefont {J.}~\bibnamefont {Kang}}, \ and\
  \bibinfo {author} {\bibfnamefont {R.~M.}\ \bibnamefont {Fernandes}},\ }\href
  {\doibase 10.1103/PhysRevB.103.024506} {\bibfield  {journal} {\bibinfo
  {journal} {Phys. Rev. B}\ }\textbf {\bibinfo {volume} {103}},\ \bibinfo
  {pages} {024506} (\bibinfo {year} {2021})}\BibitemShut {NoStop}%
\bibitem [{\citenamefont {Potasz}\ \emph {et~al.}(2021)\citenamefont {Potasz},
  \citenamefont {Xie},\ and\ \citenamefont {MacDonald}}]{Potasz2021}%
  \BibitemOpen
  \bibfield  {author} {\bibinfo {author} {\bibfnamefont {P.}~\bibnamefont
  {Potasz}}, \bibinfo {author} {\bibfnamefont {M.}~\bibnamefont {Xie}}, \ and\
  \bibinfo {author} {\bibfnamefont {A.~H.}\ \bibnamefont {MacDonald}},\
  }\href@noop {} {\bibfield  {journal} {\bibinfo  {journal} {arXiv:2102.02256}\
  } (\bibinfo {year} {2021})}\BibitemShut {NoStop}%
\bibitem [{\citenamefont {Khalaf}\ \emph {et~al.}(2021)\citenamefont {Khalaf},
  \citenamefont {Chatterjee}, \citenamefont {Bultinck}, \citenamefont
  {Zaletel},\ and\ \citenamefont {Vishwanath}}]{Khalaf21}%
  \BibitemOpen
  \bibfield  {author} {\bibinfo {author} {\bibfnamefont {E.}~\bibnamefont
  {Khalaf}}, \bibinfo {author} {\bibfnamefont {S.}~\bibnamefont {Chatterjee}},
  \bibinfo {author} {\bibfnamefont {N.}~\bibnamefont {Bultinck}}, \bibinfo
  {author} {\bibfnamefont {M.~P.}\ \bibnamefont {Zaletel}}, \ and\ \bibinfo
  {author} {\bibfnamefont {A.}~\bibnamefont {Vishwanath}},\ }\href {\doibase
  10.1126/sciadv.abf5299} {\bibfield  {journal} {\bibinfo  {journal} {Science
  Advances}\ }\textbf {\bibinfo {volume} {7}} (\bibinfo {year} {2021}),\
  10.1126/sciadv.abf5299}\BibitemShut {NoStop}%
\bibitem [{\citenamefont {Kang}\ \emph {et~al.}(2021)\citenamefont {Kang},
  \citenamefont {Bernevig},\ and\ \citenamefont {Vafek}}]{Vafek21}%
  \BibitemOpen
  \bibfield  {author} {\bibinfo {author} {\bibfnamefont {J.}~\bibnamefont
  {Kang}}, \bibinfo {author} {\bibfnamefont {B.~A.}\ \bibnamefont {Bernevig}},
  \ and\ \bibinfo {author} {\bibfnamefont {O.}~\bibnamefont {Vafek}},\
  }\href@noop {} {\bibfield  {journal} {\bibinfo  {journal} {arXiv:2104.01145}\
  } (\bibinfo {year} {2021})}\BibitemShut {NoStop}%
\bibitem [{\citenamefont {Chichinadze}\ \emph {et~al.}(2021)\citenamefont
  {Chichinadze}, \citenamefont {Classen}, \citenamefont {Wang},\ and\
  \citenamefont {Chubukov}}]{Chichinadze21}%
  \BibitemOpen
  \bibfield  {author} {\bibinfo {author} {\bibfnamefont {D.~V.}\ \bibnamefont
  {Chichinadze}}, \bibinfo {author} {\bibfnamefont {L.}~\bibnamefont
  {Classen}}, \bibinfo {author} {\bibfnamefont {Y.}~\bibnamefont {Wang}}, \
  and\ \bibinfo {author} {\bibfnamefont {A.~V.}\ \bibnamefont {Chubukov}},\
  }\href@noop {} {\bibfield  {journal} {\bibinfo  {journal} {arXiv:2108.05334}\
  } (\bibinfo {year} {2021})}\BibitemShut {NoStop}%
\bibitem [{\citenamefont {Wu}\ \emph {et~al.}(2018)\citenamefont {Wu},
  \citenamefont {MacDonald},\ and\ \citenamefont {Martin}}]{Wu_etal}%
  \BibitemOpen
  \bibfield  {author} {\bibinfo {author} {\bibfnamefont {F.}~\bibnamefont
  {Wu}}, \bibinfo {author} {\bibfnamefont {A.~H.}\ \bibnamefont {MacDonald}}, \
  and\ \bibinfo {author} {\bibfnamefont {I.}~\bibnamefont {Martin}},\ }\href
  {\doibase 10.1103/PhysRevLett.121.257001} {\bibfield  {journal} {\bibinfo
  {journal} {Phys. Rev. Lett.}\ }\textbf {\bibinfo {volume} {121}},\ \bibinfo
  {pages} {257001} (\bibinfo {year} {2018})}\BibitemShut {NoStop}%
\bibitem [{\citenamefont {Lian}\ \emph {et~al.}(2019)\citenamefont {Lian},
  \citenamefont {Wang},\ and\ \citenamefont {Bernevig}}]{Bernevig_phonons}%
  \BibitemOpen
  \bibfield  {author} {\bibinfo {author} {\bibfnamefont {B.}~\bibnamefont
  {Lian}}, \bibinfo {author} {\bibfnamefont {Z.}~\bibnamefont {Wang}}, \ and\
  \bibinfo {author} {\bibfnamefont {B.~A.}\ \bibnamefont {Bernevig}},\ }\href
  {\doibase 10.1103/PhysRevLett.122.257002} {\bibfield  {journal} {\bibinfo
  {journal} {Phys. Rev. Lett.}\ }\textbf {\bibinfo {volume} {122}},\ \bibinfo
  {pages} {257002} (\bibinfo {year} {2019})}\BibitemShut {NoStop}%
\bibitem [{\citenamefont {Wu}\ \emph {et~al.}(2019)\citenamefont {Wu},
  \citenamefont {Hwang},\ and\ \citenamefont {Das~Sarma}}]{Wu_etal2}%
  \BibitemOpen
  \bibfield  {author} {\bibinfo {author} {\bibfnamefont {F.}~\bibnamefont
  {Wu}}, \bibinfo {author} {\bibfnamefont {E.}~\bibnamefont {Hwang}}, \ and\
  \bibinfo {author} {\bibfnamefont {S.}~\bibnamefont {Das~Sarma}},\ }\href
  {\doibase 10.1103/PhysRevB.99.165112} {\bibfield  {journal} {\bibinfo
  {journal} {Phys. Rev. B}\ }\textbf {\bibinfo {volume} {99}},\ \bibinfo
  {pages} {165112} (\bibinfo {year} {2019})}\BibitemShut {NoStop}%
\bibitem [{\citenamefont {Angeli}\ \emph {et~al.}(2019)\citenamefont {Angeli},
  \citenamefont {Tosatti},\ and\ \citenamefont {Fabrizio}}]{Fabrizio19}%
  \BibitemOpen
  \bibfield  {author} {\bibinfo {author} {\bibfnamefont {M.}~\bibnamefont
  {Angeli}}, \bibinfo {author} {\bibfnamefont {E.}~\bibnamefont {Tosatti}}, \
  and\ \bibinfo {author} {\bibfnamefont {M.}~\bibnamefont {Fabrizio}},\ }\href
  {\doibase 10.1103/PhysRevX.9.041010} {\bibfield  {journal} {\bibinfo
  {journal} {Phys. Rev. X}\ }\textbf {\bibinfo {volume} {9}},\ \bibinfo {pages}
  {041010} (\bibinfo {year} {2019})}\BibitemShut {NoStop}%
\bibitem [{\citenamefont {Lewandowski}\ \emph {et~al.}(2021)\citenamefont
  {Lewandowski}, \citenamefont {Chowdhury},\ and\ \citenamefont
  {Ruhman}}]{Lewandoski_etal}%
  \BibitemOpen
  \bibfield  {author} {\bibinfo {author} {\bibfnamefont {C.}~\bibnamefont
  {Lewandowski}}, \bibinfo {author} {\bibfnamefont {D.}~\bibnamefont
  {Chowdhury}}, \ and\ \bibinfo {author} {\bibfnamefont {J.}~\bibnamefont
  {Ruhman}},\ }\href@noop {} {\bibfield  {journal} {\bibinfo  {journal} {Phys.
  Rev. B}\ }\textbf {\bibinfo {volume} {103}},\ \bibinfo {pages} {235401}
  (\bibinfo {year} {2021})}\BibitemShut {NoStop}%
\bibitem [{\citenamefont {Cea}\ and\ \citenamefont
  {Guinea}(2021)}]{Cea_Guinea}%
  \BibitemOpen
  \bibfield  {author} {\bibinfo {author} {\bibfnamefont {T.}~\bibnamefont
  {Cea}}\ and\ \bibinfo {author} {\bibfnamefont {F.}~\bibnamefont {Guinea}},\
  }\href@noop {} {\bibfield  {journal} {\bibinfo  {journal} {arXiv:2103.01815}\
  } (\bibinfo {year} {2021})}\BibitemShut {NoStop}%
\bibitem [{\citenamefont {Fernandes}\ and\ \citenamefont
  {Venderbos}(2020)}]{moire_nematicity}%
  \BibitemOpen
  \bibfield  {author} {\bibinfo {author} {\bibfnamefont {R.~M.}\ \bibnamefont
  {Fernandes}}\ and\ \bibinfo {author} {\bibfnamefont {J.~W.~F.}\ \bibnamefont
  {Venderbos}},\ }\href {\doibase 10.1126/sciadv.aba8834} {\bibfield  {journal}
  {\bibinfo  {journal} {Science Advances}\ }\textbf {\bibinfo {volume} {6}},\
  \bibinfo {pages} {eaba8834} (\bibinfo {year} {2020})}\BibitemShut {NoStop}%
\bibitem [{\citenamefont {Stepanov}\ \emph {et~al.}(2020)\citenamefont
  {Stepanov}, \citenamefont {Das}, \citenamefont {Lu}, \citenamefont
  {Fahimniya}, \citenamefont {Watanabe}, \citenamefont {Taniguchi},
  \citenamefont {Koppens}, \citenamefont {Lischner}, \citenamefont {Levitov},\
  and\ \citenamefont {Efetov}}]{exp1}%
  \BibitemOpen
  \bibfield  {author} {\bibinfo {author} {\bibfnamefont {P.}~\bibnamefont
  {Stepanov}}, \bibinfo {author} {\bibfnamefont {I.}~\bibnamefont {Das}},
  \bibinfo {author} {\bibfnamefont {X.}~\bibnamefont {Lu}}, \bibinfo {author}
  {\bibfnamefont {A.}~\bibnamefont {Fahimniya}}, \bibinfo {author}
  {\bibfnamefont {K.}~\bibnamefont {Watanabe}}, \bibinfo {author}
  {\bibfnamefont {T.}~\bibnamefont {Taniguchi}}, \bibinfo {author}
  {\bibfnamefont {F.~H.~L.}\ \bibnamefont {Koppens}}, \bibinfo {author}
  {\bibfnamefont {J.}~\bibnamefont {Lischner}}, \bibinfo {author}
  {\bibfnamefont {L.}~\bibnamefont {Levitov}}, \ and\ \bibinfo {author}
  {\bibfnamefont {D.~K.}\ \bibnamefont {Efetov}},\ }\href@noop {} {\bibfield
  {journal} {\bibinfo  {journal} {Nature}\ }\textbf {\bibinfo {volume} {583}},\
  \bibinfo {pages} {375} (\bibinfo {year} {2020})}\BibitemShut {NoStop}%
\bibitem [{\citenamefont {Saito}\ \emph {et~al.}(2020)\citenamefont {Saito},
  \citenamefont {Ge}, \citenamefont {Watanabe}, \citenamefont {Taniguchi},\
  and\ \citenamefont {Young}}]{exp2}%
  \BibitemOpen
  \bibfield  {author} {\bibinfo {author} {\bibfnamefont {Y.}~\bibnamefont
  {Saito}}, \bibinfo {author} {\bibfnamefont {J.}~\bibnamefont {Ge}}, \bibinfo
  {author} {\bibfnamefont {K.}~\bibnamefont {Watanabe}}, \bibinfo {author}
  {\bibfnamefont {T.}~\bibnamefont {Taniguchi}}, \ and\ \bibinfo {author}
  {\bibfnamefont {A.~F.}\ \bibnamefont {Young}},\ }\href@noop {} {\bibfield
  {journal} {\bibinfo  {journal} {Nat. Phys.}\ }\textbf {\bibinfo {volume}
  {16}},\ \bibinfo {pages} {926} (\bibinfo {year} {2020})}\BibitemShut
  {NoStop}%
\bibitem [{\citenamefont {Liu}\ \emph {et~al.}(2021)\citenamefont {Liu},
  \citenamefont {Wang}, \citenamefont {Watanabe}, \citenamefont {Taniguchi},
  \citenamefont {Vafek},\ and\ \citenamefont {Li}}]{exp3}%
  \BibitemOpen
  \bibfield  {author} {\bibinfo {author} {\bibfnamefont {X.}~\bibnamefont
  {Liu}}, \bibinfo {author} {\bibfnamefont {Z.}~\bibnamefont {Wang}}, \bibinfo
  {author} {\bibfnamefont {K.}~\bibnamefont {Watanabe}}, \bibinfo {author}
  {\bibfnamefont {T.}~\bibnamefont {Taniguchi}}, \bibinfo {author}
  {\bibfnamefont {O.}~\bibnamefont {Vafek}}, \ and\ \bibinfo {author}
  {\bibfnamefont {J.~I.~A.}\ \bibnamefont {Li}},\ }\href@noop {} {\bibfield
  {journal} {\bibinfo  {journal} {Science}\ }\textbf {\bibinfo {volume}
  {371}},\ \bibinfo {pages} {1261} (\bibinfo {year} {2021})}\BibitemShut
  {NoStop}%
\bibitem [{\citenamefont {Koshino}\ \emph {et~al.}(2018)\citenamefont
  {Koshino}, \citenamefont {Yuan}, \citenamefont {Koretsune}, \citenamefont
  {Ochi}, \citenamefont {Kuroki},\ and\ \citenamefont {Fu}}]{Koshino2018}%
  \BibitemOpen
  \bibfield  {author} {\bibinfo {author} {\bibfnamefont {M.}~\bibnamefont
  {Koshino}}, \bibinfo {author} {\bibfnamefont {N.~F.~Q.}\ \bibnamefont
  {Yuan}}, \bibinfo {author} {\bibfnamefont {T.}~\bibnamefont {Koretsune}},
  \bibinfo {author} {\bibfnamefont {M.}~\bibnamefont {Ochi}}, \bibinfo {author}
  {\bibfnamefont {K.}~\bibnamefont {Kuroki}}, \ and\ \bibinfo {author}
  {\bibfnamefont {L.}~\bibnamefont {Fu}},\ }\href {\doibase
  10.1103/PhysRevX.8.031087} {\bibfield  {journal} {\bibinfo  {journal} {Phys.
  Rev. X}\ }\textbf {\bibinfo {volume} {8}},\ \bibinfo {pages} {031087}
  (\bibinfo {year} {2018})}\BibitemShut {NoStop}%
\bibitem [{\citenamefont {Kang}\ and\ \citenamefont {Vafek}(2018)}]{Kang2018}%
  \BibitemOpen
  \bibfield  {author} {\bibinfo {author} {\bibfnamefont {J.}~\bibnamefont
  {Kang}}\ and\ \bibinfo {author} {\bibfnamefont {O.}~\bibnamefont {Vafek}},\
  }\href {\doibase 10.1103/PhysRevX.8.031088} {\bibfield  {journal} {\bibinfo
  {journal} {Phys. Rev. X}\ }\textbf {\bibinfo {volume} {8}},\ \bibinfo {pages}
  {031088} (\bibinfo {year} {2018})}\BibitemShut {NoStop}%
\bibitem [{\citenamefont {Yuan}\ and\ \citenamefont {Fu}(2018)}]{Yuan2018}%
  \BibitemOpen
  \bibfield  {author} {\bibinfo {author} {\bibfnamefont {N.~F.~Q.}\
  \bibnamefont {Yuan}}\ and\ \bibinfo {author} {\bibfnamefont {L.}~\bibnamefont
  {Fu}},\ }\href {\doibase 10.1103/PhysRevB.98.045103} {\bibfield  {journal}
  {\bibinfo  {journal} {Phys. Rev. B}\ }\textbf {\bibinfo {volume} {98}},\
  \bibinfo {pages} {045103} (\bibinfo {year} {2018})}\BibitemShut {NoStop}%
\bibitem [{\citenamefont {Zou}\ \emph {et~al.}(2018)\citenamefont {Zou},
  \citenamefont {Po}, \citenamefont {Vishwanath},\ and\ \citenamefont
  {Senthil}}]{Zou2018}%
  \BibitemOpen
  \bibfield  {author} {\bibinfo {author} {\bibfnamefont {L.}~\bibnamefont
  {Zou}}, \bibinfo {author} {\bibfnamefont {H.~C.}\ \bibnamefont {Po}},
  \bibinfo {author} {\bibfnamefont {A.}~\bibnamefont {Vishwanath}}, \ and\
  \bibinfo {author} {\bibfnamefont {T.}~\bibnamefont {Senthil}},\ }\href
  {\doibase 10.1103/PhysRevB.98.085435} {\bibfield  {journal} {\bibinfo
  {journal} {Phys. Rev. B}\ }\textbf {\bibinfo {volume} {98}},\ \bibinfo
  {pages} {085435} (\bibinfo {year} {2018})}\BibitemShut {NoStop}%
\bibitem [{\citenamefont {Po}\ \emph {et~al.}(2019)\citenamefont {Po},
  \citenamefont {Zou}, \citenamefont {Senthil},\ and\ \citenamefont
  {Vishwanath}}]{Po2019}%
  \BibitemOpen
  \bibfield  {author} {\bibinfo {author} {\bibfnamefont {H.~C.}\ \bibnamefont
  {Po}}, \bibinfo {author} {\bibfnamefont {L.}~\bibnamefont {Zou}}, \bibinfo
  {author} {\bibfnamefont {T.}~\bibnamefont {Senthil}}, \ and\ \bibinfo
  {author} {\bibfnamefont {A.}~\bibnamefont {Vishwanath}},\ }\href {\doibase
  10.1103/PhysRevB.99.195455} {\bibfield  {journal} {\bibinfo  {journal} {Phys.
  Rev. B}\ }\textbf {\bibinfo {volume} {99}},\ \bibinfo {pages} {195455}
  (\bibinfo {year} {2019})}\BibitemShut {NoStop}%
\bibitem [{\citenamefont {Koshino}\ and\ \citenamefont
  {Son}(2019)}]{Koshino_phonons}%
  \BibitemOpen
  \bibfield  {author} {\bibinfo {author} {\bibfnamefont {M.}~\bibnamefont
  {Koshino}}\ and\ \bibinfo {author} {\bibfnamefont {Y.-W.}\ \bibnamefont
  {Son}},\ }\href {\doibase 10.1103/PhysRevB.100.075416} {\bibfield  {journal}
  {\bibinfo  {journal} {Phys. Rev. B}\ }\textbf {\bibinfo {volume} {100}},\
  \bibinfo {pages} {075416} (\bibinfo {year} {2019})}\BibitemShut {NoStop}%
\bibitem [{\citenamefont {Ochoa}(2019)}]{phasons}%
  \BibitemOpen
  \bibfield  {author} {\bibinfo {author} {\bibfnamefont {H.}~\bibnamefont
  {Ochoa}},\ }\href {\doibase 10.1103/PhysRevB.100.155426} {\bibfield
  {journal} {\bibinfo  {journal} {Phys. Rev. B}\ }\textbf {\bibinfo {volume}
  {100}},\ \bibinfo {pages} {155426} (\bibinfo {year} {2019})}\BibitemShut
  {NoStop}%
\bibitem [{\citenamefont {Maity}\ \emph {et~al.}(2020)\citenamefont {Maity},
  \citenamefont {Naik}, \citenamefont {Maiti}, \citenamefont {Krishnamurthy},\
  and\ \citenamefont {Jain}}]{phasons2}%
  \BibitemOpen
  \bibfield  {author} {\bibinfo {author} {\bibfnamefont {I.}~\bibnamefont
  {Maity}}, \bibinfo {author} {\bibfnamefont {M.~H.}\ \bibnamefont {Naik}},
  \bibinfo {author} {\bibfnamefont {P.~K.}\ \bibnamefont {Maiti}}, \bibinfo
  {author} {\bibfnamefont {H.~R.}\ \bibnamefont {Krishnamurthy}}, \ and\
  \bibinfo {author} {\bibfnamefont {M.}~\bibnamefont {Jain}},\ }\href@noop {}
  {\bibfield  {journal} {\bibinfo  {journal} {Phys. Rev. Res.}\ }\textbf
  {\bibinfo {volume} {2}},\ \bibinfo {pages} {013335} (\bibinfo {year}
  {2020})}\BibitemShut {NoStop}%
\bibitem [{\citenamefont {Gaa}\ \emph {et~al.}(2021)\citenamefont {Gaa},
  \citenamefont {Palle}, \citenamefont {Fernandes},\ and\ \citenamefont
  {Schmalian}}]{Gaa2021}%
  \BibitemOpen
  \bibfield  {author} {\bibinfo {author} {\bibfnamefont {J.}~\bibnamefont
  {Gaa}}, \bibinfo {author} {\bibfnamefont {G.}~\bibnamefont {Palle}}, \bibinfo
  {author} {\bibfnamefont {R.~M.}\ \bibnamefont {Fernandes}}, \ and\ \bibinfo
  {author} {\bibfnamefont {J.}~\bibnamefont {Schmalian}},\ }\href {\doibase
  10.1103/PhysRevB.104.064109} {\bibfield  {journal} {\bibinfo  {journal}
  {Phys. Rev. B}\ }\textbf {\bibinfo {volume} {104}},\ \bibinfo {pages}
  {064109} (\bibinfo {year} {2021})}\BibitemShut {NoStop}%
\bibitem [{\citenamefont {Uri}\ \emph {et~al.}(2020)\citenamefont {Uri},
  \citenamefont {Grover}, \citenamefont {Cao}, \citenamefont {Crosse},
  \citenamefont {Bagani}, \citenamefont {Rodan-Legrain}, \citenamefont
  {Myasoedov}, \citenamefont {Watanabe}, \citenamefont {Taniguchi},
  \citenamefont {Moon}, \citenamefont {Koshino}, \citenamefont
  {Jarillo-Herrero},\ and\ \citenamefont {Zeldov}}]{Zeldov}%
  \BibitemOpen
  \bibfield  {author} {\bibinfo {author} {\bibfnamefont {A.}~\bibnamefont
  {Uri}}, \bibinfo {author} {\bibfnamefont {S.}~\bibnamefont {Grover}},
  \bibinfo {author} {\bibfnamefont {Y.}~\bibnamefont {Cao}}, \bibinfo {author}
  {\bibfnamefont {J.}~\bibnamefont {Crosse}}, \bibinfo {author} {\bibfnamefont
  {K.}~\bibnamefont {Bagani}}, \bibinfo {author} {\bibfnamefont
  {D.}~\bibnamefont {Rodan-Legrain}}, \bibinfo {author} {\bibfnamefont
  {Y.}~\bibnamefont {Myasoedov}}, \bibinfo {author} {\bibfnamefont
  {K.}~\bibnamefont {Watanabe}}, \bibinfo {author} {\bibfnamefont
  {T.}~\bibnamefont {Taniguchi}}, \bibinfo {author} {\bibfnamefont
  {P.}~\bibnamefont {Moon}}, \bibinfo {author} {\bibfnamefont {M.}~\bibnamefont
  {Koshino}}, \bibinfo {author} {\bibfnamefont {P.}~\bibnamefont
  {Jarillo-Herrero}}, \ and\ \bibinfo {author} {\bibfnamefont {E.}~\bibnamefont
  {Zeldov}},\ }\href@noop {} {\bibfield  {journal} {\bibinfo  {journal}
  {Nature}\ }\textbf {\bibinfo {volume} {581}},\ \bibinfo {pages} {47}
  (\bibinfo {year} {2020})}\BibitemShut {NoStop}%
\bibitem [{\citenamefont {Benschop}\ \emph {et~al.}(2021)\citenamefont
  {Benschop}, \citenamefont {de~Jong}, \citenamefont {Stepanov}, \citenamefont
  {Lu}, \citenamefont {Stalman}, \citenamefont {van~der Molen}, \citenamefont
  {Efetov},\ and\ \citenamefont {Allan}}]{Milan2021}%
  \BibitemOpen
  \bibfield  {author} {\bibinfo {author} {\bibfnamefont {T.}~\bibnamefont
  {Benschop}}, \bibinfo {author} {\bibfnamefont {T.~A.}\ \bibnamefont
  {de~Jong}}, \bibinfo {author} {\bibfnamefont {P.}~\bibnamefont {Stepanov}},
  \bibinfo {author} {\bibfnamefont {X.}~\bibnamefont {Lu}}, \bibinfo {author}
  {\bibfnamefont {V.}~\bibnamefont {Stalman}}, \bibinfo {author} {\bibfnamefont
  {S.~J.}\ \bibnamefont {van~der Molen}}, \bibinfo {author} {\bibfnamefont
  {D.~K.}\ \bibnamefont {Efetov}}, \ and\ \bibinfo {author} {\bibfnamefont
  {M.~P.}\ \bibnamefont {Allan}},\ }\href {\doibase
  10.1103/PhysRevResearch.3.013153} {\bibfield  {journal} {\bibinfo  {journal}
  {Phys. Rev. Research}\ }\textbf {\bibinfo {volume} {3}},\ \bibinfo {pages}
  {013153} (\bibinfo {year} {2021})}\BibitemShut {NoStop}%
\bibitem [{\citenamefont {Kazmierczak}\ \emph {et~al.}(2021)\citenamefont
  {Kazmierczak}, \citenamefont {Van~Winkle}, \citenamefont {Ophus},
  \citenamefont {Bustillo}, \citenamefont {Carr}, \citenamefont {Brown},
  \citenamefont {Ciston}, \citenamefont {Taniguchi}, \citenamefont {Watanabe},\
  and\ \citenamefont {Bediako}}]{nat_mat}%
  \BibitemOpen
  \bibfield  {author} {\bibinfo {author} {\bibfnamefont {N.~P.}\ \bibnamefont
  {Kazmierczak}}, \bibinfo {author} {\bibfnamefont {M.}~\bibnamefont
  {Van~Winkle}}, \bibinfo {author} {\bibfnamefont {C.}~\bibnamefont {Ophus}},
  \bibinfo {author} {\bibfnamefont {K.~C.}\ \bibnamefont {Bustillo}}, \bibinfo
  {author} {\bibfnamefont {S.}~\bibnamefont {Carr}}, \bibinfo {author}
  {\bibfnamefont {H.~G.}\ \bibnamefont {Brown}}, \bibinfo {author}
  {\bibfnamefont {J.}~\bibnamefont {Ciston}}, \bibinfo {author} {\bibfnamefont
  {T.}~\bibnamefont {Taniguchi}}, \bibinfo {author} {\bibfnamefont
  {K.}~\bibnamefont {Watanabe}}, \ and\ \bibinfo {author} {\bibfnamefont
  {D.~K.}\ \bibnamefont {Bediako}},\ }\href
  {https://doi.org/10.1038/s41563-021-00973-w} {\bibfield  {journal} {\bibinfo
  {journal} {Nature Materials}\ } (\bibinfo {year} {2021})}\BibitemShut
  {NoStop}%
\bibitem [{\citenamefont {Wilson}\ \emph {et~al.}(2020)\citenamefont {Wilson},
  \citenamefont {Fu}, \citenamefont {Das~Sarma},\ and\ \citenamefont
  {Pixley}}]{Pixley2019}%
  \BibitemOpen
  \bibfield  {author} {\bibinfo {author} {\bibfnamefont {J.~H.}\ \bibnamefont
  {Wilson}}, \bibinfo {author} {\bibfnamefont {Y.}~\bibnamefont {Fu}}, \bibinfo
  {author} {\bibfnamefont {S.}~\bibnamefont {Das~Sarma}}, \ and\ \bibinfo
  {author} {\bibfnamefont {J.~H.}\ \bibnamefont {Pixley}},\ }\href {\doibase
  10.1103/PhysRevResearch.2.023325} {\bibfield  {journal} {\bibinfo  {journal}
  {Phys. Rev. Research}\ }\textbf {\bibinfo {volume} {2}},\ \bibinfo {pages}
  {023325} (\bibinfo {year} {2020})}\BibitemShut {NoStop}%
\bibitem [{\citenamefont {Padhi}\ \emph {et~al.}(2020)\citenamefont {Padhi},
  \citenamefont {Tiwari}, \citenamefont {Neupert},\ and\ \citenamefont
  {Ryu}}]{Ryu2020}%
  \BibitemOpen
  \bibfield  {author} {\bibinfo {author} {\bibfnamefont {B.}~\bibnamefont
  {Padhi}}, \bibinfo {author} {\bibfnamefont {A.}~\bibnamefont {Tiwari}},
  \bibinfo {author} {\bibfnamefont {T.}~\bibnamefont {Neupert}}, \ and\
  \bibinfo {author} {\bibfnamefont {S.}~\bibnamefont {Ryu}},\ }\href {\doibase
  10.1103/PhysRevResearch.2.033458} {\bibfield  {journal} {\bibinfo  {journal}
  {Phys. Rev. Research}\ }\textbf {\bibinfo {volume} {2}},\ \bibinfo {pages}
  {033458} (\bibinfo {year} {2020})}\BibitemShut {NoStop}%
\bibitem [{\citenamefont {dos Santos}\ \emph {et~al.}(2007)\citenamefont {dos
  Santos}, \citenamefont {Peres},\ and\ \citenamefont {Neto}}]{dosSantos2007}%
  \BibitemOpen
  \bibfield  {author} {\bibinfo {author} {\bibfnamefont {J.~M. B.~L.}\
  \bibnamefont {dos Santos}}, \bibinfo {author} {\bibfnamefont {N.~M.~R.}\
  \bibnamefont {Peres}}, \ and\ \bibinfo {author} {\bibfnamefont {A.~H.~C.}\
  \bibnamefont {Neto}},\ }\href {\doibase 10.1103/PhysRevLett.99.256802}
  {\bibfield  {journal} {\bibinfo  {journal} {Phys. Rev. Lett.}\ }\textbf
  {\bibinfo {volume} {99}},\ \bibinfo {pages} {256802} (\bibinfo {year}
  {2007})}\BibitemShut {NoStop}%
\bibitem [{\citenamefont {Bistritzer}\ and\ \citenamefont
  {MacDonald}(2011)}]{BM}%
  \BibitemOpen
  \bibfield  {author} {\bibinfo {author} {\bibfnamefont {R.}~\bibnamefont
  {Bistritzer}}\ and\ \bibinfo {author} {\bibfnamefont {A.~H.}\ \bibnamefont
  {MacDonald}},\ }\href {\doibase 10.1073/pnas.1108174108} {\bibfield
  {journal} {\bibinfo  {journal} {Proceedings of the National Academy of
  Sciences}\ }\textbf {\bibinfo {volume} {108}},\ \bibinfo {pages} {12233}
  (\bibinfo {year} {2011})}\BibitemShut {NoStop}%
\bibitem [{SM()}]{SM}%
  \BibitemOpen
  \href@noop {} {\emph {\bibinfo {title} {See supplemental material for 1)
  details of the free-energy model and the relaxation problem, 2) estimation of
  $\mathcal{G}$ and disorder correlation functions, and 3) a microscopic
  expression of $\tau^{-1}$.}}}\BibitemShut {Stop}%
\bibitem [{\citenamefont {Guinea}\ and\ \citenamefont {Walet}(2019)}]{relax1}%
  \BibitemOpen
  \bibfield  {author} {\bibinfo {author} {\bibfnamefont {F.}~\bibnamefont
  {Guinea}}\ and\ \bibinfo {author} {\bibfnamefont {N.~R.}\ \bibnamefont
  {Walet}},\ }\href@noop {} {\bibfield  {journal} {\bibinfo  {journal} {Phys.
  Rev. B}\ }\textbf {\bibinfo {volume} {99}},\ \bibinfo {pages} {205134}
  (\bibinfo {year} {2019})}\BibitemShut {NoStop}%
\bibitem [{\citenamefont {Carr}\ \emph {et~al.}(2019)\citenamefont {Carr},
  \citenamefont {Fang}, \citenamefont {Zhu},\ and\ \citenamefont
  {Kaxiras}}]{relax2}%
  \BibitemOpen
  \bibfield  {author} {\bibinfo {author} {\bibfnamefont {S.}~\bibnamefont
  {Carr}}, \bibinfo {author} {\bibfnamefont {S.}~\bibnamefont {Fang}}, \bibinfo
  {author} {\bibfnamefont {Z.}~\bibnamefont {Zhu}}, \ and\ \bibinfo {author}
  {\bibfnamefont {E.}~\bibnamefont {Kaxiras}},\ }\href@noop {} {\bibfield
  {journal} {\bibinfo  {journal} {Phys. Rev. Res.}\ }\textbf {\bibinfo {volume}
  {1}},\ \bibinfo {pages} {013001} (\bibinfo {year} {2019})}\BibitemShut
  {NoStop}%
\bibitem [{\citenamefont {Koshino}\ and\ \citenamefont {Nam}(2020)}]{relax3}%
  \BibitemOpen
  \bibfield  {author} {\bibinfo {author} {\bibfnamefont {M.}~\bibnamefont
  {Koshino}}\ and\ \bibinfo {author} {\bibfnamefont {N.~N.~T.}\ \bibnamefont
  {Nam}},\ }\href@noop {} {\bibfield  {journal} {\bibinfo  {journal} {Phys.
  Rev. B}\ }\textbf {\bibinfo {volume} {101}},\ \bibinfo {pages} {195425}
  (\bibinfo {year} {2020})}\BibitemShut {NoStop}%
\bibitem [{\citenamefont {Mortazavi}\ \emph {et~al.}(2021)\citenamefont
  {Mortazavi}, \citenamefont {Silani}, \citenamefont {Podryabinkin},
  \citenamefont {Rabczuk}, \citenamefont {Zhuang},\ and\ \citenamefont
  {Shapeev}}]{ML_methods}%
  \BibitemOpen
  \bibfield  {author} {\bibinfo {author} {\bibfnamefont {B.}~\bibnamefont
  {Mortazavi}}, \bibinfo {author} {\bibfnamefont {M.}~\bibnamefont {Silani}},
  \bibinfo {author} {\bibfnamefont {E.~V.}\ \bibnamefont {Podryabinkin}},
  \bibinfo {author} {\bibfnamefont {T.}~\bibnamefont {Rabczuk}}, \bibinfo
  {author} {\bibfnamefont {X.}~\bibnamefont {Zhuang}}, \ and\ \bibinfo {author}
  {\bibfnamefont {A.~V.}\ \bibnamefont {Shapeev}},\ }\href@noop {} {\bibfield
  {journal} {\bibinfo  {journal} {Adv. Mater.}\ }\textbf {\bibinfo {volume}
  {33}},\ \bibinfo {pages} {21022807} (\bibinfo {year} {2021})}\BibitemShut
  {NoStop}%
\bibitem [{\citenamefont {Zakharchenko}\ \emph {et~al.}(2009)\citenamefont
  {Zakharchenko}, \citenamefont {Katsnelson},\ and\ \citenamefont
  {Fasolino}}]{elastic_constants}%
  \BibitemOpen
  \bibfield  {author} {\bibinfo {author} {\bibfnamefont {K.~V.}\ \bibnamefont
  {Zakharchenko}}, \bibinfo {author} {\bibfnamefont {M.~I.}\ \bibnamefont
  {Katsnelson}}, \ and\ \bibinfo {author} {\bibfnamefont {A.}~\bibnamefont
  {Fasolino}},\ }\href@noop {} {\bibfield  {journal} {\bibinfo  {journal}
  {Phys. Rev. Lett.}\ }\textbf {\bibinfo {volume} {102}},\ \bibinfo {pages}
  {046808} (\bibinfo {year} {2009})}\BibitemShut {NoStop}%
\bibitem [{\citenamefont {Carr}\ \emph {et~al.}(2018)\citenamefont {Carr},
  \citenamefont {Massatt}, \citenamefont {Torrisi}, \citenamefont {Cazeaux},
  \citenamefont {Luskin},\ and\ \citenamefont {Kaxiras}}]{Carr}%
  \BibitemOpen
  \bibfield  {author} {\bibinfo {author} {\bibfnamefont {S.}~\bibnamefont
  {Carr}}, \bibinfo {author} {\bibfnamefont {D.}~\bibnamefont {Massatt}},
  \bibinfo {author} {\bibfnamefont {S.~B.}\ \bibnamefont {Torrisi}}, \bibinfo
  {author} {\bibfnamefont {P.}~\bibnamefont {Cazeaux}}, \bibinfo {author}
  {\bibfnamefont {M.}~\bibnamefont {Luskin}}, \ and\ \bibinfo {author}
  {\bibfnamefont {E.}~\bibnamefont {Kaxiras}},\ }\href@noop {} {\bibfield
  {journal} {\bibinfo  {journal} {Phys. Rev. B}\ }\textbf {\bibinfo {volume}
  {98}},\ \bibinfo {pages} {224102} (\bibinfo {year} {2018})}\BibitemShut
  {NoStop}%
\bibitem [{\citenamefont {Imry}\ and\ \citenamefont {Ma}(1975)}]{Imry-Ma}%
  \BibitemOpen
  \bibfield  {author} {\bibinfo {author} {\bibfnamefont {Y.}~\bibnamefont
  {Imry}}\ and\ \bibinfo {author} {\bibfnamefont {S.-k.}\ \bibnamefont {Ma}},\
  }\href@noop {} {\bibfield  {journal} {\bibinfo  {journal} {Phys. Rev. Lett.}\
  }\textbf {\bibinfo {volume} {35}},\ \bibinfo {pages} {1399} (\bibinfo {year}
  {1975})}\BibitemShut {NoStop}%
\bibitem [{\citenamefont {Fukuyama}\ and\ \citenamefont
  {Lee}(1978)}]{collective_pinning}%
  \BibitemOpen
  \bibfield  {author} {\bibinfo {author} {\bibfnamefont {H.}~\bibnamefont
  {Fukuyama}}\ and\ \bibinfo {author} {\bibfnamefont {P.~A.}\ \bibnamefont
  {Lee}},\ }\href@noop {} {\bibfield  {journal} {\bibinfo  {journal} {Phys.
  Rev. B}\ }\textbf {\bibinfo {volume} {17}},\ \bibinfo {pages} {535} (\bibinfo
  {year} {1978})}\BibitemShut {NoStop}%
\bibitem [{\citenamefont {Forster}(2019)}]{Forster}%
  \BibitemOpen
  \bibfield  {author} {\bibinfo {author} {\bibfnamefont {D.}~\bibnamefont
  {Forster}},\ }\href@noop {} {\emph {\bibinfo {title} {Hydrodynamic
  Fluctuations, Broken Symmetry, and Correlation Functions}}}\ (\bibinfo
  {publisher} {CRC Press},\ \bibinfo {address} {Boca Raton},\ \bibinfo {year}
  {2019})\BibitemShut {NoStop}%
\bibitem [{\citenamefont {Cano}\ and\ \citenamefont {Levanyuk}(2004)}]{Cv}%
  \BibitemOpen
  \bibfield  {author} {\bibinfo {author} {\bibfnamefont {A.}~\bibnamefont
  {Cano}}\ and\ \bibinfo {author} {\bibfnamefont {A.~P.}\ \bibnamefont
  {Levanyuk}},\ }\href@noop {} {\bibfield  {journal} {\bibinfo  {journal}
  {Phys. Rev. Lett.}\ }\textbf {\bibinfo {volume} {93}},\ \bibinfo {pages}
  {245902} (\bibinfo {year} {2004})}\BibitemShut {NoStop}%
\bibitem [{\citenamefont {Baggioli}\ and\ \citenamefont {Zaccone}(2021)}]{BP}%
  \BibitemOpen
  \bibfield  {author} {\bibinfo {author} {\bibfnamefont {M.}~\bibnamefont
  {Baggioli}}\ and\ \bibinfo {author} {\bibfnamefont {A.}~\bibnamefont
  {Zaccone}},\ }\href@noop {} {\bibfield  {journal} {\bibinfo  {journal} {Int.
  J. Mod. Phys. B}\ }\textbf {\bibinfo {volume} {35}},\ \bibinfo {pages}
  {2130002} (\bibinfo {year} {2021})}\BibitemShut {NoStop}%
\bibitem [{\citenamefont {Bi}\ \emph {et~al.}(2019)\citenamefont {Bi},
  \citenamefont {Yuan},\ and\ \citenamefont {Fu}}]{Fu_heterostrain}%
  \BibitemOpen
  \bibfield  {author} {\bibinfo {author} {\bibfnamefont {Z.}~\bibnamefont
  {Bi}}, \bibinfo {author} {\bibfnamefont {N.~F.~Q.}\ \bibnamefont {Yuan}}, \
  and\ \bibinfo {author} {\bibfnamefont {L.}~\bibnamefont {Fu}},\ }\href
  {\doibase 10.1103/PhysRevB.100.035448} {\bibfield  {journal} {\bibinfo
  {journal} {Phys. Rev. B}\ }\textbf {\bibinfo {volume} {100}},\ \bibinfo
  {pages} {035448} (\bibinfo {year} {2019})}\BibitemShut {NoStop}%
\bibitem [{\citenamefont {Yudhistira}\ \emph {et~al.}(2019)\citenamefont
  {Yudhistira}, \citenamefont {Chakraborty}, \citenamefont {Sharma},
  \citenamefont {Ho}, \citenamefont {Laksono}, \citenamefont {Sushkov},
  \citenamefont {Vignale},\ and\ \citenamefont {Adam}}]{scattering1}%
  \BibitemOpen
  \bibfield  {author} {\bibinfo {author} {\bibfnamefont {I.}~\bibnamefont
  {Yudhistira}}, \bibinfo {author} {\bibfnamefont {N.}~\bibnamefont
  {Chakraborty}}, \bibinfo {author} {\bibfnamefont {G.}~\bibnamefont {Sharma}},
  \bibinfo {author} {\bibfnamefont {D.~Y.~H.}\ \bibnamefont {Ho}}, \bibinfo
  {author} {\bibfnamefont {E.}~\bibnamefont {Laksono}}, \bibinfo {author}
  {\bibfnamefont {O.~P.}\ \bibnamefont {Sushkov}}, \bibinfo {author}
  {\bibfnamefont {G.}~\bibnamefont {Vignale}}, \ and\ \bibinfo {author}
  {\bibfnamefont {S.}~\bibnamefont {Adam}},\ }\href@noop {} {\bibfield
  {journal} {\bibinfo  {journal} {Phys. Rev. B}\ }\textbf {\bibinfo {volume}
  {99}},\ \bibinfo {pages} {140302(R)} (\bibinfo {year} {2019})}\BibitemShut
  {NoStop}%
\bibitem [{\citenamefont {Ishizuka}\ \emph {et~al.}(2020)\citenamefont
  {Ishizuka}, \citenamefont {Fahimniya}, \citenamefont {Guinea},\ and\
  \citenamefont {Levitov}}]{scattering2}%
  \BibitemOpen
  \bibfield  {author} {\bibinfo {author} {\bibfnamefont {H.}~\bibnamefont
  {Ishizuka}}, \bibinfo {author} {\bibfnamefont {A.}~\bibnamefont {Fahimniya}},
  \bibinfo {author} {\bibfnamefont {F.}~\bibnamefont {Guinea}}, \ and\ \bibinfo
  {author} {\bibfnamefont {L.}~\bibnamefont {Levitov}},\ }\href@noop {}
  {\bibfield  {journal} {\bibinfo  {journal} {arXiv:2011.01701}\ } (\bibinfo
  {year} {2020})}\BibitemShut {NoStop}%
\bibitem [{\citenamefont {Rubio-Verd{\'u}}\ \emph {et~al.}(2020)\citenamefont
  {Rubio-Verd{\'u}}, \citenamefont {Turkel}, \citenamefont {Song},
  \citenamefont {Klebl}, \citenamefont {Samajdar}, \citenamefont {Scheurer},
  \citenamefont {Venderbos}, \citenamefont {Watanabe}, \citenamefont
  {Taniguchi}, \citenamefont {Ochoa} \emph {et~al.}}]{Carmen_etal}%
  \BibitemOpen
  \bibfield  {author} {\bibinfo {author} {\bibfnamefont {C.}~\bibnamefont
  {Rubio-Verd{\'u}}}, \bibinfo {author} {\bibfnamefont {S.}~\bibnamefont
  {Turkel}}, \bibinfo {author} {\bibfnamefont {L.}~\bibnamefont {Song}},
  \bibinfo {author} {\bibfnamefont {L.}~\bibnamefont {Klebl}}, \bibinfo
  {author} {\bibfnamefont {R.}~\bibnamefont {Samajdar}}, \bibinfo {author}
  {\bibfnamefont {M.~S.}\ \bibnamefont {Scheurer}}, \bibinfo {author}
  {\bibfnamefont {J.~W.}\ \bibnamefont {Venderbos}}, \bibinfo {author}
  {\bibfnamefont {K.}~\bibnamefont {Watanabe}}, \bibinfo {author}
  {\bibfnamefont {T.}~\bibnamefont {Taniguchi}}, \bibinfo {author}
  {\bibfnamefont {H.}~\bibnamefont {Ochoa}},  \emph {et~al.},\ }\href@noop {}
  {\bibfield  {journal} {\bibinfo  {journal} {arXiv:2009.11645}\ } (\bibinfo
  {year} {2020})}\BibitemShut {NoStop}%
\end{thebibliography}%

\clearpage
\onecolumngrid

\setcounter{equation}{0}
\renewcommand{\theequation}{S\arabic{equation}}

\setcounter{figure}{0}
\renewcommand{\thefigure}{S\arabic{figure}}

\appendix

\section{Supplemental Material}
\maketitle
\onecolumngrid

\subsubsection{Stacking order and lattice relaxation}

Formally, we can define the stacking order function $\boldsymbol{\phi}(\mathbf{r})$ as a mapping between coordinate space (i.e., a lateral position $\mathbf{r}$ in the bilayer) and the continuous two-dimensional manifold formed by all physically distinct commensurate structures generated by a rigid translation of one layer with respect to the other:\begin{align}
\boldsymbol{\phi}\left(\mathbf{r}\right):\,\mathbf{r}\longrightarrow \boldsymbol{\phi}\in\mathcal{M}.
\end{align}
We will refer to $\mathcal{M}$ as the \textit{configuration space}.

Let us construct this function from the mass distribution in the lattices. Consider first the mass density in a single layer, written in Fourier components as\begin{align}
\varrho\left(\mathbf{r}\right)=\sum_{\left\{\boldsymbol{g}\right\}}\varrho_{\boldsymbol{g}}\,e^{i\boldsymbol{g}\cdot\left(\mathbf{r}-\mathbf{u}\left(\mathbf{r}\right)\right)}.
\end{align}
The component with $\boldsymbol{g}=0$ corresponds to the parameter $\varrho$ introduced in the main text. The distribution of carbon masses within the unit cell are described by the remaining harmonics. We allow for smooth, in-phase distortions of the mass density parametrized by a displacement field $\mathbf{u}(\mathbf{r})$. The positions of the centers of mass of the unit cells are defined by the condition\begin{align}
\boldsymbol{g}\cdot\left(\mathbf{r}-\mathbf{u}\left(\mathbf{r}\right)\right)=2\pi\Longrightarrow \mathbf{r}_i=\mathbf{R}_i+\mathbf{u}\left(\mathbf{r}_i\right),
\end{align}
where $\mathbf{R}_i$ are the vectors of the Bravais lattice. This last equation defines implicitly the position of the unit cells in the distorted crystal, $\mathbf{r}_i=\mathbf{r}\left(\mathbf{R}_i\right)$, through the displacement field $\mathbf{u}(\mathbf{r}_i)$ in Eulerian coordinates, i.e., labelled by the actual position in the deformed lattice. The harmonic elastic energy of the crystal is
\begin{align}
\label{eq:elastic_energy}
F_{\textrm{el}}\left[\mathbf{u}\left(\mathbf{r}\right)\right]=\frac{1}{2}\int d^2\mathbf{r}\left[\lambda\left(u_{ii}\right)^2+\mu\, u_{ij}u_{ij}\right],
\end{align}
with the harmonic strain tensor defined as usual, $u_{ij}=(\partial_i u_j+\partial_j u_i)/2$.

Consider now the superposition of the mass densities of the two layers prior to the twist,\begin{subequations}\begin{align}
\varrho_{\textrm{t}}\left(\mathbf{r}\right)+\varrho_{\textrm{b}}\left(\mathbf{r}\right)=\sum_{\left\{\boldsymbol{g}\right\}}2\,\varrho_{\boldsymbol{g}}\,\cos\left(\frac{\boldsymbol{g}\cdot\boldsymbol{u}\left(\mathbf{r}\right)}{2}\right)e^{i\boldsymbol{g}\cdot\left(\mathbf{r}-\mathbf{u}_{\textrm{cm}}\left(\mathbf{r}\right)\right)},
\end{align}
where we have introduced relative and center of mass coordinates for the two layers,\begin{align}
& \boldsymbol{u}\left(\mathbf{r}\right)\equiv\mathbf{u}_{\textrm{t}}\left(\mathbf{r}\right)-\mathbf{u}_{\textrm{b}}\left(\mathbf{r}\right),\\
& \mathbf{u}_{\textrm{cm}}\left(\mathbf{r}\right)\equiv\frac{\mathbf{u}_{\textrm{t}}\left(\mathbf{r}\right)+\mathbf{u}_{\textrm{b}}\left(\mathbf{r}\right)}{2}.
\end{align}
\end{subequations}
A rigid translation of both layers, $\mathbf{u}_{\textrm{cm}}(\mathbf{r})\rightarrow \mathbf{u}_{\textrm{cm}}(\mathbf{r})+\mathbf{u}$, just translates the origin of the mass density. As the energy does not depend on the global position of the system in space, there are two soft modes (acoustic phonons) associated with in-phase oscillations of the layers. However, a rigid relative translation, $\boldsymbol{u}(\mathbf{r})\rightarrow \boldsymbol{u}(\mathbf{r})+\boldsymbol{u}$, modifies the amplitude of the density wave and hence the adhesion energy (see below). These are optical phonons.

We can identify the stacking order from the argument of the squared amplitude as $4\cos^2(\boldsymbol{g}\cdot\boldsymbol{u}/2)\equiv2+2\cos(\boldsymbol{g}\cdot\boldsymbol{\phi})$; this is the function represented in Fig.~\ref{fig:textures}~(a)~and~(c) of the main text (summed over the first harmonics, $\boldsymbol{g}=\boldsymbol{g}_1,\boldsymbol{g}_2,\boldsymbol{g}_3$). In this case, we just have $\boldsymbol{\phi}(\mathbf{r})=\boldsymbol{u}(\mathbf{r})$, in accordance with our formal definition. The differences in energy between different stacking configurations can be written as a Landau-like expansion in powers of the amplitude of the Fourier harmonics of the mass density; the model in Eq.~\eqref{eq:V_adhesion} contains only vectors in the first star. It follows then that $\boldsymbol{\phi}$ and $\boldsymbol{\phi}+\mathbf{R}$, where $\mathbf{R}$ is a vector of the graphene Bravais lattice, cost the same energy and can be identified as the same; therefore, $\mathcal{M}$ possesses the topology of a torus, $\mathcal{M}\cong S_1\times S_1$. Non-trivial loops are classified according to the fundamental group $\pi_1(\mathcal{M})=\mathbb{Z}\times\mathbb{Z}$; the two integers define the burgers vectors of misfit dislocations.

Let us consider now two graphene layers rotated with respect to each other by an angle $\theta$. As in the previous equations, we only allow for smooth, in-phase distortions of the positions of the carbon atoms with respect to the crystalline order on each layer; we can write then\begin{align}
\varrho_{\textrm{t}}\left(\mathbf{r}\right)+\varrho_{\textrm{b}}\left(\mathbf{r}\right)=\sum_{\left\{\boldsymbol{g}_{\textrm{t}}\right\}}\varrho_{\boldsymbol{g}_{\textrm{t}}}\,e^{i\boldsymbol{g}_{\textrm{t}}\cdot\left(\mathbf{r}-\mathbf{u}_{\textrm{t}}\left(\mathbf{r}\right)\right)}+\sum_{\left\{\boldsymbol{g}_{\textrm{b}}\right\}}\varrho_{\boldsymbol{g}_{\textrm{b}}}\,e^{i\boldsymbol{g}_{\textrm{b}}\cdot\left(\mathbf{r}-\mathbf{u}_{\textrm{b}}\left(\mathbf{r}\right)\right)}.
\end{align}
In this prescription, $\mathbf{u}_{\textrm{b}}(\mathbf{r})=\mathbf{u}_{\textrm{t}}(\mathbf{r})=0$ would correspond to the superposition of the mass densities generated by a rigid rotation of the layers, therefore\begin{subequations}
\begin{align}
& \boldsymbol{g}_{\textrm{t},\textrm{b}}=\hat{R}_{\pm\frac{\theta}{2}}\cdot\boldsymbol{g},\\
& \varrho_{\boldsymbol{g}_{\textrm{t}}}=\varrho_{\boldsymbol{g}_{\textrm{b}}}=\varrho_{\boldsymbol{g}}.
\end{align}
\end{subequations}
The same manipulations as before lead to\begin{align}
\varrho_{\textrm{t}}\left(\mathbf{r}\right)+\varrho_{\textrm{b}}\left(\mathbf{r}\right)\approx \sum_{\left\{\boldsymbol{g}\right\}}2\,\varrho_{\boldsymbol{g}}\,\cos\left(\frac{\mathbf{G}_{\boldsymbol{g}}\cdot\left(\mathbf{r}-\mathbf{u}_{\textrm{cm}}\left(\mathbf{r}\right)\right)+\boldsymbol{g}\cdot\boldsymbol{u}\left(\mathbf{r}\right)}{2}\right)e^{i\boldsymbol{g}\cdot\left(\mathbf{r}-\mathbf{u}_{\textrm{cm}}\left(\mathbf{r}\right)\right)},
\end{align}
with $\mathbf{G}_{\boldsymbol{g}}=\boldsymbol{g}_{\textrm{b}}-\boldsymbol{g}_{\textrm{t}}=-2\sin\frac{\theta}{2}\,\hat{\mathbf{z}}\times\boldsymbol{g}$, and where we have dropped subleading terms in the relative orientation/position of the layers, i.e. terms $\mathcal{O}(\theta^2)$ and $\mathcal{O}(\theta\boldsymbol{u})$. Just as before, $\mathbf{u}_{\textrm{cm}}$ parametrizes translations of the whole density wave. The novelty is in the amplitude, which is now modulated in space with vectors in the moir\'e reciprocal lattice. Following the same prescription as before, and ignoring $\mathbf{u}_{\textrm{cm}}$ (which translates the whole density rather than modify its amplitude) we have in this case $\boldsymbol{\phi}(\mathbf{r})=2\sin\frac{\theta}{2}\,\mathbf{\hat{z}}\times\mathbf{r}+\boldsymbol{u}(\mathbf{r})$.
The positions of local AA stacking are defined implicitly by the formula $\boldsymbol{\phi}(\mathbf{r}_i)=\mathbf{R}_i$. The beating pattern defines a superlattice of period $L_{m}$, which is only commensurate with the microscopic lattice for a discrete set of angles. Alternatively, the moir\'e pattern can be envioned as a two-dimensional array of misftit dislocations.

The modulation of the density amplitude introduces an energy cost that the system tries to minimize by introducing some lattice deformation. Let us define $\boldsymbol{u}_0(\mathbf{r})$ as the \textit{equilibrium} solution corresponding to an average angle $\theta\rightarrow \overline{\theta}$. In our local description, the adhesion potential landscape is defined by Eq.~\eqref{eq:V_adhesion}, understood now as a functional in configuration space, \begin{align}
\mathcal{V}_{\textrm{ad}}\left(\boldsymbol{\phi}\right)\longrightarrow \mathcal{V}_{\textrm{ad}}\left[\boldsymbol{\phi}\left(\mathbf{r}\right)\right]
=\frac{V_{\textrm{AA}}}{3}+\frac{2V_{\textrm{AA}}}{9}\sum_{i=1}^{3}\cos\left[\mathbf{G}_i\cdot\mathbf{r}+\boldsymbol{g}_i\cdot\boldsymbol{u}_0\left(\mathbf{r}\right)\right],
\end{align}
with $\mathbf{G}_i\equiv\mathbf{G}_{\boldsymbol{g}_i}$. The periodicity in configuration space is translated to a modulation with the beating pattern in real space. This term competes with the cost in elastic energy of heterostrain fields, which can be evaluated from the functional in Eq.~\eqref{eq:elastic_energy}. The minimization of the total energy leads to the equilibrium of lateral forces expressed as\begin{align}
\label{eq:relaxation}
\frac{\lambda+\mu}{2}\,\boldsymbol{\nabla}\left(\boldsymbol{\nabla}\cdot\boldsymbol{u}_0\right)+\frac{\mu}{2}\,\nabla^2\boldsymbol{u}_0-\frac{\partial \mathcal{V}_{\textrm{ad}}}{\partial\boldsymbol{\phi}}|_{\delta\boldsymbol{\phi}=0}=0.
\end{align} Note that the Lam\'e coefficients enter divided by 2, as we are concerned only about the relative displacements of the unit cells of both layers.

\begin{figure}[t!]
\begin{center}
\includegraphics[width=\columnwidth]{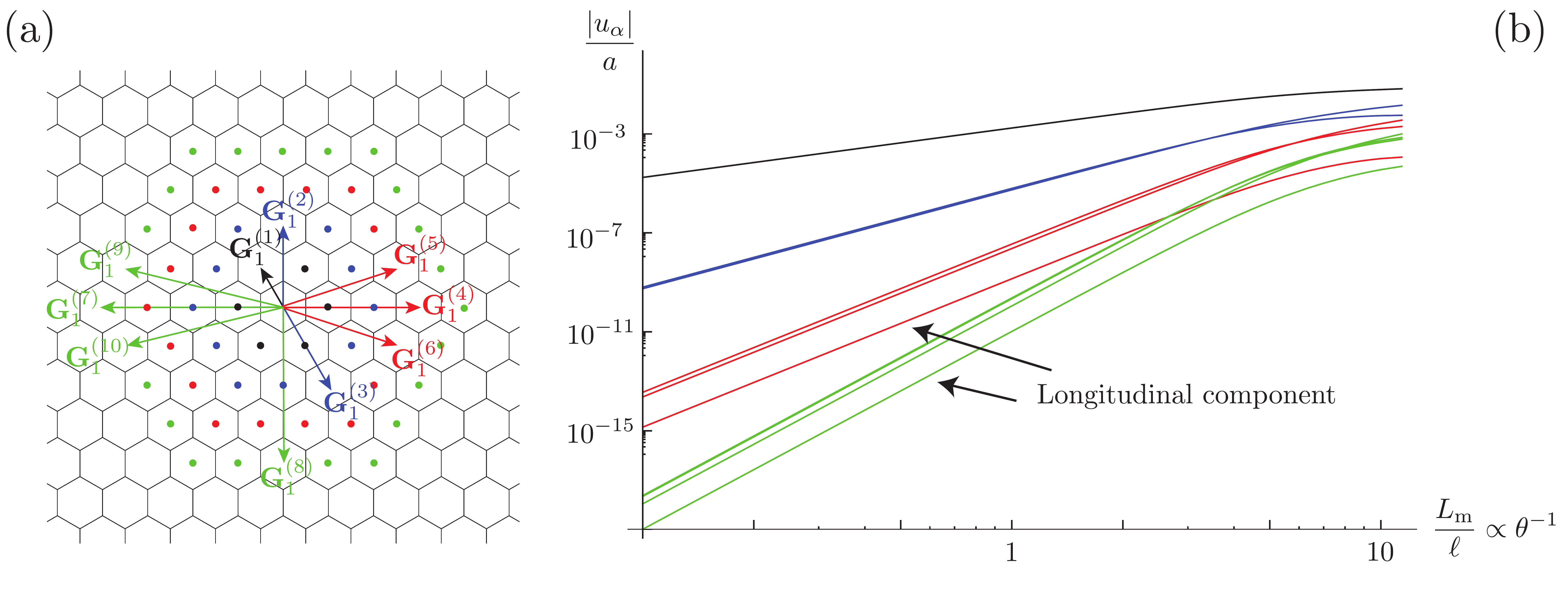}
\caption{\textbf{Fourier components of \textit{equilibrium} displacements, $\boldsymbol{u}_0(\mathbf{r})$}. (a) Reciprocal lattice vectors in the first 10 stars of a moir\'e pattern. Stars 5-6 and 9-10 are related by in-plane C$_2$ rotations; the others are invariant under this symmetry. (b) Self-consistent solution of the Fourier harmonics $u_{T,L}^{\alpha}$ in units of graphene's lattice constant $a$ as a function of $L_{\textrm{m}}/\ell\propto\theta^{-1}$ in double logarithmic scale. Colors label the momentum star as in (a). Note that the longitudinal components (indicated by arrows) are always two orders of magnitude smaller than the transverse components of the same star.} 
\label{fig:lattice}
\end{center}
\end{figure}

\begin{figure}[t!]
\begin{center}
\includegraphics[width=\columnwidth]{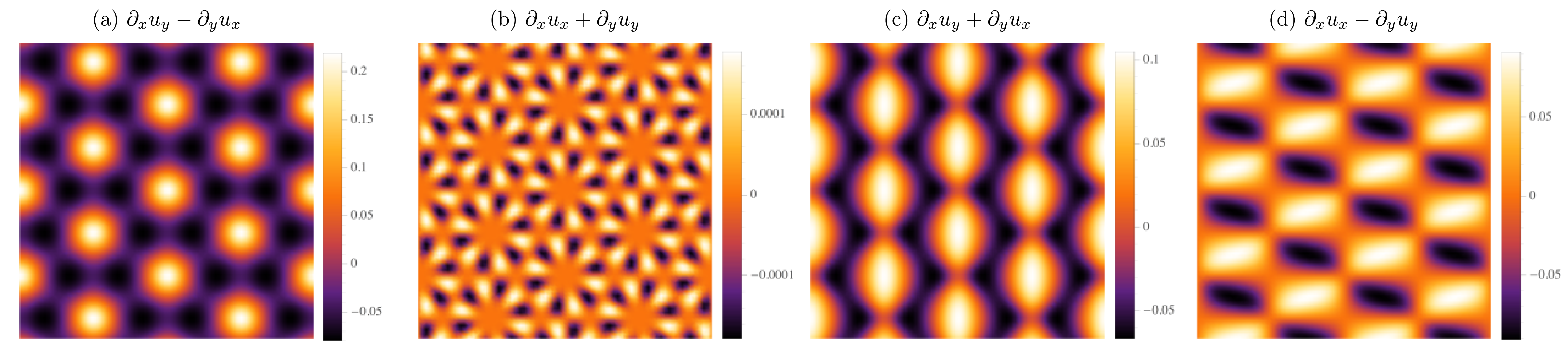}
\caption{\textbf{Symmetry-adapted components of $\boldsymbol{\partial_iu_j}$ at the magic angle}. The relaxation of adhesion forces respecting the symmetries of the moir\'e pattern amplifies the local rotation angle (panel a) in order to minimize the areas of AA stacking and maximize Bernal stacking. This process introduces large shear components of heterostrain (panels c and d) localized along the stacking domain walls and a small longitudinal component (panel b) localized in AB and BA regions.} 
\label{fig:strain}
\end{center}
\end{figure}

In order to solve Eq.~\eqref{eq:relaxation}, we assume that the relaxed structure preserves the symmetries of the undistorted moir\'e pattern. In particular, the 6-fold rotational symmetry imposes the following Fourier expansion with momenta in the moir\'e reciprocal lattice: \begin{align}
\label{eq:general_expansion}
\boldsymbol{u}_0\left(\mathbf{r}\right)=\sum_{\left\{\mathbf{G}_i^{\alpha}\right\}} e^{i\mathbf{G}_i^{\alpha}\cdot\mathbf{r}}\left[\frac{i\mathbf{G}_i^{\alpha}}{\left|\mathbf{G}_i^{\alpha}\right|}\,u_L^{\alpha}+\frac{i\mathbf{G}_i^{\alpha}\times\mathbf{\hat{z}}}{\left|\mathbf{G}_i^{\alpha}\right|}\,u_T^{\alpha}\right].
\end{align}
Here $\alpha$ labels momentum stars in reciprocal space, see Fig.~\ref{fig:lattice}~(a), where $i=1...6$ runs on the corresponding reciprocal lattice vectors, and $u_{L,T}^{\alpha}$ are real numbers with units of length. In-plane C$_{2}$ rotations further restrict this expansion. For stars whose vectors lie along these axes (e.g., the first moir\'e star $\alpha=1$, represented by the black arrow in Fig.~\ref{fig:lattice}(a)), this symmetry implies $u_{L}^{\alpha}=0$, i.e., the corresponding Fourier harmonic introduces transverse displacements only. For those pairs of stars $\alpha_{1,2}$ not aligned with these axes and related by C$_2$ rotations (e.g., stars $\alpha=5,6$ in red or $\alpha=9,10$ in green), symmetry constrains their coefficients as $u_T^{\alpha_1}=u_T^{\alpha_2}$, $u_L^{\alpha_1}=-u_L^{\alpha_2}$. Hence, the full $D_6$ symmetry group restricts the number of independent coefficient to only one per star. These are determined by the following equations derived from~\eqref{eq:relaxation}:\begin{subequations}
\label{eq:self-consistent}
\begin{align}
& \frac{4\pi}{\sqrt{3}a}u_{T}^{\alpha}=\frac{4 f_T^{\alpha}}{3\ell^2\left|\mathbf{G}^{\alpha}\right|^2},\\
& \frac{4\pi}{\sqrt{3}a}u_{L}^{\alpha}=\frac{2(1-\nu) f_L^{\alpha}}{3\ell^2\left|\mathbf{G}^{\alpha}\right|^2},
\end{align}
\end{subequations}
where we have introduced the Poison ratio $\nu=\lambda/(\lambda+2\mu)$, the domain wall width $\ell$ in Eq.~\eqref{eq:width}, and the dimensionless adhesion forces with momentum in star $\alpha$, \begin{subequations}\begin{align}
& f_T^{\alpha}=\frac{i\sqrt{3}a}{4\pi A}\sum_{j=1}^3\frac{\left(\mathbf{G}^{\alpha}\times\boldsymbol{g}_j\right)_z}{\left|\mathbf{G}^{\alpha}\right|}\int d\mathbf{r}\,\sin\left[\boldsymbol{g}_j\cdot\boldsymbol{\phi}_0\left(\mathbf{r}\right)\right] e^{-i\mathbf{G}^{\alpha}\cdot\mathbf{r}},\\
& f_L^{\alpha}=\frac{-i\sqrt{3}a}{4\pi A}\sum_{j=1}^3\frac{\mathbf{G}^{\alpha}\cdot\boldsymbol{g}_j}{\left|\mathbf{G}^{\alpha}\right|}\int d\mathbf{r}\,\sin\left[\boldsymbol{g}_j\cdot\boldsymbol{\phi}_0\left(\mathbf{r}\right)\right] e^{-i\mathbf{G}^{\alpha}\cdot\mathbf{r}}.
\end{align}
\end{subequations}
We have omitted the index $i$ as the calculation can be performed for any of the six vectors. Note that the forces are in fact functionals of the equilibrium stacking configuration. The procedure is to start from an ansatz for these coefficients, determine the forces, then compute a new set of parameters and so forth until we reach convergence. In the calculations, we have restricted the number of harmonics up to 10 stars. The results are displayed in Fig.~\ref{fig:lattice}~(b). The corresponding values for the magic angle are used to generate Fig.~\ref{fig:textures}~(b)~and~(c) in the main text. Notice the change in the slope of the curves above the magic angle ($L_{\textrm{m}}/\ell\approx 4.67$). A couple of additional points are worth emphasizing: 
\begin{itemize}
\item The pre-factor in the right-hand side of Eqs.~\eqref{eq:self-consistent} is basically the ratio $(L_{\textrm{m}}/\ell)^2$, which control the amount of heterostrain and local changes of twist angle generated by the relaxation process. The four components of the tensor formed by derivatives of the displacement field are represented in Fig.~\ref{fig:strain} for the magic angle. Note that these are classified in terms of the irreducible representations of $D_6$: the antisymmetric component (or local variation of the twist angle, panel~a) belongs to $A_1$, the trace (b) belongs to $A_2$, and the other two symmetric combinations in (c) and (d) form an $E_2$ doublet.

\item As inferred from the previous discussion and the numerical results shown in Fig.~\ref{fig:lattice}~(b), where $u_{L}^{\alpha}$ is several orders of magnitude smaller than $u_{T}^{\alpha}$ (if not $0$ by symmetry), the heterostrain profile consists mostly of shear components; around the magic angle, the longitudinal component of the heterostrain tensor, Fig.~\ref{fig:strain}~(b), is three orders of magnitude smaller than the rest. Note that $\boldsymbol{\nabla}\cdot\boldsymbol{u}_0$ is mostly concentrated in AB and BA regions. The antisymmetric component dominates over the rest, leading to a vortex-like texture around AA stacked regions, see Fig.~\ref{fig:textures}~(b) in the main text. The vorticity is determined by the direction of layer rotation. Note that under mirror reflection along the $\mathbf{\hat{z}}$ axis the layers are exchanged, hence $\boldsymbol{u}_0\rightarrow-\boldsymbol{u}_0$, which is the solution corresponding to twisting the layers in opposite direction. Mirror-reflecting the structure is the same as rotating the layers in opposite direction, provided that both layers are physically equivalent. Twisting the layers introduces a notion of chirality that is inherited by the stacking texture.
\end{itemize}

In addition to the acoustic phonons associated with in-phase translations of the layers, we expect another set of acoustic modes. Neglecting for a moment relaxation, $\boldsymbol{u}_0(\mathbf{r})=0$, a rigid relative translation of the layers, $\boldsymbol{\phi}_0(\mathbf{r})\rightarrow \boldsymbol{\phi}_0(\mathbf{r})+\boldsymbol{u}$, translates the beating pattern as $\boldsymbol{\phi}_0(\mathbf{r})\rightarrow\boldsymbol{\phi}_0(\mathbf{r}-\boldsymbol{\tilde{u}})$, with $\boldsymbol{\tilde{u}}=\mathbf{\hat{z}}\times\boldsymbol{u}/2\sin\frac{\overline{\theta}}{2}$. The new mass density is physically distinct but energetically equivalent to the original one as long as the system explores all the stacking configurations, i.e., if the moir\'e pattern is incommensurate with the underlying microscopic lattice. Therefore, $\boldsymbol{\tilde{u}}$ parametrizes another set of soft modes whose energy should go to zero as $\mathbf{q}\rightarrow0$. This argument is a bit oversimplified, since neglecting $\boldsymbol{u}_0$ is basically the same as neglecting the interaction between layers, so we just recovered the original acoustic phonons of the individual layers recast in new variables. However, even in the presence of interlayer interactions, there is not just one texture but a whole family of degenerate solutions of the relaxation problem, $\boldsymbol{\phi}_0(\mathbf{r}-\boldsymbol{\tilde{u}})=2\sin\frac{\overline{\theta}}{2}\,\mathbf{\hat{z}}\times(\mathbf{r}-\boldsymbol{\tilde{u}})+\boldsymbol{u}_0(\mathbf{r}-\boldsymbol{\tilde{u}})$, parametrized by an arbitrary $\boldsymbol{\tilde{u}}$. Therefore, displacements of the origin of the beating pattern (the amplitude of the density wave) with respect to the center of mass of the structure are soft. These displacements, however, no longer correspond to a uniform translation of one layer with respect to the other, which is no longer a symmetry of the problem. Rigid relative translations involve soft modes but also optical modes that distort the density amplitude, the weight of which is determined by the amount of lattice relaxation (see next section). Intuitively, we can understand this result from the fact that lattice relaxation tries always to expand regions of partial commensuration; in those areas, a rigid shift of the layers must introduce an energy cost just as in the example analyzed before. The new soft modes describe more complex atomic re-arrangements describing the sliding of the stacking texture with respect to the microscopic lattice, akin to the phason in the Frenkel-Kontorowa model.

\subsubsection{Hamiltonian of stacking fluctuations and harmonic oscillations}

Next, we consider deviation from the relaxed structure, $\boldsymbol{u}(\mathbf{r})=\boldsymbol{u}_0(\mathbf{r})+\delta\boldsymbol{\phi}(\mathbf{r})$. We can expand the previous energy functionals in powers of $\delta\boldsymbol{\phi}(\mathbf{r})$ in order to evaluate the free-energy cost of stacking fluctuations:\begin{align}
\label{eq:model}
F\left[\delta\boldsymbol{\phi}\left(\mathbf{r}\right)\right]=F_0+\int d^2\mathbf{r}\left[\frac{\lambda}{4}\left(\boldsymbol{\nabla}\cdot\delta\boldsymbol{\phi}\right)^2+\frac{\mu}{4}\, \left(\partial_i\delta\phi_j\partial_i\delta\phi_j+\partial_i\delta\phi_j\partial_j\delta\phi_i\right)+\sum_{n=2}^{\infty}\frac{1}{n!}\frac{\partial^n\mathcal{V}_{\textrm{ad}}}{\partial\phi_i...\partial\phi_j}|_{\delta\boldsymbol{\phi}=0}\,\delta\phi_i...\delta\phi_j\right].
\end{align}
The first term represents the free energy of the relaxed texture, $\boldsymbol{\phi}_0(\mathbf{r})$. Linear terms in $\delta\boldsymbol{\phi}(\mathbf{r})$ cancel due to the equilibrium of forces in Eq.~\eqref{eq:relaxation}. The other terms quantify the energy cost of stacking fluctuations with respect to the minimum energy solution. The static susceptibility is related to the functional derivative of the free energy as
\begin{align} 
\left[\hat{\chi}_0^{-1}(\mathbf{r},\mathbf{r}')\right]_{ij}=\frac{\delta^{2}F\left[\delta\boldsymbol{\phi}\left(\mathbf{r}\right)\right]}{\delta \phi_i\left(\mathbf{r}\right)\delta \phi_j\left(\mathbf{r}'\right)}|_{\delta\boldsymbol{\phi}=0}.
\end{align}

It is convenient to introduce Fourier series for the stacking fluctuations. In order to exploit the moir\'e translational invariance, we separate momenta in a component $\mathbf{q}$ within the moir\'e Brillouin zone (mBZ) and a reciprocal lattice vector:\begin{align}
\delta\phi_i\left(\mathbf{r}\right)=\frac{1}{\sqrt{A}}\sum_{\mathbf{q}\in\textrm{mBZ}}\sum_{\left\{\mathbf{G}\right\}}\phi_{i,\mathbf{G}}\left(\mathbf{q}\right) e^{i\left(\mathbf{q}+\mathbf{G}\right)\cdot\mathbf{r}}.
\end{align}
The free energy of the Fourier components of the fluctuation fields reads then (summation over repeated latin indices is assumed)
\begin{subequations}
\begin{align}
F= & F_0+\frac{1}{2}\sum_{\mathbf{q}\in\textrm{mBZ}}\sum_{\left\{\mathbf{G}\right\}}\left[\frac{\mu}{2}\left|\mathbf{q}+\mathbf{G}\right|^2\delta_{ij}+\frac{\mu+\lambda}{2}\left(\mathbf{q}+\mathbf{G}\right)_i\left(\mathbf{q}+\mathbf{G}\right)_j\right]\phi_{i,-\mathbf{G}}\left(-\mathbf{q}\right)\phi_{j,\mathbf{G}}\left(\mathbf{q}\right)
\\
& +\sum_{n=2}^{\infty}\sum_{\mathbf{q}_1\in\textrm{mBZ}}\sum_{\left\{\mathbf{G}_1\right\}}...\sum_{\mathbf{q}_n\in\textrm{mBZ}}\sum_{\left\{\mathbf{G}_n\right\}}\frac{1}{n!}\mathcal{V}_{i...j}^{(n)}\left(\mathbf{q}_1+\mathbf{G}_1+...+\mathbf{q}_n+\mathbf{G}_n\right)\phi_{i,\mathbf{G}_1}\left(\mathbf{q}_1\right)...\,\phi_{j,\mathbf{G}_n}\left(\mathbf{q}_n\right),
\nonumber
\end{align}
with\begin{align}
\mathcal{V}_{i...j}^{(n)}\left(\mathbf{k}\right)=\frac{1}{A^{\frac{n}{2}}}\int d^2\mathbf{r}\,e^{i\mathbf{k}\cdot\mathbf{r}}\, \frac{\partial^n\mathcal{V}_{\textrm{ad}}}{\partial\phi_i...\partial\phi_j}|_{\delta\boldsymbol{\phi}=0}.
\end{align}
\end{subequations}
The moir\'e translational symmetry of the equilibrium stacking texture limits the Fourier harmonics $\mathbf{k}$ in this last expression to vectors in the moir\'e reciprocal lattice. The components of the static susceptibility in momentum space follow from inverting the matrix\begin{align}
\left[\hat{\chi}_0^{-1}\left(\mathbf{q}\right)\right]_{i,j;\mathbf{G}_1,\mathbf{G}_2}=\frac{\partial^2 F}{\partial\phi_{i,-\mathbf{G}_1}\left(-\mathbf{q}\right)\partial\phi_{j,\mathbf{G}_2}\left(\mathbf{q}\right)}|_{\delta\boldsymbol{\phi}=0},
\end{align}
which reduces to Eqs.~\eqref{eq:harmonic_spectrum} of the main text for the present model.

The total Hamiltonian is $H=T+F$, where $T$ is the kinetic energy of stacking fluctuations, $T=\varrho^{-1}\int d^2\mathbf{r}\,\boldsymbol{\pi}^2(\mathbf{r})$. Here $\boldsymbol{\pi}(\mathbf{r})$ is the momentum density conjugate to the staking fluctuation field, $\{\delta\phi_i(\mathbf{x}),\pi_j(\mathbf{y})\}=\delta_{ij}\,\delta^{(2)}(\mathbf{x}-\mathbf{y})$, which can be identified with the relative linear momentum of the layers. Up to quadratic order, the Hamiltonian reads\begin{align}
H^{(2)}=\sum_{\mathbf{q}\in\textrm{mBZ}}\left[\frac{\boldsymbol{\pi}^{\dagger}\left(\mathbf{q}\right)\cdot\boldsymbol{\pi}\left(\mathbf{q}\right)}{\varrho}+\frac{1}{2}\boldsymbol{\phi}^{\dagger}\left(\mathbf{q}\right)\cdot\hat{\chi}_0^{-1}\left(\mathbf{q}\right)\cdot\boldsymbol{\phi}\left(\mathbf{q}\right)\right],
\end{align}
where we have written the fields in the vector notation introduced in the main text, e.g., $\boldsymbol{\phi}(\mathbf{q})=\sum_{\{\mathbf{G}\}}\phi_{i,\mathbf{G}}(\mathbf{q})\hat{\boldsymbol{e}}_{i,\mathbf{G}}$. The equations of motion deduced from this Hamiltonian can be recast as\begin{align}
\ddot{\boldsymbol{\phi}}\left(\mathbf{q}\right)+\frac{2}{\varrho}\,\hat{\chi}_0^{-1}\left(\mathbf{q}\right)\cdot\boldsymbol{\phi}\left(\mathbf{q}
\right)=0.\end{align}
Therefore, the frequencies and polarization vectors of harmonic oscillations follow from the diagonalization of $2\hat{\chi}_0^{-1}(\mathbf{q})/\varrho$. Note also that in normal coordinates, a generic stacking fluctuation can be written as \begin{subequations}\begin{align}
\boldsymbol{\phi}\left({\mathbf{q}}\right)=\sum_{n}\phi_n\left(\mathbf{q}\right)\hat{\boldsymbol{e}}_{n}\left(\mathbf{q}\right),
\end{align}
or back in real space,\begin{align}
\delta\phi_i\left({\mathbf{r}}\right)=\frac{1}{\sqrt{A}}\sum_{n}\sum_{\mathbf{q}\in\textrm{mBZ}}\sum_{\left\{\mathbf{G}\right\}}\phi_n\left(\mathbf{q}\right)c_{i,\mathbf{G}}^{n}\left(\mathbf{q}\right)e^{i\left(\mathbf{q}+\mathbf{G}\right)\cdot\mathbf{r}}.
\end{align}
\end{subequations}

To proceed, we first neglect the interaction between layers, $V_{\textrm{AA}}=0$. The harmonic spectrum corresponds to the original graphene phonons folded back onto the mBZ. In this reduced zone scheme, the acoustic branches corresponds to the longitudinal and transverse acoustic phonons with momentum within the first mBZ (i.e., $\mathbf{G}=0$); we have\begin{subequations}
\begin{align}
& c_{L}=\sqrt{\frac{\lambda+2\mu}{\varrho}},\,\,\, c_{i,\mathbf{G}}^L\left(\mathbf{q}\right)=\frac{i\,q_i}{|\mathbf{q}|}\,\delta_{\mathbf{G},\mathbf{0}},\\
& c_{T}=\sqrt{\frac{\mu}{\varrho}},\,\,\, c_{i,\mathbf{G}}^T\left(\mathbf{q}\right)=\frac{i\,(\mathbf{q}\times\mathbf{\hat{z}})_i}{|\mathbf{q}|}\,\delta_{\mathbf{G},\mathbf{0}}.
\end{align}
\end{subequations}

Let us now introduce the coupling between layers. Its first effect is to relax the structure, as we have seen. The sound velocity of the new acoustic branches does not change much as the lower energy of the soliton system is compensated by the smaller inertia of the sliding motion \cite{phasons,Koshino_phonons}. The polarization vector must change, however, reflecting the formation of a sharper texture. We can estimate these coefficients in perturbation theory in the Fourier coefficients of $\boldsymbol{u}_0(\mathbf{r})$. The result is that a stacking fluctuation related to the sliding motion of the soliton system can be related to a fluctuation of the collective coordinate $\boldsymbol{\tilde{u}}$ as\begin{align}
\label{eq:previous_eq}
\frac{1}{\sqrt{A}}\sum_{n=L,T}\sum_{\mathbf{q}\in\textrm{mBZ}}\sum_{\left\{\mathbf{G}\right\}}\phi_n\left(\mathbf{q}\right)c_{i,\mathbf{G}}^{n}\left(\mathbf{q}\right)e^{i\left(\mathbf{q}+\mathbf{G}\right)\cdot\mathbf{r}}\approx2\sin\frac{\overline{\theta}}{2}\left(\delta\boldsymbol{\tilde{u}}(\mathbf{r})\times\mathbf{\hat{z}}\right)_i-\delta\boldsymbol{\tilde{u}}(\mathbf{r})\cdot\boldsymbol{\nabla}\left[\boldsymbol{u}_0\right]_i\left(\mathbf{r}\right).
\end{align}
In the left-hand side we truncated the summation to the acoustic branches only. The zeroth-order approximation corresponds to the previous solution and establishes the following relation between infinitesimal phason fluctuations and stacking fluctuations:\begin{align}
\frac{1}{\sqrt{A}}\sum_{n=L,T}\sum_{\mathbf{q}\in\textrm{mBZ}}\phi_n\left(\mathbf{q}\right)c_{i,\mathbf{0}}^{n}\left(\mathbf{q}\right)e^{i\mathbf{q}\cdot\mathbf{r}}\approx2\sin\frac{\overline{\theta}}{2}\left(\delta\boldsymbol{\tilde{u}}(\mathbf{r})\times\mathbf{\hat{z}}\right)_i\Longrightarrow \delta\boldsymbol{\tilde{u}}(\mathbf{r})\approx i\sum_{\mathbf{q}\in\textrm{mBZ}}\frac{\frac{\mathbf{q}}{|\mathbf{q}|}\phi_T\left(\mathbf{q}\right)+\frac{\mathbf{\hat{z}}\times\mathbf{q}}{|\mathbf{q}|}\phi_L\left(\mathbf{q}\right)}{2\sin\frac{\overline{\theta}}{2}\sqrt{A}}e^{i\mathbf{q}\cdot\mathbf{r}}.
\end{align}
Note that longitudinal (transverse) phason fluctuations correspond to transverse (longitudinal) stacking fluctuations. Plugging this result into the right-hand side of Eq.~\eqref{eq:previous_eq} along with the general expansion in Eq.~\eqref{eq:general_expansion}, we obtain the first-order correction for the $\mathbf{G}\neq0$ components,\begin{subequations}
\begin{align}
& c_{i,\mathbf{G}_j^{\alpha}}^L\left(\mathbf{q}\right)\approx -\frac{i\mathbf{q}\cdot\boldsymbol{g}_j^{\alpha}}{|\mathbf{q}|}\left[\frac{\mathbf{G}_j^{\alpha}}{\left|\mathbf{G}_j^{\alpha}\right|}u_L^{\alpha}+\frac{\mathbf{G}_j^{\alpha}\times\mathbf{\hat{z}}}{\left|\mathbf{G}_j^{\alpha}\right|}u_T^{\alpha}\right]_i,\\
& c_{i,\mathbf{G}_j^{\alpha}}^T\left(\mathbf{q}\right)\approx \frac{i\left(\mathbf{q}\times\boldsymbol{g}_j^{\alpha}\right)_z}{|\mathbf{q}|}\left[\frac{\mathbf{G}_j^{\alpha}}{\left|\mathbf{G}_j^{\alpha}\right|}u_L^{\alpha}+\frac{\mathbf{G}_j^{\alpha}\times\mathbf{\hat{z}}}{\left|\mathbf{G}_j^{\alpha}\right|}u_T^{\alpha}\right]_i.
\end{align}
\end{subequations}

\subsubsection{Disorder correlation functions}

As explained in the main text, we assume Gaussian distributions for the random potentials, coarse-grained on the scale $\zeta_{\alpha}$:
\begin{align}
\left\langle V_{\alpha}\left(\mathbf{x}\right) V_{\beta}\left(\mathbf{y}\right)\right\rangle_{\textrm{dis}}=\overline{V_{\alpha}^2}\zeta_{\alpha}^2\,\delta^{(2)}\left(\mathbf{x}-\mathbf{y}\right)\delta_{\alpha,\beta}.
\end{align}
The Fourier components of $\mathcal{G}_{ij}\left(\mathbf{x},\mathbf{y}\right)$ do not mix moir\'e reciprocal lattice vectors,\begin{align}
\mathcal{G}_{ij;\mathbf{G},\mathbf{G}}\left(\mathbf{q}\right) = \zeta_1^2\overline{V_1^2}\left[\boldsymbol{\hat{z}}\times\left(\mathbf{q}+\mathbf{G}\right)\right]_i\left[\boldsymbol{\hat{z}}\times\left(\mathbf{q}+\mathbf{G}\right)\right]_j
 + \zeta_2^2\overline{V_2^2}\left[\mathbf{q}+\mathbf{G}\right]_i\left[\mathbf{q}+\mathbf{G}\right]_j.
\end{align}
The long-wavelength limit of the force correlations projected onto the phason subspace reduces to\begin{align}
\mathcal{G}_{n_1,n_2}\left(0\right)=\zeta_1^2\overline{V_1^2}\sum_{i,j,\left\{\mathbf{G}\right\}}\left[c_{i,\mathbf{G}}^{n_1}\left(0\right)\right]^*c_{j,\mathbf{G}}^{n_2}\left(0\right)\left(\mathbf{\hat{z}}\times\mathbf{G}\right)_i\left(\mathbf{\hat{z}}\times\mathbf{G}\right)_j+\zeta_2^2\overline{V_2^2}\sum_{i,j,\left\{\mathbf{G}\right\}}\left[c_{i,\mathbf{G}}^{n_1}\left(0\right)\right]^*c_{j,\mathbf{G}}^{n_2}\left(0\right)G_iG_j.
\end{align}
Averaging over the disorder distribution recovers the 6-fold rotational symmetry of the moir\'e pattern, leading to a diagonal tensor $\mathcal{G}_{n_1,n_2}=\mathcal{G}\delta_{n_1,n_2}$. From the perturbative calculation above, we get\begin{align}
\mathcal{G}\approx 3\sum_{\alpha}\left| \boldsymbol{g}^{\alpha} \right|^2\left| \mathbf{G}^{\alpha} \right|^2\left[\zeta_1^2\overline{V_1^2}\left(u_T^{\alpha}\right)^2+\zeta_2^2\overline{V_2^2}\left(u_L^{\alpha}\right)^2\right],
\end{align}
where the summation is in momentum stars. Figure~\ref{fig:textures}~(d) in the main text shows $\mathcal{G}$ calculated from this formula and the self-consistent solution of the relaxation problem including up to 10 momentum stars, Fig.~\ref{fig:lattice}. The deviation from a quadratic growth results from the change in slope of $u_{T,L}^{\alpha}$ as a function of $L_{\textrm{m}}/\ell$.

The disorder correlator in Eq.~\eqref{eq:disorder_C} can be recast as\begin{align}
\label{eq:corr_general}
C_{ij}\left(\mathbf{x},\mathbf{y}\right)=\sum_{n_1,n_2}\sum_{\left\{\mathbf{G}_1\right\}}\sum_{\left\{\mathbf{G}_2\right\}}e^{i\mathbf{G}_1\cdot\mathbf{x}-i\mathbf{G}_2\cdot\mathbf{y}}\int \frac{d^2\mathbf{q}}{\left(2\pi\right)^2}\,e^{i\mathbf{q}\cdot\left(\mathbf{x}-\mathbf{y}\right)}\frac{4\,\mathcal{G}_{n_1,n_2}\left(\mathbf{q}\right)}{\varrho^2\omega_{n_1}^2\left(\mathbf{q}\right)\omega_{n_2}^2\left(\mathbf{q}\right)}c_{i,\mathbf{G}_1}^{n_1}\left(\mathbf{q}\right)\left[c_{j,\mathbf{G}_2}^{n_2}\left(\mathbf{q}\right)\right]^*.
\end{align}
The correlation function at long distances is dominated by the soft modes yielding a diverging integrand in the $\mathbf{q}\rightarrow 0$ limit. Lattice relaxation already introduces a modulation on the moir\'e scale through $c_{i,\mathbf{G}}^{n}(\mathbf{q})$, so even if we retain only the phason contribution, $C_{ij}(\mathbf{x},\mathbf{y})$ is no longer a function of the difference in the arguments, $\mathbf{x}-\mathbf{y}$. Yet, this modulation is subleading in the equilibrium displacement $\boldsymbol{u}_0(\mathbf{r})$, so for the sake of simplicity we can just ignore it and approximate (summation over repeated latin indices is assumed)\begin{align}
\label{eq:approx_corr}
C_{ij}\left(\mathbf{x},\mathbf{y}\right)\approx\frac{4\mathcal{G}}{\varrho^2}\left(\frac{\delta_{ik}\delta_{jl}}{c_L^4}+\frac{\varepsilon_{ik}\varepsilon_{jl}}{c_T^4}\right)\int \frac{d^2\mathbf{q}}{\left(2\pi\right)^2}\, e^{i\mathbf{q}\cdot\left(\mathbf{x}-\mathbf{y}\right)}\frac{q_kq_l}{|\mathbf{q}|^6},
\end{align}
where $\varepsilon_{ij}$ the 2D Levi-Civita symbol. The rapid growth of disorder fluctuations with distance implies that stacking correlations are exponentially lost,\begin{align}
\left\langle e^{i\boldsymbol{g}\cdot\left(\delta\boldsymbol{\phi}(\mathbf{x})-\delta\boldsymbol{\phi}(\mathbf{y})\right)} \right\rangle_{\textrm{dis}}\approx \exp\left[-\frac{4\mathcal{G}|\mathbf{x}-\mathbf{y}|^2}{\varrho^2}\left(\frac{g_ig_j}{c_L^4}+\frac{(\hat{\mathbf{z}}\times\boldsymbol{g})_i(\hat{\mathbf{z}}\times\boldsymbol{g})_j}{c_T^4}\right)I_{ij}(\mathbf{x}-\mathbf{y})\right],
\end{align}
where we have introduced the dimensionless integral\begin{align}
I_{ij}(\mathbf{r})=\int \frac{d^2\boldsymbol{\xi}}{\left(2\pi\right)^2}\,\frac{\xi_i\xi_j}{|\boldsymbol{\xi}|^6}\left(1-e^{\frac{i\boldsymbol{\xi}\cdot\mathbf{r}}{|\mathbf{r}|}}\right)=\int\frac{d\xi}{2\pi}\,\xi^{-3}\left[\left(\frac{1}{2}-\frac{J_{1}(\xi)}{\xi}\right)\delta_{ij}+\frac{r_ir_j}{|\mathbf{r}|^2}J_2(\xi)\right].
\end{align}
Here $J_i(\xi)$ are Bessel functions and the integral is performed in $\xi\equiv |\mathbf{r}||\mathbf{q}|$. In order to avoid the infrared divergence we can cut-off the available momenta by the size of the sample $L$, $\xi_c\sim |\mathbf{r}|/L$, which leads to the scaling in Eq.~\eqref{eq:scaling_pinning}; specifically,\begin{align}
\left\langle e^{i\boldsymbol{g}\cdot\left(\delta\boldsymbol{\phi}(\mathbf{x})-\delta\boldsymbol{\phi}(\mathbf{y})\right)} \right\rangle_{\textrm{dis}}\approx\left(\frac{L}{|\mathbf{x}-\mathbf{y}|}\right)^{-\frac{\mathcal{G}}{8\pi\varrho^2}\left[\frac{|\boldsymbol{g}|^2|\mathbf{x}-\mathbf{y}|^2+2|\boldsymbol{g}\cdot(\mathbf{x}-\mathbf{y})|^2}{c_L^4}+\frac{|\boldsymbol{g}|^2|\mathbf{x}-\mathbf{y}|^2+2|\boldsymbol{g}\times(\mathbf{x}-\mathbf{y})|^2}{c_T^4}\right]}.
\end{align}
The length scale $L_c$ in Eq.~\eqref{eq:pinning} is the characteristic length at which stacking fluctuations become of the order of graphene's lattice constant, which translates to fluctuations of the order of the moir\'e period for the center of the mass-density amplitude, $\langle \delta\boldsymbol{\tilde{u}}^2 \rangle_{\textrm{dis}}\sim L_{\textrm{m}}^2$.

Another interesting correlation function is the fluctuation of the twist angle, $\delta\theta=\partial_x\delta\phi_y-\partial_y\delta\phi_x$. From Eq.~\eqref{eq:approx_corr}, we have\begin{align}
\label{eq:twit_angle_fluctuations}
\left\langle \left(\delta\theta\left(\mathbf{r}\right) - \delta\theta\left(0\right)\right)^2 \right\rangle_{\textrm{dis}}\approx\frac{8\mathcal{G}}{\varrho^2c_T^4}\int \frac{d^2\mathbf{q}}{\left(2\pi\right)^2}\frac{1-e^{i\mathbf{q}\cdot\mathbf{r}}}{\left|\mathbf{q}\right|^2}\approx\frac{4\mathcal{G}}{\pi\varrho^2c_T^4}\,\ln\frac{|\mathbf{r}|}{L_{\textrm{m}}},
\end{align}
where momenta are cut-off by $1/L_{\textrm{m}}$ in order to avoid the ultraviolet divergence.

Fluctuations of the twist angle grow logarithmically. However, this growth must be stopped at $L_c$, where the centers of the mass amplitudes cease to be correlated, so the moir\'e superlattice does not respond elastically to external perturbations anymore. In other words, the cumulative effect of microscopic forces acting on stacking configurations stops at this length scale. Hence, fluctuations of the twist angle are characterized by the value of this correlation function at $|\mathbf{r}|=L_c$, which leads to Eq.~\eqref{eq:variance}; specifically,\begin{align}
    \left\langle \delta\theta^2 \right\rangle_{\textrm{dis}}\approx\frac{a^2}{\pi^2L_c^2}\frac{c_L^4}{c_L^4+c_T^4}\ln\left(\frac{L_c}{L_{\textrm{m}}}\right).
\end{align}
In the expression used in the main text, we dropped the material-dependent factor $c_L^4/(c_L^4+c_T^4)\sim 1$.

Let us consider now the same correlation functions if we neglected lattice relaxation or, equivalently, the interaction between layers. In that case, we have for the soft modes\begin{align}
\mathcal{G}_{n_1,n_2}\left(\mathbf{q}\right)=\zeta_1^2\overline{V_1^2}|\mathbf{q}|^2\delta_{n_1,T}\delta_{n_2,T}+\zeta_2^2\overline{V_2^2}|\mathbf{q}|^2\delta_{n_1,L}\delta_{n_2,L}.
\end{align}
From Eq.~\eqref{eq:corr_general} and following the same approximations as before, we have now (summation over repeated indices is assumed):\begin{align}
C_{ij}\left(\mathbf{x},\mathbf{y}\right)\approx\frac{4}{\varrho^2}\left(\frac{\zeta_1^2\overline{V_1^2}}{c_L^4}\delta_{ik}\delta_{jl}+\frac{\zeta_2^2\overline{V_2^2}}{c_T^4}\varepsilon_{ik}\varepsilon_{jl}\right)\int \frac{d^2\mathbf{q}}{\left(2\pi\right)^2}\, e^{i\mathbf{q}\cdot\left(\mathbf{x}-\mathbf{y}\right)}\frac{q_kq_l}{|\mathbf{q}|^4}.
\end{align}
The growth of disorder fluctuations is strongly attenuated with respect to the previous case. Consequently, stacking correlations only decay algebraically. Specifically, we can write\begin{align}
\left\langle e^{i\boldsymbol{g}\cdot\left(\delta\boldsymbol{\phi}(\mathbf{x})-\delta\boldsymbol{\phi}(\mathbf{y})\right)} \right\rangle_{\textrm{dis}}\approx e^{-\frac{4}{\varrho^2}\left(\frac{\zeta_2^2 \overline{V_2^2}}{c_L^4}g_ig_j+\frac{\zeta_1^2\overline{V_1^2}}{c_T^4}(\hat{\mathbf{z}}\times\boldsymbol{g})_i(\hat{\mathbf{z}}\times\boldsymbol{g})_j\right)\tilde{I}_{ij}(\mathbf{x}-\mathbf{y})},
\end{align}
where the dimensionless integral reads now\begin{align}
\tilde{I}_{ij}(\mathbf{r})=\int \frac{d\xi}{2\pi}\,\xi^{-1}\left[\left(\frac{1}{2}-\frac{J_1(\xi)}{\xi}\right)\delta_{ij}+\frac{r_ir_j}{|\mathbf{r}|^2}J_2(\xi)\right]\sim\frac{1}{4\pi}\ln\left(\frac{|\mathbf{r}|}{L_{\textrm{m}}}\right),
\end{align}
leading to slow (algebraical) decay of stacking correlations,\begin{align}
\left\langle e^{i\boldsymbol{g}\cdot\left(\delta\boldsymbol{\phi}(\mathbf{x})-\delta\boldsymbol{\phi}(\mathbf{y})\right)} \right\rangle_{\textrm{dis}}\sim \left(\frac{L_{\textrm{m}}}{\left|\mathbf{x}-\mathbf{y}\right|}\right)^{\frac{16\pi}{3a^2}\left(\frac{\zeta_1^2\overline{V_1^2}}{\mu^2}+\frac{\zeta_2^2\overline{V_2^2}}{(\lambda+2\mu)^2}\right)}.
\end{align}

For the fluctuations in twist angle, we have\begin{align}
\left\langle \left(\delta\theta\left(\mathbf{r}\right) - \delta\theta\left(0\right)\right)^2 \right\rangle_{\textrm{dis}}\approx\frac{8\zeta_1^2\overline{V_1^2}}{\varrho^2c_T^4}\int \frac{d^2\mathbf{q}}{\left(2\pi\right)^2}\left(1-e^{i\mathbf{q}\cdot\mathbf{r}}\right)\approx\frac{2\zeta_1^2\overline{V_1^2}}{\pi\varrho^2c_T^4L_{\textrm{m}}^2}\left[1-\frac{2L_{\textrm{m}}}{|\mathbf{r}|}J_{1}\left(\frac{|\mathbf{r}|}{L_{\textrm{m}}}\right)\right].
\end{align}
There is no criterion to cut-off this correlation function, but there is no need: in this case fluctuations in the twist angle saturate quickly to the pre-factor in this last expression. Comparing this value with the pre-factor in Eq.~\eqref{eq:twit_angle_fluctuations} and the calculation of $\mathcal{G}$ in Fig.~\ref{fig:textures}~(d), we see that the latter contribution is always larger as long as $L_{\textrm{m}}> \ell$. Therefore, for a model of random tensions, regardless of their microscopic origin, lattice relaxation and the mutual interaction between layers ultimately determine the twist-angle landscape for the regime of small angles pertinent to the experiments.

\subsubsection{Anharmonic fluctuations}

We now consider the time-dependent thermal correlation function,\begin{align}
C_{ij}\left(\mathbf{x},\mathbf{y},t-t'\right)=\left\langle \delta\boldsymbol{\phi}_i\left(\mathbf{x},t\right) \delta\boldsymbol{\phi}_j\left(\mathbf{y},t'\right) \right\rangle_T.
\end{align}
At equal times, this is just the thermal average $C_{ij}(\mathbf{x},\mathbf{y},0)=k_B T\hat{\chi}(\mathbf{x},\mathbf{y})$. The fluctuation-dissipation theorem establishes a more general relation with the dynamical response of the system; in particular, for the Laplace transformed quantities (with complex frequency $z$), we have \begin{align}
\hat{C}\left(\mathbf{x},\mathbf{y},z\right)=\frac{k_B T}{iz}\left[\hat{\chi}\left(\mathbf{x},\mathbf{y},z\right)-\hat{\chi}\left(\mathbf{x},\mathbf{y}\right)\right],
\end{align}
where $\hat{\chi}(\mathbf{x},\mathbf{y},z)$ is the analytical continuation of the stacking susceptibility such that $\hat{\chi}(\mathbf{x},\mathbf{y},\omega)=\lim_{\epsilon\rightarrow 0}\hat{\chi}(\mathbf{x},\mathbf{y},z)|_{z=\omega+i\epsilon}$. In the harmonic approximation, we have simply \begin{align}
\hat{C}^{(2)}\left(\mathbf{q},z\right)=\frac{2i zk_B T}{\varrho}\sum_n\frac{\hat{\boldsymbol{e}}_n\left(\mathbf{q}\right)\otimes \hat{\boldsymbol{e}}_n^{\dagger}\left(\mathbf{q}\right)}{\left(z^2-\omega_{n}^2\left(\mathbf{q}\right)\right)\omega_{n}^2\left(\mathbf{q}\right)},
\end{align}
which, in real-frequency domain, becomes \begin{align}
\label{eq:harmonic_corr}
\hat{C}^{(2)}\left(\mathbf{q},\omega\right)=2\lim_{\epsilon\rightarrow 0}\Re\,\hat{C}^{(2)}\left(\mathbf{q},\omega+i\epsilon\right)=\sum_n\frac{2\pi k_BT}{\varrho\,\omega\,\omega_n\left(\mathbf{q}\right)}\left[\delta\left(\omega-\omega_{n}\left(\mathbf{q}\right)\right)-\delta\left(\omega+\omega_{n}\left(\mathbf{q}\right)\right)\right]\hat{\boldsymbol{e}}_n\left(\mathbf{q}\right)\otimes \hat{\boldsymbol{e}}_n^{\dagger}\left(\mathbf{q}\right).
\end{align}

It is useful to consider the time-dependent stacking fluctuations as formal solutions of the equations of motion, $\delta\boldsymbol{\phi}(\mathbf{r},t)=e^{it\hat{L}}\delta\boldsymbol{\phi}(\mathbf{r},0)$, where $\hat{L}$ is the Liouville operator defined from the microscopic Hamiltonian. This can be understood as a Hermitian operator acting on the Hilbert space spanned by the dynamical variables of the theory \cite{Forster}; for the previous Hamiltonian, these are the stacking fluctuations and the conjugate momenta. The correlation function in, for example, normal coordinates, can be written as the matrix elements of the propagator $i(z-\hat{L})^{-1}$, \begin{align}
C_{n_1,n_2}\left(\mathbf{q},z\right)=\hat{\boldsymbol{e}}_{n_1}^{\dagger}\left(\mathbf{q}\right)\cdot\hat{C}\left(\mathbf{q},z\right)\cdot \hat{\boldsymbol{e}}_{n_2}\left(\mathbf{q}\right)\equiv\left\langle \phi_{n_1}\left(\mathbf{q}\right)\left|\frac{i}{z-\hat{L}}\right| \phi_{n_2}\left(\mathbf{q}\right)\right\rangle.
\end{align}
We can go back to real space just by changing the vectors of the basis,\begin{align}
\left| \delta\phi_i\left(\mathbf{r}\right)\right\rangle=\frac{1}{\sqrt{A}}\sum_n\sum_{\mathbf{q}\in\textrm{mBZ}}\sum_{\left\{\mathbf{G}\right\}}\left[c_{i,\mathbf{G}}^n\left(\mathbf{q}\right)\right]^*e^{-i\left(\mathbf{q}+\mathbf{G}\right)\cdot\mathbf{r}}\left|\phi_n\left(\mathbf{q}\right)\right\rangle.
\end{align}
The memory-matrix function introduced in the main text is related to the self-energy by $\hat{\Pi}(\mathbf{q},z)=k_B T\hat{\sigma}(\mathbf{q},z)$. The latter is given by \begin{subequations}\begin{align}
\Pi_{n_1,n_2}\left(\mathbf{q},z\right)=\left\langle \dot{\pi}_{n_1}\left(\mathbf{q}\right)\left|\hat{Q}\frac{i}{z-\hat{Q}\hat{L}\hat{Q}}\hat{Q}\right| \dot{\pi}_{n_2}\left(\mathbf{q}\right)\right\rangle,
\end{align}
or, equivalently, as the Laplace transform of the memory function\begin{align}
\Pi_{n_1,n_2}\left(\mathbf{q},t\right)=\left\langle \dot{\pi}_{n_1}\left(\mathbf{q}\right)\left|\hat{Q}e^{-it\hat{Q}\hat{L}\hat{Q}}\hat{Q}\right| \dot{\pi}_{n_2}\left(\mathbf{q}\right)\right\rangle.
\end{align}
In these last expressions $\hat{Q}$ is a Mori operator that projects out the intrinsic fluctuations of the normal modes \cite{Forster}.
\end{subequations}

Albeit abstract in form, these last expressions contain important information. Let us first neglect the interaction between layers. The relative linear momentum density is then locally conserved, hence the equations of motion are of the form (summation over repeated indices is assumed)\begin{subequations}\begin{align}
& \dot{\pi}_L(\mathbf{q})=\frac{q_iq_j}{|\mathbf{q}|}w_{ij}(\mathbf{q}),\\
& \dot{\pi}_T(\mathbf{q})=\frac{\left(\mathbf{q}\times\hat{\mathbf{z}}\right)_iq_j}{|\mathbf{q}|}w_{ji}(\mathbf{q}),
\end{align}
\end{subequations}
where we have introduced Fourier components of a symmetric \textit{heterostress} tensor, $w_{ij}\equiv \partial \overline{F}/\partial(\partial_i\delta\phi_j)$ ($\overline{F}$ is the free-energy density). It is clear that in this case $\sigma_{n_1,n_2}(\mathbf{q},z)=\beta\Pi_{n_1,n_2}(\mathbf{q},z)\propto|\mathbf{q}|^2$. However, \begin{align}
\lim_{z\rightarrow 0}\lim_{\mathbf{q}\rightarrow 0}\frac{1}{|\mathbf{q}|^2}\Pi_{n_1,n_2}\left(\mathbf{q},z\right)=k_BT\,\eta_{n_1}\delta_{n_1,n_2}
\end{align}
is finite. The parameters $\eta_{L,T}$ are bulk and shear viscosities, $\eta_{L}=\eta_{xxyy}+2\,\eta_{xyxy}$, $\eta_{T}=\eta_{xyxy}$, related to the fluctuations of the heterostress tensor via Kubo formulae of the form\begin{subequations}
\begin{align}
& \eta_{xxyy}=\frac{1}{k_BT}\lim_{\epsilon\rightarrow 0}\lim_{\mathbf{q}\rightarrow 0}\int_0^{\infty}dt\,e^{-\epsilon t}\left\langle w_{xx}\left(\mathbf{q},t\right)w_{yy}\left(-\mathbf{q},0\right) \right\rangle_T,\\
& \eta_{xyxy}=\frac{1}{k_BT}\lim_{\epsilon\rightarrow 0}\lim_{\mathbf{q}\rightarrow 0}\int_0^{\infty}dt\,e^{-\epsilon t}\left\langle w_{xy}\left(\mathbf{q},t\right)w_{xy}\left(-\mathbf{q},0\right) \right\rangle_T.
\end{align}
\end{subequations}
Note that these are the only independent components of the viscosity tensor due to the 6-fold symmetry of the system. As for the reactive forces, the stacking viscosities introduced here can be related to the viscosities of individual graphene layers. 

\begin{figure}[t!]
\begin{center}
\includegraphics[width=0.7\columnwidth]{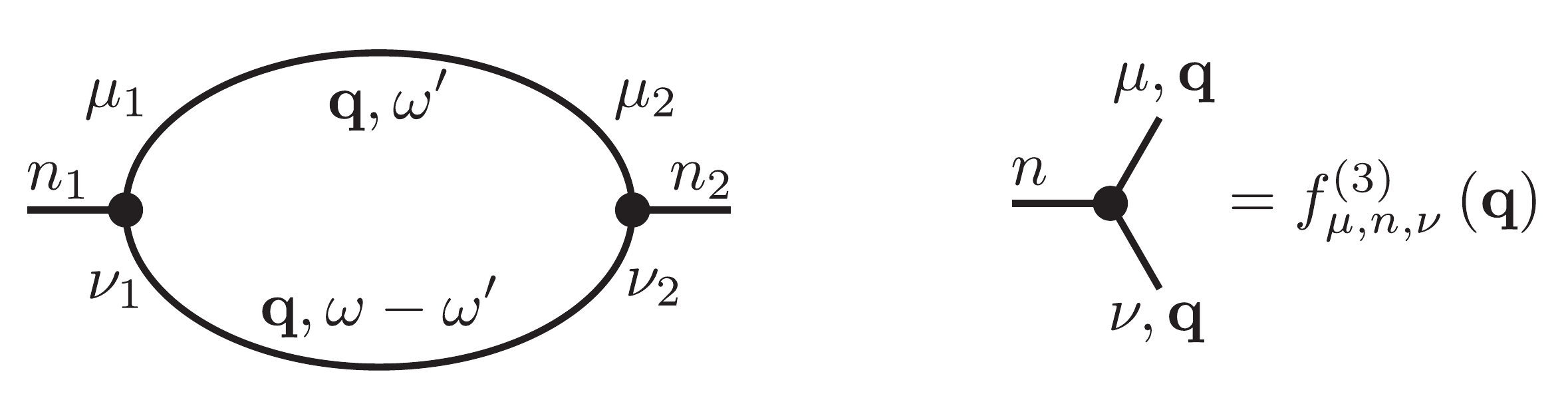}
\caption{\textbf{Phason self-energy due to third-order anharmonic forces}. Straight lines represent thermal correlation functions related to the dynamical susceptibility via the fluctuation-dissipation theorem. The vertex is defined in Eq.~\eqref{eq:kernel}.} 
\label{fig:diagram}
\end{center}
\end{figure}

The inclusion of forces between the layers implies that $\boldsymbol{\pi}$ is no longer conserved and $\lim_{\mathbf{q}\rightarrow 0}\Pi_{n_1,n_2}(\mathbf{q},z)$ is finite in general. We can estimate the self-energy of the soft modes as follows. Let us take $H$ introduced before as our \textit{microscopic} Hamiltonian. The Liouville operator in normal coordinates reads\begin{subequations}\begin{align}
\hat{L}=\sum_{\mu}\sum_{\mathbf{q}\in\textrm{mBZ}}\left[\frac{2\pi_{\mu}(\mathbf{q})}{\varrho}\left(-i\frac{\partial}{\partial\phi_{\mu}\left(\mathbf{q}\right)}\right)+f_{\mu}\left(\mathbf{q}\right)\left(-i\frac{\partial}{\partial\pi_{\mu}\left(\mathbf{q}\right)}\right)\right],
\end{align}
where we have introduced the force on mode $\mu$,
\begin{align}
f_{\mu}\left(\mathbf{q}\right)=\hat{\boldsymbol{e}}_{\mu}^{\dagger}\left(\mathbf{q}\right)\cdot\hat{\chi}_0^{-1}\left(\mathbf{q}\right)\cdot\boldsymbol{\phi}\left(\mathbf{q}\right)+\sum_{n=3}^{\infty}f_{\mu}^{(n)}\left(\mathbf{q}\right).
\end{align}
\end{subequations}
The fist term contains the forces produced by harmonic fluctuations, the rest are corrections given by higher-order terms in the expansion of the adhesion potential. Note also that\begin{align}
\left|\dot{\pi}_{\mu}\left(\mathbf{q}\right)\right\rangle=i\hat{L}\left|\pi_{\mu}\left(\mathbf{q}\right)\right\rangle=\left|f_{\mu}\left(\mathbf{q}\right)\right\rangle=\frac{\varrho}{2}\omega_{\mu}^2\left(\mathbf{q}\right)\left|\phi_{\mu}\left(\mathbf{q}\right)\right\rangle+\sum_{n=3}^{\infty}\left|f_{\mu}^{(n)}\left(\mathbf{q}\right)\right\rangle.
\end{align}
The basic idea of the calculation is to separate the soft modes from the rest of the spectrum, provided that there is always a gap between the phasons and the optical modes once we include lattice relaxation. Optical modes act then as a dissipative bath for phasons. This is implemented in the operator formalism by introducing the following Mori projector:\begin{align}
\hat{Q}=\hat{1}-\frac{1}{k_BT}\sum_{\mathbf{q}\in\textrm{mBZ}}\sum_{\mu=L,T}\left[\frac{\varrho}{2}\omega_{\mu}^2\left(\mathbf{q}\right)\left|\phi_{\mu}\left(\mathbf{q}\right)\right\rangle\left\langle \phi_{\mu}\left(\mathbf{q}\right)\right|+\frac{2}{\varrho}\left|\pi_{\mu}\left(\mathbf{q}\right)\right\rangle\left\langle \pi_{\mu}\left(\mathbf{q}\right)\right|\right].
\end{align}
It is clear from this definition that $\hat{Q}$ acting on $|\dot{\pi}_{\mu}(\mathbf{q})\rangle$ removes the harmonic component of the force. The first contribution comes from $n=3$, $\hat{Q}|\dot{\pi}_{\mu}(\mathbf{q})\rangle\approx |f_{\mu}^{(3)}(\mathbf{q})\rangle$. The self-energy can be written as a Green's function describing the fluctuations of anharmonic forces,\begin{align}
\label{eq:Green_function}
\hat{\Pi}_{n_1,n_2}^{(3)}\left(\mathbf{0},z\right)=\lim_{\mathbf{q}\rightarrow0}\int_0^{\infty} dt\,e^{izt}\left\langle f_{n_1}^{(3)}\left(\mathbf{q},t\right) f_{n_2}^{(3)}\left(-\mathbf{q},0\right) \right\rangle_T,
\end{align}
diagramatically depicted in Fig.~\ref{fig:diagram}. For the forces, we have (hereafter repeated indices are summed) \begin{align}
f_{n}^{(3)}\left(\mathbf{q}\right) & =-\hat{\boldsymbol{e}}_{n}^{\dagger}\left(\mathbf{q}\right)\cdot\frac{\partial F^{(3)}}{\partial\boldsymbol{\phi}^{\dagger}\left(\mathbf{q}\right)}\\
& =-\sum_{\mathbf{q}_1,\mathbf{q}_2\in\textrm{mBZ}}\sum_{\left\{\mathbf{G}_1\right\},\left\{\mathbf{G}_2\right\},\left\{\mathbf{G}_3\right\}}\left[c_{k,\mathbf{G}_3}^{n}\left(\mathbf{q}\right)\right]^{*}\frac{1}{2}\mathcal{V}_{ijk}^{(3)}\left(\mathbf{q}_1+\mathbf{q}_2+\mathbf{G}_1+\mathbf{G}_2-\mathbf{q}-\mathbf{G}_3\right)\phi_{i,\mathbf{G}_1}\left(\mathbf{q}_1\right)\phi_{j,\mathbf{G}_2}\left(\mathbf{q}_2\right).
\nonumber
\end{align} 
In the long-wavelength limit, this expression reduces to\begin{subequations}\begin{align}
f_{n}^{(3)}\left(\mathbf{0}\right)=-\frac{1}{2\sqrt{A}}\sum_{\mathbf{q}\in\textrm{mBZ}}\sum_{\mu,\nu} f_{\mu,n,\nu}^{(3)}\left(\mathbf{q}\right)\phi_{\mu}\left(-\mathbf{q}\right)\phi_{\nu}\left(\mathbf{q}\right),
\end{align}
with the vertex given by\begin{align}
\label{eq:kernel}
f_{\mu,n,\nu}^{(3)}\left(\mathbf{q}\right)=\frac{1}{A}\sum_{\left\{\mathbf{G}_1\right\}}\sum_{\left\{\mathbf{G}_2\right\}}\sum_{\left\{\mathbf{G}_3\right\}}\int d\mathbf{r}\,e^{-i\left(\mathbf{G}_3-\mathbf{G}_1-\mathbf{G}_2\right)} \,\frac{\partial^3 \mathcal{V}_{\textrm{ad}}}{\partial\phi_i\partial\phi_j\partial\phi_k}|_{\delta\boldsymbol{\phi}=0}\, c_{i,\mathbf{G}_1}^{\mu}\left(-\mathbf{q}\right)c_{j,\mathbf{G}_2}^{\nu}\left(\mathbf{q}\right)\left[c_{k,\mathbf{G}_3}^{n}\left(\mathbf{0}\right)\right]^*.
\end{align}
\end{subequations}
From the diagram in Fig.~\ref{fig:diagram}, we have\begin{align}
\Pi_{n_1,n_2}^{(3)}(\mathbf{0},\omega)=\frac{1}{A}\sum_{\mathbf{q}\in\textrm{mBZ}}\sum_{\mu_1,\mu_2}\sum_{\nu_1,\nu_2}\int\frac{d\omega '}{2\pi}f_{\mu_1,n_1,\nu_1}\left(\mathbf{q}\right)f_{\mu_2,n_2,\nu_2}\left(-\mathbf{q}\right)C_{\mu_1,\mu_2}\left(\mathbf{q},\omega'\right)C_{\nu_1,\nu_2}\left(\mathbf{q},\omega-\omega'\right).
\end{align}
Neglecting anharmonic forces in the correlation functions (Eq.~\ref{eq:harmonic_corr}), the final result for the scattering rate reads\begin{align}
\tau^{-1}=\frac{2}{\varrho\, k_B T}\lim_{\omega\rightarrow 0}\Pi^{(3)}_{n,n}\left(\mathbf{0},\omega\right)=\frac{2k_BT}{\pi\varrho^3}\sum_{\mu,\nu}\int d^2\mathbf{q}\,\frac{\left|f_{\mu,n,\nu}^{(3)}\left(\mathbf{q}\right)\right|^2}{\omega_{\mu}^2\left(\mathbf{q}\right)\omega_{\nu}^2\left(\mathbf{q}\right)}\,\delta\left(\omega_{\mu}\left(\mathbf{q}\right)-\omega_{\nu}\left(\mathbf{q}\right)\right).
\end{align}

For an order of magnitude estimate, we can neglect the reconstruction of the spectrum of vibrations, which is not an unreasonable approximation for relatively large angles, $\overline{\theta}>3^{\textrm{o}}$. In an extended zone scheme, phasons at the zone center $\mathbf{q}=0$ are scattered into vibrations with momentum $\mathbf{G}$ with amplitude $|f^{(3)}|^2\sim (V_{\textrm{AA}})^2|\mathbf{g}_1|^6$. The density of these high momenta phonons goes as $|\mathbf{G}_1|/c_n$, so we can estimate\begin{align}
    \tau^{-1}\sim\frac{2k_B T\, \left(V_{\textrm{AA}}\right)^2}{\pi\varrho^3 c_n^5}\left(\frac{4\pi}{\sqrt{3}a}\right)^3\left(\overline{\theta}\right)^{-3}.
\end{align}
As for the disorder strength, this scaling with twist angle is expected to saturate once stacking domain walls are well defined.

The low-frequency limit of the stacking correlation function then reads (hereafter $n=L,T$):\begin{align}
\label{eq:anharmonic_function}
C_{n,n}\left(\mathbf{q},\omega\right)=\frac{2k_B T}{\omega}\,\Im\chi_{n,n}\left(\mathbf{q},\omega\right)=\frac{4k_B T}{\varrho}\frac{\tau^{-1}}{\left(\omega^2-c_n^2|\mathbf{q}|^2\right)^2+\omega^2\tau^{-2}}.
\end{align}
The poles of this correlation function are located at\begin{align}
\omega_n\left(\mathbf{q}\right)=-\frac{i\tau^{-1}}{2}\pm\sqrt{c_n^2|\mathbf{q}|^2-\frac{\tau^{-2}}{4}}\approx\begin{cases}
\pm c_{n}|\mathbf{q}|-\frac{i\tau^{-1}}{2} & \textrm{if}\,\,|\mathbf{q}|\gg\frac{1}{2\tau c_n},\\
-i\tau^{-1}, -ic_n^2\tau|\mathbf{q}|^2 & \textrm{if}\,\,|\mathbf{q}|\ll\frac{1}{2\tau c_n}.
\end{cases}
\end{align}
At short wavelengths we have damped propagating modes leading to a well defined peak in the frequency response. Note that, contrary to acoustic phonons in a crystalline lattice, damping is finite at $\mathbf{q}\rightarrow 0$. Hence, in this regime, low-frequency modes are overdamped. Phasons evolve into a relaxation and a diffusive mode. The former corresponds to the relaxation of the relative linear momentum of the layers, although its spectral weight is much smaller than the diffusive pole (the ratio of their quasiparticle residues goes as $|Z_{\textrm{relax}}/Z_{\textrm{diff}}|=\tau^2c_n^2|\mathbf{q}|^2$). In this limit, the correlation function presents a broad peak at $\omega=0$, leading to a diffusive response of the form\begin{align}
\chi_{n,n}\left(\mathbf{q}\rightarrow 0,\omega\right)\approx\frac{2\varrho^{-1}\,\tau}{-i\omega+c_n^2\tau|\mathbf{q}|^2}.
\end{align}
The physical picture that emerges from this analysis is the following: Against dynamical perturbations on stacking configurations with characteristic wavelengths longer than the moir\'e period but smaller than the \textit{relaxation length} $L_{\tau}=2\tau c_n$, the system responds by carrying out damped collective oscillations of the soliton system. For longer wavelengths, the system relaxes into the equilibrium stacking configuration $\boldsymbol{\phi}_0$ via internal diffusive processes involving anharmonic couplings with higher-frequency vibration modes. 

\end{document}